\documentclass[lettersize,journal]{IEEEtran}

\usepackage{subcaption,caption}

\usepackage{amsfonts,amssymb}
\usepackage{amsmath}
\usepackage{bm}
\usepackage{graphicx}
\usepackage{epstopdf}
\usepackage{ragged2e}

\usepackage{cite}
\usepackage{soul}
\usepackage{color,xcolor}
\usepackage{verbatim}
\usepackage{stfloats}

\begin{document}

\title{Joint  Device Activity Detection, Channel Estimation and Signal Detection for Massive Grant-free Access via BiGAMP }

\author{Shanshan~Zhang,
	Ying Cui,~\IEEEmembership{Member,~IEEE,}
	and Wen Chen,~\IEEEmembership{Senior Member,~IEEE}
	\thanks{
	This work is supported by National key project 2018YFB1801102, NSFC 62071296, Shanghai 22JC1404000, 20JC1416502, PKX2021-D02, and 20ZR1425300.{\it (Corresponding author: Wen Chen.)}
		
	Shanshan Zhang is with the Department of Electronic Engineering, Shanghai Jiao Tong University, Shanghai 200240, China (e-mail: shansz@sjtu.edu.cn).
		
	Ying Cui is with the IoT Thrust, Hong Kong University of Science and Technology Guangzhou, Guangzhou 511400, China, and also with the Department of Electronic Engineering, Hong Kong University of Science and Technology, Hong Kong, China (e-mail: yingcui@ust.hk).
		
	Wen Chen is with the Department of Electronic Engineering, Shanghai Jiao Tong University, Shanghai 200240, China (e-mail: wenchen@sjtu.edu.cn).
	}

}

\maketitle

\begin{abstract}
Massive access has been challenging for the fifth generation (5G) and beyond since the abundance of devices causes communication overload to skyrocket. In an uplink massive access scenario, device traffic is sporadic in any given coherence time. Thus, channels across the antennas of each device exhibit correlation, which can be characterized by the row sparse channel matrix structure. In this work, we develop a bilinear generalized approximate message passing (BiGAMP) algorithm based on the row sparse channel matrix structure. This algorithm can jointly detect device activities, estimate channels, and detect signals in massive multiple-input multiple-output (MIMO) systems by alternating updates between channel matrices and signal matrices. The signal observation provides additional information for performance improvement compared to the existing algorithms. We further analyze state evolution (SE) to measure the performance of the proposed algorithm and characterize the convergence condition for SE. Moreover, we perform theoretical analysis on the error probability of device activity detection, the mean square error of channel estimation, and the symbol error rate of signal detection. The numerical results demonstrate the superiority of the proposed algorithm over the state-of-the-art methods in DAD-CE-SD, and the numerical results are relatively close to the theoretical analysis results.

\end{abstract}

\begin{IEEEkeywords}
Massive grant-free access, device activity detection, signal detection, bilinear generalized approximate message passing (BiGAMP), state evolution.
\end{IEEEkeywords}

\IEEEpeerreviewmaketitle

\section{Introduction}

\IEEEPARstart{}{} The cellular Internet of Things (IoT) accelerates the expansion of the number of devices connected to base stations (BSs). Meanwhile, massive machine-based communication (mMTC) emerges as one of the critical application scenarios for wireless communication networks. As a result, massive access has become an urgent problem for the current generation of wireless communication. The main characteristics of massive access include low power, massive connectivity, and broad coverage \cite{RN175}. In massive access scenarios, many devices exist, but the device activity patterns are typically sporadic so that only a small subset of potential devices are active at any given instant \cite{RN56}. Therefore, it is a challenge to perform device activity detection, channel estimation, and signal detection (DAD-CE-SD) from a large number of devices in an efficient and timely manner.

\subsection{Related Work and Motivation}
In the existing long-term evolution (LTE), the communication system mainly adopts the grant-based random access protocol, designed for human-to-human (H2H) communication scenarios with few active devices and high transmission rate requirements. In the grant-based random access protocol, the device must connect with the BS before signal transmission. \cite{chap4_mtm,chap4_mimo} studied a contention-based protocol where each active device utilizes a signature preamble and the favorable propagation of massive multiple-input multiple-output (MIMO) channels to achieve collision detection. If any other device does not choose the selected preamble, the active device can access the BS. However, contention-based protocols suffer from potential conflicts due to many potential devices, and the contention phase may lead to excessive overhead for control signaling. Therefore, for limited pilot sequences and physical uplink shared channel (PUSCH) resources, the grant-based random access protocol is not practical in mMTC scenarios.

To support mMTC scenarios, 3GPP proposed the grant-free protocol in 2016 \cite{3gpp.36.331}. In grant-free protocol, active devices freely access the BS without waiting for any scheduling grant. In contrast to the existing grant-based protocols where pilot sequences are randomly selected at each coherence time, in grant-free protocols, each device is assigned a unique pilot sequence used for all coherence times \cite{RN56}. So the grant-free random access scheme significantly reduces the scheduling signaling overhead to support mMTC requirements. However, since the pilot sequence length is restricted by the coherence time and the number of devices, it is impossible to pre-assign orthogonal pilot sequences, as conventional orthogonal multiple access (OMA), to all the potential devices. To this end, non-orthogonal multiple access (NOMA) is proposed to combine with grant-free protocols to meet the requirements of massive access \cite{RN244, RN186, RN307}. In the grant-free NOMA scheme, devices are assigned non-orthogonal pilot sequences to reduce the pilot overhead caused by a large number of devices, and they send pilots and signals to the BS simultaneously. Then the BS identifies active devices, estimates channels, and/or detects signals in each coherence time. As MIMO is another essential technology supporting future mMTC scenarios, combining the grant-free NOMA scheme with massive MIMO can better meet mMTC's requirements. However, this will undoubtedly bring higher complexity to communication systems and such problems are usually cast into sparse signal recovery problems.

Currently, compressed sensing (CS) techniques have been widely used in signal recovery problems in communication. One of the approaches is optimization-based via convex programming, such as the least absolute shrinkage and selection operator (LASSO) \cite{RN245} and group LASSO \cite{RN281}. The alternating direction method of multipliers (ADMM) algorithm \cite{RN133} is studied to solve the LASSO problem. \cite{9374476} proposes an optimization method based on the Maximum Likelihood (ML) algorithm to detect active devices.
Apart from this, approximation algorithms are extensively used in CS and there are kinds of approximate algorithms developed to solve sparse signal recovery problems. \cite{RN164} and \cite{RN106} propose approximate message passing (AMP) algorithms to solve multiple measurement vector (MMV) problems, which consider device activity detection and channel estimation. Orthogonal AMP (OAMP) \cite{RN114} and vector AMP (VAMP) \cite{RN251} are proposed for non-independent and identically distributed (non-i.i.d) Gaussian sensing matrices. \cite{2010arXiv1010.5141R} proposes generalized AMP (GAMP) for systems with generalized output channels. Deep learning architectures are recently proposed by combining traditional CS methods and deep learning methods to design effective sparse signal recovery methods \cite{8550778, RN148, RN225}.

Although all of the above are studied to solve massive access problems, most of them divide DAD-CE-SD into two or three phases. Specifically, \cite{RN164,RN106,RN114,RN251,2010arXiv1010.5141R} first detect active devices and estimate the channels, then \cite{RN166} studies the signal detection.
\cite{RN108} develops a joint DAD-CE-SD algorithm by leveraging AMP. However, it only works for single antenna BSs. Algorithms that jointly detect device activity and data are proposed by embedding information bits into pilot sequences\cite{RN270,RN69}. But they require a lot of pilot resources and have limited data load capacity. \cite{RN157} proposes a bilinear generalized AMP (BiGAMP) algorithm, which allows for joint DAD-CE-SD. Under the assumption that all devices are activated, \cite{RN83} utilizes BiGAMP to estimate channels and detect signals jointly with constructing independent sparse signals. However, the constructed sparse signals will reduce the efficiency of receiving valid signals and increase the delay of processing signals in BS. Therefore, it is unpractical in existing systems. Since BiGAMP in \cite{RN157} is difficult to reconstruct row sparse matrices, the joint DAD-CE-SD is still an open problem.

Furthermore, extensive numerical experiments tested that the behavior of the AMP algorithm is accurately described by a formalism called ``state evolution" (SE) \cite{RN45}, which is crucial for guiding the adaptive selection of the pilot sequence length. Donoho et al. analyzed the constraint relationship between SE and AMP reconstruction accuracy \cite{RN42}. \cite{7457269} presents heuristic SE for BiGAMP based on random variables. Our work aims to describe the performance of the BiGAMP with correlation in the sparse matrix. Therefore, we construct the SE for BiGAMP based on random vectors.

\subsection{Main Contributions}
This paper focuses on the joint DAD-CE-SD in the uplink massive grant-free access system for the multi-antenna BS. By formulating the joint DAD-CE-SD as a generalized bilinear inference problem, we propose a BiGAMP algorithm to address the joint DAD-CE-SD in massive access scenarios. Different from the variable-based BiGAMP algorithm in \cite{RN157, RN83}, to obtain more information from the correlated channels caused by the sporadic device activity pattern, the proposed algorithm is constructed and derived based on random vectors. We apply the central limit theorem (CLT) and Taylor series arguments to approximate the minimum mean-squared error (MMSE) estimation of channels and signals for handling the NP-hard problem in this algorithm. Compared to the conventional algorithms that divide the DAD-CE-SD problem into two phases, we utilize the statistics and observation of the transmitted signals, which helps estimate channels and detect signals more accurately.

Then, we construct the SE of the proposed BiGAMP algorithm, which can be used to characterize the convergence performance of the algorithm. We also analyze the convergence conditions of SE for optimal performance.
Based on the analysis of SE, we study the theoretical performance of the proposed algorithm for joint DAD-CE-SD, including the error probability of device activity detection (DAD), the mean square error (MSE) of channel estimation (CE), and the symbol error rate (SER) of signal detection (SD).

Finally, we design simulations to verify the performance of the algorithm.  The numerical results demonstrate that the proposed algorithm performs better in DAD-CE-SD than the existing algorithms \cite{RN133,9374476, RN164, RN106} in general. In addition, the numerical results are close to the theoretical analysis, which shows that the theoretical analysis can characterize the performance of DAD-CE-SD to a certain extent.

\subsection{Organization}
The rest of this paper is organized as follows. Section \ref{SM} formulates the DAD-CE-SD problem as a row sparse bilinear problem. Section \ref{III} outlines an algorithm to solve the bilinear matrix estimation problem and presents the details of applying the algorithm to solve the DAD-CE-SD problem proposed in Section \ref{SM}.  Section \ref{Per} constructs SE to describe the performance of the algorithm and analyzes the theoretical performance for DAD-CE-SD. Section \ref{Num} provides the numerical results. Finally, Section \ref{Con} concludes the findings of this work.
\subsection{Notation}
Throughout this paper, random scalar variables are denoted by the normal lowercases (e.g., $\mathsf x$) and the italic lowercases (e.g., $x$) for the common scalars. Bold lowercases (e.g., $\mathbf x$) denote random vectors and bold italic lowercases (e.g., $\bm x$) for the common vectors. In the case of no ambiguity, we do not distinguish between random matrices and common matrices, and use bold uppercase (e.g., $\mathbf X$) to denote matrices. Let $\mathbf I$ denote the unit matrix. Use calligraphy uppercases (e.g., $\mathcal{N})$ to represent sets. $|\mathcal{N}|$ is the number of elements in set $\mathcal{N}$. $x_{ij}=[\mathbf X]_{i,j}$ denotes the $(i, j)$-th element of matrix $\mathbf X$. Hadamard product is denoted by $\odot$. $\propto$ denotes a positive correlation. The transpose, complex conjugate, and conjugate transpose operators are denoted by $(\cdot)^T$, $(\cdot)^*$, and $(\cdot)^H$, respectively. $\mathsf{Re}(\cdot)$ and $\mathsf{Tr}(\cdot)$ denote the real part and trace of the term, respectively. $\|\cdot\|_2$ and $\|\cdot\|_F$ denote the 2-norm and Frobenius norm, respectively. $\mathcal{CN}(x;u,v)$ denotes that the variable $x$ follows a complex Gaussian distribution with mean $u$ and variance $v$. $g(\cdot)$ and $\bm g(\cdot)$ denote functions whose output is a scalar and a vector, respectively.

\section{System Model}
\label{SM}
We consider a single-cell cellular network consisting of $N$ single-antenna IoT devices and one BS equipped with $M$ antennas. This paper adopts a narrow-band block-fading model where channels follow independent quasi-static flat-fading in each coherence time. The fading coefficient of the channel from device $n$ to the BS is denoted by $\bm{h}_n = [h_{n1},h_{n2},\ldots,h_{nM}]^T\in \mathbb{C}^{M\times1}$,  where $n\in\mathcal{N}$ and $\mathcal{N}\triangleq \{1,2,\ldots,N\}$ denotes the potential device set. We model the channel $\bm{h}_n = \sqrt{\beta_n}\bm{g}_n$, where $\beta_n$ denotes the path-loss and shadowing component. $\bm{g}_n$ is the Rayleigh fading component generated by complex Gaussian distribution $\mathcal{CN}(\mathbf 0,\mathbf I)$.

This paper considers a massive access scenario, where only a small fraction of $N$ potential devices are active and access the BS in each coherence time. Assume that all devices have the same probability $\varepsilon \in (0, 1)$ to access the BS in each coherence time with an i.i.d. manner. We use $\mathcal{K}$ ($\mathcal{K}\subset \mathcal{N}$) to denote the set of active devices and $|\mathcal{K}|=K$. For all $n\in \mathcal{N}$, let $\alpha_n\in\{0,1\}$ denote the activity indicator of device $n$, where $\alpha_n=1$ if device $n$ is active, and $\alpha_n=0$ otherwise. Thus, $
\Pr(\alpha_n=1) =
\varepsilon$, and $\Pr(\alpha_n=0) =
1-\varepsilon$.

\begin{figure}[t]
	\centering
	\begin{center}
		\includegraphics[scale=.8]{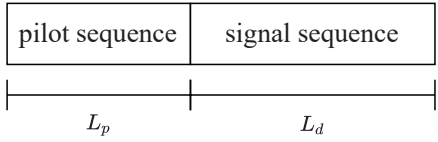}
	\end{center}
	\caption{\scriptsize{The transmitted sequence structure.}}
	\label{Fig1}
	\vspace{-0.5cm}
\end{figure}

We adopt a grant-free access scheme. Specifically, each device $n\in\mathcal{N}$ is preassigned a unique pilot sequence of length $L_p$, denoted by $\bm c_n \in\mathbb{C}^{L_p\times1} $. We set $ L_p\ll N$, then all pilot sequences are non-orthogonal. In each coherence time, each device $n$ transmits its pilot sequence and signal sequence of length $L_d$, denoted by $\bm d_n \in\mathbb{C}^{L_d\times1} $, as shown in Fig. \ref{Fig1}.
The length of the overall transmitted sequence is $L=L_p+L_d$. The sequence transmitted by device $n$ is denoted by $\bm a_n = [\bm c_{n}^T, \bm d_{n}^T]^T \in \mathbb{C}^{L\times1}$.
We assume that the signal symbols of $\bm a_n$ are uncorrelated and the entries of $\bm c_n$ are generated by i.i.d complex Gaussian distribution with zero mean and variance $1/L$.  For the Gaussian codebook \cite{RN286,RN287,RN239}, without loss of generality, we assume the signal symbol $d_{ln}$ is generated by $\mathcal{CN}(0,1/L)$.
\footnote{Other distributions on signal symbols could be estimated by the BiGAMP algorithm proposed in this paper. The numerical results in Fig. \ref{fig:4} reveal that the proposed algorithm also applies to discrete codewords in existing communication systems.}

The overall channel input-output relationship can be modeled as
 \begin{eqnarray}
 \mathbf Y = \sum_{n=1}^{N}  \bm a_n \alpha_n \bm h_{n}^T + \mathbf W,
 \label{eq:chap22}
 \end{eqnarray}
 where $\mathbf Y \in \mathbb{C}^{L\times M}$ is the received signal across $M$ antennas at the BS, and $\mathbf W\in \mathbb{C}^{L\times M}$ is the additive white Gaussian noise (AWGN) with $\mathbf w_m\sim\mathcal{CN}(\bm 0, \sigma^2\mathbf I), m=1,2,\ldots,M$. We can transform the system output (\ref{eq:chap22}) into
\begin{eqnarray}
\mathbf Y = \mathbf A \mathbf X + \mathbf W,
\label{eq:chap24}
\end{eqnarray}
where $\mathbf A = [\bm a_1,\bm a_2,\ldots,\bm a_N]\in \mathbb{C}^{L\times N}$ is the transmitted symbol. The product of activity indicators and channels are denoted by $\mathbf X = [\bm x_1,\bm x_2,\ldots,\bm x_n]^T\in \mathbb{C}^{N\times M}$, where $\bm x_n = \alpha_n \bm h_{n}$, i.e.,
\begin{eqnarray}
\bm x_n = \left\{ \begin{array}{l}
\bm h_n,\quad \hfill \alpha_n = 1\\
\bm 0,\quad \hfill \alpha_n = 0
\end{array}, n\in \mathcal{N}. \right.
\label{eq:chap25}
\end{eqnarray}
Thus, channel matrix $\mathbf X$ is a  row sparse matrix correlated in rows. Each row of $\mathbf X$ follows a Bernoulli Gaussian distribution. The probability distribution function (pdf) of $\bm x_n$ is
\begin{eqnarray}
p_{\mathbf x_n}(\bm x_n)=(1-\varepsilon)\delta_0(\bm x_n) + \varepsilon p_{\mathbf h_n}(\bm x_n),
\label{eq:chap26}
\end{eqnarray}
where $\delta_0$ denotes the point mass measured at zero, and $p_{\mathbf h_n}$ is the pdf of device $n$'s channel $\mathbf h_n \sim \mathcal{CN}(\bm 0,\beta_n \mathbf I)$.

To estimate $\mathbf X$ and signal symbols in $\mathbf A$, we develop a BiGAMP-based algorithm, which exploits the statistical characteristics of random vectors for channels and random variables for signal symbols. The proposed algorithm can implement joint DAD-CE-SD. Considering the situation of massive access scenarios, this paper studies an asymptotic regime as claim 1.

{\it Claim 1:}
The asymptotic regime means that $L,N,M\to \infty$, and $M/N$ and $L/N$  are fixed. Therefore, the number of active devices $K\to \varepsilon N$ as $N\to \infty$.

\section{The BiGAMP-based Joint Device Activity Detection, Channel Estimation, and Signal Detection}
\label{III}
\subsection{Problem Formulation}

For the above system statistical model, the pdfs of $\mathbf A$ and $\mathbf X$ are
\begin{eqnarray}
\begin{aligned}
p_{\bm{\mathsf A}}(\mathbf A)&=\prod_{l=1}^L \prod_{n=1}^N p_{\mathsf{a}_{ln}}(a_{ln}),\\
p_{\bm{\mathsf X}}(\mathbf X)&=\prod_{n=1}^{N}p_{{\mathbf{x}}_n}({\boldsymbol{x}}_n),
\label{eq:chap28}
\end{aligned}
\end{eqnarray}
and the posterior distribution of $\mathbf A$ and $\mathbf X$ is (\ref{eq:poster}),
\begin{figure*}[!t]
	\vspace*{-8pt}
	\normalsize
\begin{eqnarray}
\begin{aligned}
p_{\bm{\mathsf{X,A|Y}}}(\mathbf{X,A|Y}) \propto  p_{\bm{\mathsf{Y|X,A}}}(\mathbf{Y|X,A})p_{\bm{\mathsf X}}(\mathbf{X})p_{\bm{\mathsf A}}(\mathbf{A}) = \prod_{l=1}^{L}p_{{\mathbf{y}}_l|[\mathbf A]_{l,:},\bm{\mathsf X} }({\boldsymbol{y}}_l|\sum_{n=1}^N a_{ln}\boldsymbol{x}_n)\prod_{n=1}^{N}p_{{\mathbf{x}}_n}({\boldsymbol{x}}_n)\prod_{l=1}^L \prod_{n=1}^N p_{\mathsf{a}_{ln}}(a_{ln}).
\label{eq:poster}
\end{aligned}
\end{eqnarray}
\hrulefill
\vspace*{-8pt}
\end{figure*}
where $\mathbf{Y}=[\bm{y}_1,\bm{y}_2,\ldots,\bm{y}_L]^T$ with $\bm{y}_l\in \mathbb{C}^{M\times1}$, and $[\mathbf A]_{l,:}$ denotes the $l$-th row of $\mathbf A$.

\begin{figure}[h]
	
	\centering
	\begin{center}
		\includegraphics[scale=0.55]{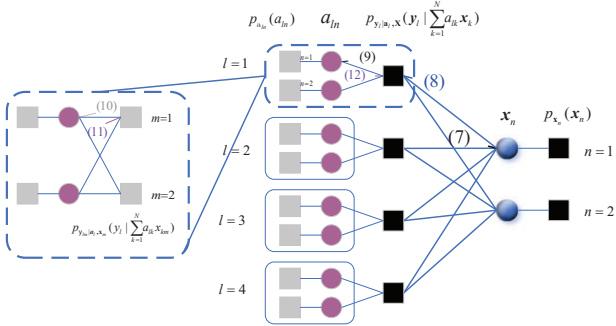}
	\end{center}
	\caption{\scriptsize{The bilinear factor graph for problem dimensions $L=4$, $M=2$, and $N=2$. The function nodes are described as ``factor nodes" denoted by squares. The random variable $a_{l,n}\in\mathbb{C}$ is described as ``variable node" denoted by a circle. The random vector $\bm x_n\in \mathbb{C}^{M\times1}$ is described as ``vector node" denoted by a ball. The update rules for the propagation of messages (\ref{eq:4B1})-(\ref{eq:palln}) are shown in the factor graph.}}
	\label{Fig2}
	\vspace{-8pt}
\end{figure}

This work aims to obtain MMSE estimates of $\mathbf X$ and $\mathbf A$ which are the means of the marginal posteriors $p_{a_{ln}|\mathbf Y}(\cdot |\mathbf Y)$ and $p_{\mathbf x_n|\mathbf Y}(\cdot |\mathbf Y)$ \cite[Section 11.4]{RN313}. Although it is generally prohibitive to compute the marginal posteriors through integrating on (\ref{eq:poster}), the marginal posteriors can be efficiently approximated by loopy belief propagation (LBP) \cite{RN221}. In LBP, the posterior distribution is usually figured with a factor graph, as shown in Fig. \ref{Fig2}. Messages of the random variables (vectors) are propagated between factor nodes and variable (vector) nodes until converging. The standard way to compute these messages is known as the sum-product algorithm (SPA) which obtains exact marginal posteriors when the factor graph has no loops \cite{RN199}. Unfortunately, it is an NP-hard problem for the loopy factor graph, so LBP can't guarantee the correct posterior pdfs. But empirical studies demonstrate that the loopy beliefs often converge and give good approximations to the correct marginals \cite{10.5555/2073796.2073849}. In high-dimensional inference problems, the complexity of the exact implementation of SPA is high, and approximations of the SPA have been applied to solve the generalized CS problem, like \cite{RN157, RN43,2010arXiv1010.5141R, RN140}.
The proposed BiGAMP algorithm employs approximations to the vector-based SPA on the bilinear factor graph in Fig. \ref{Fig2}, where we use vector node $\bm x_n$ instead of variable nodes $x_{1n}, x_{n2},\ldots,x_{nM}$ to characterize the correlation of $\bm x_n$. As we shall see, these approximations are fundamentally established by the CLT and Taylor-series arguments.

\subsection{Sum-Product Algorithm}

Since BiGAMP derives from approximations of SPA, let's first show the propagation process of messages between factor nodes and variable nodes at iteration $t\in\mathbb{Z}$. By applying the SPA to the bilinear factor graph in Fig. \ref{Fig2}, the update rules for the propagation of messages\footnote{The messages mentioned here essentially refer to probabilistic information.  Messages in (\ref{eq:4B1})-(\ref{eq:posx}) are developed from SPA that operates in Fig. \ref{Fig2}. Interested readers can refer to \cite{RN199,RN209},\cite[Section IV.26]{RN218} for more details about SPA.} are as follows:

\noindent 1) Messages between factor nodes and vector nodes:

Message from factor node $p_{{\mathbf{y}}_l|[\mathbf A]_{l,:},\mathbf{X}}\left({\boldsymbol{y}}_l|\sum_{k=1}^N a_{lk}\boldsymbol{x}_k\right)$ to vector node $\bm x_n$ can be expressed as (\ref{eq:4B1}).
\begin{figure*}[!t]
	\vspace*{-8pt}
	\normalsize
\begin{eqnarray}
\begin{aligned}
I_{l\to n}^{\mathbf{x}}(\boldsymbol{x}_n,t)\propto & \int_{[\mathbf A]_{l,:},\{{\boldsymbol{x}_r}\}_{r\ne n}} p_{{\mathbf{y}}_l|[\mathbf A]_{l,:},\mathbf{X}}\left({\boldsymbol{y}}_l|\sum_{k=1}^N a_{lk}\boldsymbol{x}_k\right) \prod_{r=1,r\ne n}^{N} I_{l\leftarrow r}^{\mathbf{x}}(\boldsymbol{x}_r,t)\prod_{k=1}^N I_{l\leftarrow lk}^{\mathsf{a}}({a}_{lk},t).
\label{eq:4B1}		
\end{aligned}
\end{eqnarray}
\vspace*{-8pt}
\end{figure*}
Message from vector node $\bm x_n$ to factor node $p_{{\mathbf{y}}_l|[\mathbf A]_{l,:},\mathbf{X}}\left({\boldsymbol{y}}_l|\sum_{k=1}^N a_{lk}\boldsymbol{x}_k\right)$ is
\begin{eqnarray}
\begin{aligned}
I_{l\leftarrow n}^{\mathbf{x}}(\boldsymbol{x}_n,t+1)\propto p_{{\mathbf{x}}_n}({\boldsymbol{x}}_n)\prod_{k=1,k\ne l}^{L} I_{k\to n}^{\mathbf{x}}(\boldsymbol{x}_n,t),
\label{eq:xntol}	
\end{aligned}
\end{eqnarray}
where $p_{{\mathbf{x}}_n}({\boldsymbol{x}}_n)$ is prior probability of $\bm {x}_n$.

\noindent 2) Messages between factor nodes and variable nodes:

Message from factor node $p_{{\mathbf{y}}_l|[\mathbf A]_{l,:},\mathbf{X}}({\boldsymbol{y}}_l|\sum_{k=1}^N a_{lk}\boldsymbol{x}_k)$ to variable node $a_{ln}$ is (\ref{eq:Iltoln}).
\begin{figure*}[!t]
	\vspace*{-8pt}
	\normalsize
\begin{eqnarray}
\begin{aligned}
I_{l\to ln}^{\mathsf{a}}({a}_{ln},t)\propto& \int_{{\{a_{lr}\}}_{r\ne n},{\mathbf{X}}} p_{{\mathbf{y}}_l|[\mathbf A]_{l,:},\mathbf{X}}\left({\boldsymbol{y}}_l|\sum_{k=1}^N a_{lk}\boldsymbol{x}_k\right) \prod_{k=1}^{N} I_{l\leftarrow k}^{\mathbf{x}}(\boldsymbol{x}_k,t)\prod_{r=1,r\ne n}^N I_{l\leftarrow lr}^{\mathsf{a}}({a}_{lr},t).	
\end{aligned}\label{eq:Iltoln}
\end{eqnarray}
\vspace*{-8pt}
\begin{eqnarray}
\begin{aligned}
I_{lm\to ln}^{\mathsf{a}}({a}_{ln},t)\propto \int_{\{a_{lr}\}_{r\ne n},\mathbf{X}_{\sim m}} p_{\mathbf{y}_{l/ m}|[\mathbf A]_{l,:},\mathbf{X}_{\sim m}}\left(\boldsymbol{y}_{l/ m}|\sum_{k=1}^N a_{lk}\boldsymbol{x}_{k/ m}\right)  \prod_{k=1}^{N} I_{l\leftarrow k}^{\mathbf{x}}(\boldsymbol{x}_{k/ m},t)\prod_{r=1,r\ne n}^N I_{l\leftarrow lr}^{\mathsf{a}}({a}_{lr},t)	.
\label{eq:l/mtoln}
\end{aligned}
\end{eqnarray}
\hrulefill
\vspace*{-8pt}
\end{figure*}
Message from variable node $a_{ln}$ to factor node $p_{{\mathbf{y}}_l|[\mathbf A]_{l,:},\mathbf{X}}({\boldsymbol{y}}_l|\sum_{k=1}^N a_{lk}\boldsymbol{x}_k)$ is slightly more complicated. According to the typical SPA, there is $I_{l\leftarrow ln}^{\mathsf{a}}({a}_{ln},t+1)\propto p_{{\mathsf{a}}_{ln}}({a}_{ln})$, which means messages from variable nodes to factor nodes cannot be updated as iterations. The above problem is caused by ignoring that each element in $\bm y_l$ may propagate different messages to $a_{ln}$ as shown in Fig. \ref{Fig2}. To this end, we assume the joint message from $p_{\mathbf{y}_{l/ m}|[\mathbf A]_{l,:},\mathbf{X}_{\sim m}}\left(\boldsymbol{y}_{l/ m}|\sum_{k=1}^N a_{lk}\boldsymbol{x}_{k/ m}\right)$ to $a_{ln}$ is (\ref{eq:l/mtoln}),
where $\bm {y}_{l/m}=[y_{l1},\ldots,y_{l(m-1)},y_{l(m+1)},\ldots,y_{lM}]^T, {\bm {x}}_{n/m}=[{x}_{n1},\ldots,x_{n(m-1)},x_{n(m+1)},\ldots,x_{nM}]^T$, $\mathbf{X}_{\sim m}=[{\mathbf{x}}_{1/m},\ldots,{\mathbf{x}}_{N/m}]^T$ and $I_{l\leftarrow n}^{\mathbf{x}}(\boldsymbol{x}_{n/ m})= \int_{x_{nm}} I_{l\leftarrow n}^{\mathbf{x}}(\boldsymbol{x}_{n})$.
Then the message from $a_{ln}$ to $p_{{y}_{lm}|{\mathbf{a}}_{l}^T\mathbf{x}_{m}}\left({y}_{lm}|\sum_{k=1}^N a_{lk}{x}_{km}\right)$ is
\begin{eqnarray}
\begin{aligned}
I_{lm\leftarrow ln}^{\mathsf{a}}({a}_{ln},t+1)\propto p_{{\mathsf{a}}_{ln}}({a}_{ln})I_{lm\to ln}^{\mathsf{a}}({a}_{ln},t),		
\label{eq:palmln}
\end{aligned}
\end{eqnarray}
where $p_{{\mathsf{a}}_{ln}}({a}_{ln})$ is the prior probability of ${{a}}_{ln}$.
Finally, we take the geometric mean of $I_{lm\leftarrow ln}^{\mathsf{a}}({a}_{ln},t+1)$ as message from variable node $a_{ln}$ to factor node $p_{{\mathbf{y}}_l|[\mathbf A]_{l,:},\mathbf{X}}({\boldsymbol{y}}_l|\sum_{k=1}^N a_{lk}\boldsymbol{x}_k)$, i.e.,
\begin{eqnarray}
\begin{aligned}
I_{l\leftarrow ln}^{\mathsf{a}}({a}_{ln},t+1)\propto \left(\prod_{m=1}^MI_{lm\leftarrow ln}^{\mathsf{a}}({a}_{ln},t)\right)^{1/M}.	
\label{eq:palln}	
\end{aligned}
\end{eqnarray}

\noindent 3) The posterior probabilities of $\bm x_n$ and $a_{ln}$ can be approximated as:
\begin{subequations}\label{eq:posx}
\begin{align}
I_{n}^{\mathbf{x}}(\boldsymbol{x}_n,t+1)\propto & p_{{\mathbf{x}}_n}({\boldsymbol{x}}_n)\prod_{k=1}^{L} I_{k\to n}^{\mathbf{x}}(\boldsymbol{x}_n,t+1) 	,	\\
I_{ln}^{\mathsf{a}}({a}_{ln},t+1)\propto & p_{{\mathsf{a}}_{ln}}({a}_{ln})I_{l\to ln}^{\mathsf{a}}({a}_{ln},t+1).		
\end{align}
\end{subequations}

Due to high-dimensional integrations, the computations of (\ref{eq:4B1})-(\ref{eq:posx}) are generally intractable.
Thus, we apply CLT and Taylor series arguments to approximate the SPA updates (\ref{eq:4B1})-(\ref{eq:posx}). These approximations will be exact in the asymptotic regime.

\subsection{Messages Approximated from Factor Nodes to Variable Nodes (F-to-V)}
\label{ftov}

Define $\mathbf{Z}\triangleq \mathbf{AX}$. Without loss of generality, we assume that $\mathbb{E}[z_{lm}^2]$ and $\mathbb{E}[x_{nm}^2]$ scale as $O(1)$. Since $\mathsf z_{lm}=\sum_{k=1}^{N}\mathsf a_{lk}\mathsf x_{km}$, $\mathbb{E}[a_{ln}^2]$ must scale as $O(1/N)$ as $N\to\infty$. So $\mathbb{E}[x_{nm}]$ scales as $O(1)$ and $\mathbb{E}[a_{ln}]$ scales as $O(1/\sqrt N)$. These assumptions hold for random variables $\mathsf a_{ln}$, $\mathsf x_{nm}$ and $\mathsf z_{lm}$ according to the prior pdfs and (\ref{eq:4B1})-(\ref{eq:posx}).

Assume $\mathbf x_{l,n}\in \mathbb{C}^{M\times1}$ is a random vector whose probability distribution is $I_{l\leftarrow n}^{\mathbf{x}}$ and its mean and covariance matrix are denoted by $\hat{\bm x}_{l,n}$ and $\bm v_{l,n}^{\mathbf x}$, respectively. Similarly, assume that $\mathsf a_{l,ln}$ is a random variable whose probability distribution is $I_{l\leftarrow ln}^{\mathsf{a}}$ with the mean $\hat{a}_{l,ln}$ and variance $v_{l,ln}^{\mathsf a}$. According to the CLT, we can characterize the pdf of $\mathbf{z}_l$ as Gaussian distribution. First, define the estimated mean ${\hat {\boldsymbol p}_{l}(t)}$ and covariance matrix $\boldsymbol{v}_{l}^{\mathsf{p}}(t)$ as

\begin{subequations}\label{eq:pmvar}
	\begin{align}
{\hat {\boldsymbol p}_{l}(t)}&=\sum_{k=1}^{N}\hat{a}_{l,lk}(t)\hat{\boldsymbol{x}}_{l,k}(t),\label{eq:pmvara}\\
\boldsymbol{v}_{l}^{\mathsf{p}}(t)&=\sum_{k=1}^{N}|\hat{a}_{l,lk}(t)|^2{\boldsymbol{v}}_{l,k}^\mathbf{x}(t)+v_{l,lk}^{\mathsf{a}}(t)\hat{\boldsymbol{x}}_{l,k}(t)\hat{\boldsymbol{x}}_{l,k}^H(t)\nonumber\\&\quad +v_{l,lk}^{\mathsf{a}}(t) {\boldsymbol{v}}_{l,k}^\mathbf{x}(t), \label{eq:pmvarb}
\end{align}
\end{subequations}
where ${\hat {\boldsymbol p}_{l}(t)}$ and $\boldsymbol{v}_{l}^{\mathsf{p}}(t)$ scale as $O(1)$. Then, define the conditional pdf
\begin{equation}
p_{\mathbf{z}_l|\mathbf{p}_l}(\boldsymbol{z}_l|\hat{\boldsymbol p}_l(t);\boldsymbol{v}_l^{\mathsf p}(t))\triangleq \frac{1}{C_{\bm z}}p_{\mathbf{y}_l|\mathbf{z}_l}(\boldsymbol{y}_l|{\boldsymbol z}_l)\mathcal{CN}(\boldsymbol{z}_l;\hat{\boldsymbol p}_l(t),\boldsymbol{v}_l^{\mathsf p}(t)),
\label{eq:pp}
\end{equation}
where $C_{\bm z}=\int_{\boldsymbol z}p_{\mathbf{y}_l|\mathbf{z}_l}(\boldsymbol{y}_l|{\boldsymbol z}_l)\mathcal{CN}(\boldsymbol{z}_l;\hat{\boldsymbol p}_l(t),\boldsymbol{v}_l^{\mathsf p}(t)) $. 
After approximating $\mathbf z_l$ as Gaussian distribution, the estimated mean and covariance matrix under the observation $\bm y_l$ are
\begin{subequations}\label{eq:pvar}
\begin{align}
\hat{\boldsymbol{z}}_l(t)& = \mathbb{E}[\boldsymbol{z}_l|\hat{\boldsymbol p}_l(t);\boldsymbol{v}_l^{\mathsf p}(t)]\triangleq \bm g_{\mathbf{z}}(\hat{\boldsymbol p}_l(t),\boldsymbol{v}_l^{\mathsf p}(t)),\\
\boldsymbol{v}_l^{\mathsf z}(t)& = \mathsf{var}[\boldsymbol{z}_l|\hat{\boldsymbol p}_l(t);\boldsymbol{v}_l^{\mathsf p}(t)]= \boldsymbol{v}_l^{\mathsf p}(t)\nabla_{ \boldsymbol u} \boldsymbol g_{\mathbf{z}}(\hat{\boldsymbol p}_l(t),\boldsymbol{v}_l^{\mathsf p}(t)),
\end{align}
\end{subequations}
where $\nabla_{ \boldsymbol u} \boldsymbol g_{\mathbf{z}}({\boldsymbol u},\boldsymbol \Sigma)$ is the gradient of $\bm g_{\mathbf z}$ with respect to the first parameter term.
Through Gaussian approximations and a Taylor expansion at point $\hat {\bm x}_n(t)$, ${I}_{l\to n}^{\mathbf{x}}(\boldsymbol{x}_n,t)$ can be approximated as
\begin{eqnarray}
\begin{aligned}
&{I}_{l\to n}^{\mathbf{x}}(\boldsymbol{x}_n,t)
\approx const\\&\cdot\exp\left(  {\mathsf{Re}\left[{2\boldsymbol{x}_n^H\left({\hat a}_{l,ln}^{*}(t)\hat{\boldsymbol s}_l(t) +|\hat{a}_{ln}(t)|^2 \boldsymbol{v}_{l}^\mathsf{s}(t)\hat{\boldsymbol{x}}_{n}(t)\right)+}\right.}\right.\\&\left.{\left.{ {\boldsymbol{x}_n^H}\left(v_{ln}^{\mathsf{a}}(t)\left(\hat{\boldsymbol s}_l(t)\hat{\boldsymbol s}_l^H(t) -\boldsymbol{v}_{l}^\mathsf{s}(t)\right)-|\hat{a}_{ln}(t)|^2\boldsymbol{v}_{l}^\mathsf{s}(t)\right)\boldsymbol{x}_n}\right]}\right),
\label{eq:21}
\end{aligned}
\end{eqnarray}
where
\vspace*{-8pt}
\begin{subequations}\label{eq:s}
	\begin{align}
	\hat{\boldsymbol s}_l(t)&= \boldsymbol{v}_{l}^\mathsf{p}(t)^{-1}(\hat{\boldsymbol{z}}_l(t)-\hat {\boldsymbol p}_{l}(t)),\label{eq:s1}\\ \boldsymbol{v}_{l}^\mathsf{s}(t)&= \boldsymbol{v}_{l}^\mathsf{p}(t)^{-1}(\mathbf I-\boldsymbol{v}_{l}^\mathsf{z}(t)\boldsymbol{v}_{l}^\mathsf{p}(t)^{-1}).
	\end{align}
\end{subequations}
$\hat{\boldsymbol s}_l(t)$ is the scaled residual for the posterior estimate $\hat{\boldsymbol{z}}_l(t)$ and $\boldsymbol{v}_{l}^\mathsf{s}(t)$ is the inverse-residual-covariance.
The $const$ represents a constant such that the integral of the pdf is $1$. The detailed derivation of ${I}_{l\to n}^{\mathbf{x}}(\boldsymbol{x}_n,t)$  is presented in Appendix \ref{x}.

The derivation of $I_{l\to ln}^{\mathsf{a}}(a_{ln},t)$
is similar to the derivation of ${I}_{l\to n}^{\mathbf{x}}(\boldsymbol{x}_n,t)$. In particular, using Gaussian approximations according to CLT and Taylor-series expansions, ${I}_{l\to ln}^{\mathsf{a}}({a}_{ln},t)$ is approximated as (\ref{eq:ltoln}).
\begin{equation}
\begin{aligned}
&{I}_{l\to ln}^{\mathsf{a}}({a}_{ln},t)\approx const \\&\cdot\exp (
\mathsf{Re}[2a_{ln}^*\left(\hat{\boldsymbol s}_l^T(t)\hat{\boldsymbol{x}}_{l,n}^*(t) +\mathsf{Tr}\left(\boldsymbol{v}_{l}^\mathsf{s}(t)\hat{\boldsymbol{x}}_{n}^*(t)\hat{\boldsymbol{x}}_{n}^T(t)\right)\hat{a}_{ln}(t)\right) \\&  -|a_{ln}|^2\mathsf{Tr}(\boldsymbol{v}_{l}^\mathsf{s}(t)\hat{\boldsymbol{x}}_{n}^*(t)\hat{\boldsymbol{x}}_{n}^T(t)-\left(\hat{\boldsymbol s}_l(t)\hat{\boldsymbol s}_l^H(t)-\boldsymbol{v}_{l}^\mathsf{s}(t)\right)^T\boldsymbol v_{n}^{\mathbf{x}}(t))]) .
\label{eq:ltoln}	
\end{aligned}
\end{equation}

\subsection{Messages Approximated from Variable Nodes to Factor Nodes (V-to-F)}

In Section \ref{ftov}, we obtain the approximation of ${I}_{l\to n}^{\mathbf{x}}(\boldsymbol{x}_n,t)$. Now, we try to approximate ${I}_{l\leftarrow n}^{\mathbf{x}}(\boldsymbol{x}_n,t)$ according to (\ref{eq:xntol}) and  (\ref{eq:21}). The ${I}_{l\leftarrow n}^{\mathbf{x}}(\boldsymbol{x}_n,t)$ can be written as
\begin{eqnarray}
\begin{aligned}
&{I}_{l\leftarrow n}^{\mathbf{x}}(\boldsymbol{x}_n,t+1) 
\\&\approx p_{\mathbf{x}_n}\left(\boldsymbol x_n\right) \cdot \exp\left({\mathsf{Re}\left[{-\left(\boldsymbol x_n - \hat{\boldsymbol r}_{l,n}(t)\right)^H}\right.}\right.\\&\qquad\qquad\qquad\left.{\left.{\boldsymbol v_{l,n}^{\mathsf r}(t)^{-1}\left(\boldsymbol x_n - \hat{\boldsymbol r}_{l,n}(t)\right)}\right]}\right)\cdot const
\\&=p_{\mathbf{x}_n}(\boldsymbol x_n)\mathcal{CN}\left(\boldsymbol x_n; \hat{\boldsymbol r}_{l,n}\left(t\right),\boldsymbol v_{l,n}^{\mathsf r}\left(t\right)\right)\cdot const,
\end{aligned}\
\end{eqnarray}
where
\begin{subequations}\label{eq:Ermn}
\begin{align}
\boldsymbol v_{l,n}^{\mathsf r}(t) &\triangleq \left({\sum_{k=1,k\ne l}^{L} |\hat{a}_{kn}(t)|^2\boldsymbol{v}_{k}^\mathsf{s}(t)}\right.\nonumber\\&\qquad\qquad\left.{-v_{kn}^{\mathsf{a}}(t)\left(\hat{\boldsymbol s}_k(t)\hat{\boldsymbol s}_k^H(t) -\boldsymbol{v}_{k}^\mathsf{s}(t)\right)}\right)^{-1},
\\
\hat{\boldsymbol r}_{l,n}(t) &\triangleq \boldsymbol v_{l,n}^{\mathsf r}(t) \left({\sum_{k=1,k\ne l}^{L}{\hat a}_{k,kn}^{*}(t)\hat{\boldsymbol s}_k(t) }\right.\nonumber\\&\qquad\qquad\qquad\left.{+ |\hat{a}_{kn}(t)|^2\boldsymbol{v}_{k}^\mathsf{s}(t)\hat{\boldsymbol{x}}_{n}(t)}\right).
\end{align}
\end{subequations}
By adopting a MMSE denoiser, we have
\begin{subequations}
\begin{align}
\hat{\boldsymbol x}_{l,n}(t+1) &= \boldsymbol g_{\mathbf{x}}\left(\hat{\boldsymbol r}_{l,n}\left(t\right),\boldsymbol v_{l,n}^{\mathsf r}\left(t\right)\right) \\& \triangleq \frac{1}{C_{\bm x_{l,n}}}\int_{\boldsymbol x}\boldsymbol x p_{\mathbf{x}_n}(\boldsymbol x)\mathcal{CN}\left(\boldsymbol x; \hat{\boldsymbol r}_{l,n}(t),\boldsymbol v_{l,n}^{\mathsf r}(t)\right), \nonumber\\
\boldsymbol v_{l,n}^{\mathbf{x}}(t+1) &\triangleq \frac{1}{C_{\bm x_{l,n}}}\int_{\boldsymbol x} (\boldsymbol x-\hat{\boldsymbol x}_{l,n}(t+1))(\boldsymbol x-\hat{\boldsymbol x}_{l,n}(t+1))^H \notag\\&\qquad\qquad\cdot p_{\mathbf{x}_n}(\boldsymbol x)\mathcal{CN}(\boldsymbol x; \hat{\boldsymbol r}_{l,n}(t),\boldsymbol v_{l,n}^{\mathsf r}(t))\nonumber\\& = \boldsymbol v_{l,n}^{\mathsf r}(t)\nabla_{ \boldsymbol u} \boldsymbol g_{\mathbf{x}}(\hat{\boldsymbol r}_{l,n}(t),\boldsymbol v_{l,n}^{\mathsf r}(t)),
\label{eq:xvar}
\end{align}
\end{subequations}
where ${C_{\bm x_{l,n}}}=\int_{\boldsymbol x} p_{\mathbf{x}_n}(\boldsymbol x)\mathcal{CN}(\boldsymbol x; \hat{\boldsymbol r}_{l,n}(t),\boldsymbol v_{l,n}^{\mathsf r}(t))$. 
Like (\ref{eq:Ermn}), define
\begin{subequations}\label{eq:Ern}
\begin{align}
\boldsymbol v_{n}^{\mathsf r}(t) &\triangleq \left({\sum_{k=1}^{L} |\hat{a}_{kn}(t)|^2\boldsymbol{v}_{k}^\mathsf{s}(t)}\right.\nonumber\\&\qquad\qquad\left.{-v_{kn}^{\mathsf{a}}(t)\left(\hat{\boldsymbol s}_k(t)\hat{\boldsymbol s}_k^H(t) -\boldsymbol{v}_{k}^\mathsf{s}(t)\right)}\right)^{-1},\label{eq:Erna}
\\
\hat{\boldsymbol r}_{n}(t) &\triangleq \boldsymbol v_{n}^{\mathsf r}\left(t\right) \sum_{k=1}^{L}\left({\hat a}_{k,kn}^{*}(t)\hat{\boldsymbol s}_k\left(t\right) + |\hat{a}_{kn}\left(t\right)|^2\boldsymbol{v}_{k}^\mathsf{s}\left(t\right)\hat{\boldsymbol{x}}_{n}\left(t\right)\right).\label{eq:Ernb}
\end{align}
\end{subequations}
Comparing (\ref{eq:Ern}) with (\ref{eq:Ermn}), there is
\begin{subequations}
\begin{align}
\boldsymbol v_{l,n}^{\mathsf r}(t) =& \boldsymbol v_{n}^{\mathsf r}(t)+O( {1/N}),\\
\hat{\boldsymbol r}_{l,n}(t) = &\hat{\boldsymbol r}_{n}(t)-\boldsymbol v_{l,n}^{\mathsf r}(t)\hat{a}_{ln}^*(t) \hat{\boldsymbol s}_l(t)+O({1/N}).
\end{align}
\end{subequations}
Expanding $\hat{\boldsymbol{x}}_{l,n}(t+1)$ at $\hat{\boldsymbol{r}}_n(t)$ by Taylor series, it shows
\begin{eqnarray}
\begin{aligned}
&\hat{\boldsymbol x}_{l,n}(t+1) \\& =\boldsymbol g_{\mathbf{x}}\left(\hat{\boldsymbol r}_{l,n}(t),\boldsymbol v_{l,n}^{\mathsf r}(t)\right)\\&=\boldsymbol g_{\mathbf{x}}\left({\hat{\boldsymbol r}_{n}(t)-\boldsymbol v_{l,n}^{\mathsf r}(t)\hat{a}_{ln}^*(t) \hat{\boldsymbol s}_l(t)+O\left({1/N}\right),}\right.\\&\qquad\left.{\boldsymbol v_{n}^{\mathsf r}(t)+O\left({1/N}\right)}\right)\\& \approx \boldsymbol g_{\mathbf{x}}\left(\hat{\boldsymbol r}_{n}(t),\boldsymbol v_{n}^{\mathsf r}(t)\right)-\\&\qquad 2\mathsf{Re}\left[\left(\boldsymbol v_{n}^{\mathsf r}(t)\hat{a}_{ln}^*(t) \hat{\boldsymbol s}_l(t)\right)^H\nabla_{ \boldsymbol u^*} \boldsymbol g_{\mathbf{x}}\left(\hat{\boldsymbol r}_{n}(t),\boldsymbol v_{n}^{\mathsf r}(t)\right)\right]^H\\& =\hat{\boldsymbol x}_{n}(t+1)- 2\mathsf{Re}\left[\left(\hat{a}_{ln}^*(t) \hat{\boldsymbol s}_l(t)\right)^H\boldsymbol v_{n}^{\mathbf x}(t+1)\right]^H,
\label{eq:xn}
\end{aligned}
\end{eqnarray}
where
\begin{subequations}\label{eq:Ex}
\begin{align}
\hat{\boldsymbol x}_{n}(t+1)&\triangleq \boldsymbol g_{\mathbf{x}}(\hat{\boldsymbol r}_{n}(t),\boldsymbol v_{n}^{\mathsf r}(t))\\&=\frac{1}{C_{\bm x}}\int_{\boldsymbol x}\boldsymbol x p_{\mathbf{x}_n}(\boldsymbol x)\mathcal{CN}(\boldsymbol x; \hat{\boldsymbol r}_{n}(t),\boldsymbol v_{n}^{\mathsf r}(t)),
\\
\boldsymbol v_{n}^{\mathbf{x}}(t+1) &\triangleq \boldsymbol v_{n}^{\mathsf r}(t)\nabla_{ \boldsymbol u} \boldsymbol g_{\mathbf{x}}\left(\hat{\boldsymbol r}_{n}(t),\boldsymbol v_{n}^{\mathsf r}(t)\right),
\end{align}
\end{subequations}
and ${C_{\bm x}}=\int_{\boldsymbol x} p_{\mathbf{x}_n}(\boldsymbol x)\mathcal{CN}(\boldsymbol x; \hat{\boldsymbol r}_{n}(t),\boldsymbol v_{n}^{\mathsf r}(t))$.
$\hat{\boldsymbol x}_{n}(t+1)$ and $\boldsymbol v_{n}^{\mathbf{x}}(t+1)$ are obtained by the MMSE denoiser $\bm g_{\mathbf x}$. 
Eq. (\ref{eq:xn}) confirms that $\hat{\boldsymbol x}_{n}(t)-\hat{\boldsymbol x}_{l,n}(t)$ scales as $O(1/\sqrt N)$. Similarly, using Taylor series expansion for $\boldsymbol v_{l,n}^{\mathbf{x}}(t+1)$ in (\ref{eq:xvar}) at $\hat{\boldsymbol r}_{n}(t)$ in the first argument and $\boldsymbol v_{n}^{\mathsf r}(t)$ in the second argument, the result confirms that $\boldsymbol v_{n}^{\mathbf{x}}(t)-\boldsymbol v_{l,n}^{\mathbf{x}}(t)$ scales as $O(1/\sqrt N)$ .

Similar to the above procedure to derive an approximation to $ {I}_{l\leftarrow ln}^{\mathsf{a}}({a}_{ln},t+1)$, whose corresponding mean is then further approximated as
\begin{eqnarray}
\begin{aligned}
&\hat{a}_{l,ln}(t+1)\\ &  \approx  \hat{a}_{ln}(t+1) - 2\mathsf{Re}\left[\frac{1}{M}\hat{\bm x}_{n}^H(t)\hat{\bm s}_{l}(t)v_{ln}^{\mathsf a}(t+1)\right].
\label{eq:xn2}
\end{aligned}
\end{eqnarray}
for
\begin{subequations}\label{eq:avar}
\begin{align}
&\hat{a}_{ln}(t+1)= g_{\mathsf{a}}(\hat{ q}_{ln}(t),v_{ln}^{\mathsf q}(t)) \\&\qquad\qquad\triangleq \frac{1}{C_a}\int_{a}a p_{\mathsf a_{ln}}(a)\mathcal{CN}(a; \hat{ q}_{ln}(t),v_{ln}^{\mathsf q}(t)),\\
&v_{ln}^{a}(t+1) = v_{ln}^{\mathsf q}(t) \nabla_{u}g_{\mathsf{a}}(\hat{ q}_{ln}(t),v_{ln}^{\mathsf q}(t)) \\&\triangleq \frac{1}{C_a}\int_{a}|a-\hat{a}_{ln}(t+1)|^2 p_{\mathsf a_{ln}}(a)\mathcal{CN}(a; \hat{ q}_{ln}(t),v_{ln}^{\mathsf q}(t)),
\end{align}
\end{subequations}
where $C_a=\int_{a} p_{\mathsf a_{ln}}(a)\mathcal{CN}(a; \hat{ q}_{ln}(t),v_{ln}^{\mathsf q}(t))$ and
\begin{subequations}\label{Eq:qvar}
\begin{align}
v_{ln}^{\mathsf q}(t)&=\mathsf{Tr}\left({\boldsymbol{v}_{l}^\mathsf{s}(t)\hat{\boldsymbol{x}}_{n}^*(t)\hat{\boldsymbol{x}}_{n}^T(t)}\right.\nonumber\\&\qquad\left.{-\left(\hat{\boldsymbol s}_l(t)\hat{\boldsymbol s}_l^H(t)-\boldsymbol{v}_{l}^\mathsf{s}(t)\right)^T \boldsymbol v_{n}^{\mathbf{x}}(t)}\right)^{-1},
\\
\hat{ q}_{ln}\left(t\right)&=v_{ln}^{\mathsf q}\left(t\right)\left({\hat{\boldsymbol s}_l^T\left(t\right)\hat{\boldsymbol{x}}_{l,n}^*\left(t\right)}\right.\nonumber\\&\qquad\left.{ +\mathsf{Tr}\left(\boldsymbol{v}_{l}^\mathsf{s}\left(t\right)\hat{\boldsymbol{x}}_{n}\left(t\right)^*\hat{\boldsymbol{x}}_{n}^T\left(t\right)\right)\hat{a}_{ln}\left(t\right)}\right).
\end{align}
\end{subequations}
According to (\ref{Eq:qvar}),  $v_{ln}^{\mathsf q}$ scales as $O(1/N)$. Hence, the difference of $\hat a_{ln}(t)-\hat a_{l,ln}$ scales $O\left(1/N\right)$. Likewise, it can be verified that $v_{ln}^{\mathsf a}(t)-v_{l,ln}^{\mathsf a}(t)$ scales $O\left(1/N^{2/3}\right)$.

\subsection{Message Passing Loop}
Finally, we try to close the message passing loop to achieve iterations. Plugging (\ref{eq:xn}) and (\ref{eq:xn2}) into (\ref{eq:pmvar}) in Appendix \ref{simplify}, we have
\begin{subequations}\label{eq:vphat}
\begin{align}
{\hat {\boldsymbol p}_{l}(t)}\approx &\bar {\boldsymbol p}_{l}(t) -\bar{\boldsymbol{v}}_{l}^{\mathsf{p}}(t)\hat{\boldsymbol s}_l(t-1),\\
\boldsymbol{v}_{l}^{\mathsf{p}}(t)
\approx&\bar {\boldsymbol v}_{l}^{\mathsf p}(t)+\sum_{k=1}^{N}v_{lk}^{\mathsf a}(t)\boldsymbol v_{k}^{\mathbf x}(t),	
\end{align}
\end{subequations}
where
\begin{subequations}\label{eq:barp}
\begin{align}
\bar {\boldsymbol p}_{l}(t)\triangleq& \sum_{k=1}^{N}\hat{a}_{lk}(t)\hat{\boldsymbol x}_{k}(t),\label{eq:barp1}\\\bar {\boldsymbol v}_{l}^{\mathsf p}(t)\triangleq& \sum\limits_{k=1}^{N}|\hat{a}_{lk}(t)|^2\boldsymbol v_{k}^{\mathbf x}(t)+v_{lk}^{\mathsf a}(t)\hat{\boldsymbol x}_{k}(t)\hat{\boldsymbol x}_{k}^H(t).
\end{align}
\end{subequations}
$\bar{\bm p}_l(t)$ and $\bar{\bm v}^{\mathsf p}_l(t)$ are estimates of the matrix product $[\mathbf {AX}]_{l,:}$ and the corresponding covariance matrix, respectively. Eq. (\ref{eq:vphat}) adopts Onsager correction to obtain ${\hat {\boldsymbol p}_{l}(t)}$ and $\boldsymbol{v}_{l}^{\mathsf{p}}(t)$.

Plugging (\ref{eq:xn2}) and (\ref{eq:Erna}) into (\ref{eq:Ernb}), we have
\begin{eqnarray}
\begin{aligned}
\hat{\boldsymbol r}_{n}(t) &\approx \boldsymbol v_{n}^{\mathsf r}(t) \left(\sum_{k=1}^{L}{\hat a}_{kn}^{*}\hat{\boldsymbol s}_k(t)\right) \\&+ \left(\mathbf I-\sum_{k=1}^{L}\boldsymbol v_{n}^{\mathsf r}(t)v_{kn}^{a}(t)\boldsymbol{v}_{k}^\mathsf{s}(t)\right)\hat{\boldsymbol x}_{n}(t).
\label{eq:Er}		
\end{aligned}
\end{eqnarray}
According to the definition of $\hat {\bm s}_l$ and $\bm v_l^{\mathsf s}$ in Appendix \ref{x}, Appendix B in \cite{RN157} proved that $\hat{\boldsymbol s}_l(t)\hat{\boldsymbol s}_l^H(t)-\boldsymbol{v}_{l}^\mathsf{s}(t)$ approximates to be zero-valued. Then (\ref{eq:Erna}) is simplified as
\begin{eqnarray}
\begin{aligned}
\boldsymbol v_{n}^{\mathsf r}(t) \approx \left(\sum_{k=1}^{L} |\hat{a}_{kn}(t)|^2\boldsymbol{v}_{k}^\mathsf{s}(t)\right)^{-1}.
\label{eq:xva}
\end{aligned}
\end{eqnarray}
$\hat{\bm r}_{n}(t)$ can be interpreted as the observation (i.e., $\mathbf r_n =\hat{\bm r}_{n}(t)$)  of the true $\bm x_{n}$ plus the white Gaussian  noise with covariance matrix $\bm{ v}^{\mathsf r}_{ n}(t)$. The relationship is like $\mathbf r_n=\mathbf x_n+\mathbf w_n^{\mathsf r}$, where $\mathbf w^{\mathsf r}\sim \mathcal{CN}(\bm 0,\bm v_n^{\mathsf r}(t))$. Therefore, $\bm g_{\mathbf x}$ is a MMSE denoiser that estimates $\hat {\bm x}_n(t)$ under the observation $\hat{\bm r}_n(t)$. Similarly, we can obtain
\begin{align}
\hat{ q}_{ln}(t)&\approx v_{ln}^{\mathsf q}(t)\hat{\boldsymbol s}_l^T(t)\hat{\boldsymbol{x}}_{n}^*(t)\nonumber\\&\qquad + (1-v_{ln}^{\mathsf q}(t)\mathsf{Tr}\left(\boldsymbol v_{n}^{\mathbf{x}}(t)\boldsymbol{v}_{l}^\mathsf{s}(t)\right))\hat{a}_{ln}(t),
\label{eq:Eq}
\\
v_{ln}^{\mathsf q}(t) &\approx \mathsf{Tr}\left(\boldsymbol{v}_{l}^\mathsf{s}(t)\hat{\boldsymbol{x}}_{n}^*(t)\hat{\boldsymbol{x}}_{n}^T(t)\right)^{-1}.
\label{eq:qvar}
\end{align}
$\hat{q}_{ln}(t)$ also can be interpreted as the  observation (i.e., $\mathsf q_{ln}=\hat{q}_{ln}(t)$) of the true $ a_{ln}$ plus the white Gaussian noise  with variance ${ v}^{\mathsf q}_{ln}(t)$, i.e., $\mathsf q_{ln}=\mathsf a_{ln}+\mathsf w_{ln}^{\mathsf q}$,  and $\mathsf w_{ln}^{\mathsf q}\sim \mathcal{CN}(0, v_{ln}^{\mathsf q}(t))$. $g_{\mathsf a}$ is also a MMSE denoiser which estimates $\hat a_{ln}(t)$ under the observation $\hat{q}_{ln}(t)$.

\subsection{Joint DAD-CE-SD Based on the Proposed BiGAMP}
Considering massive access scenarios and the system model in Section \ref{SM}, we can give specific forms of function $\bm g_{\mathbf z}$, $\bm g_{\mathbf x}$, and $g_{\mathsf a}$ and do some simplifications.

{\it Assumption 1:} In Section \ref{SM}, random variables $\mathsf a_{ln}$ and random vectors $\mathbf x_n$ are independent of each other for all $l, n$, and random variables $\mathsf x_{n1}, \ldots, \mathsf x_{nM}$ are i.i.d under the condition that device $n$ is active. In the asymptotic regime, the covariance matrix $\bm v_{n}^{\mathbf x}$ is a diagonal matrix with the same diagonal elements and can be expressed as $\bm v_{n}^{\mathbf x} = v_{n}^{\mathsf x}\mathbf I$.
Similarly, $\bar{\bm v_{l}}^{\mathsf p} = \bar{ v_{l}}^{\mathsf p}\mathbf I$,  $\bm v_{l}^{\mathsf p} = v_{l}^{\mathsf p}\mathbf I$,  $ \bm v_{l}^{\mathsf z} =   v_{l}^{\mathsf z}\mathbf I$, $ \bm v_{l}^{\mathsf s} = v_{l}^{\mathsf s}\mathbf I$, and $ \bm v_{n}^{\mathsf r} = v_{n}^{\mathsf r}\mathbf I$.

Considering the AWGN output channel, according to (\ref{eq:pvar}) and {\it Assumption 1}, the output estimate $\bm z_l(t)$ and variance $ v_n^{\mathsf z}(t)$ are
\begin{subequations}\label{eq:pvar1}
\begin{align}
\hat{\boldsymbol{z}}_l(t)=&\left(\sigma^2+{v}_l^{\mathsf p}(t)\right)^{-1}\left({v}_l^{\mathsf p}(t)\boldsymbol{y}_l+\sigma^2\hat{\boldsymbol p}_l(t)\right),\label{eq:pvar11}\\
{v}_l^{\mathsf z}(t) =&\sigma^2\left(\sigma^2+{v}_l^{\mathsf p}(t)\right)^{-1}{v}_l^{\mathsf p}(t).
\end{align}
\end{subequations}

According to (\ref{eq:Ex})-(\ref{eq:avar}), the $\bm g_{\mathbf x}$ and $g_{\mathsf a}$ are MMSE denoisers to estimate channels and signals. Since the pilot sequences are known at the BS, we have $\hat{a}_{ln}(t) = c_{ln}$ and $v_{ln}^{\mathsf a}(t)=0$ for $l\leq L_p$ according to (\ref{eq:avar}). For Gaussian codewords, when $l> L_p$, the estimate $\hat{a}_{ln}(t)$ and variance $v_{ln}^{\mathsf a}(t)$ are
\begin{eqnarray}
\begin{aligned}
\hat{a}_{ln}(t+1)& = \frac{\hat q_{ln}(t) } {1+Lv_{ln}^{\mathsf q}(t)},
\\
v_{ln}^{a}(t+1) & = \frac{v_{ln}^{\mathsf q}(t)}{1+Lv_{ln}^{\mathsf q}(t)}.
\label{eq:avard}
\end{aligned}
\end{eqnarray}

\textbf{Proposition 1}: For a Bernoulli Gaussian distribution like (\ref{eq:chap26}), the estimate $\hat{\bm x}_n(t+1)$ through MMSE denoiser $\bm g_{\mathbf x}$ is
\begin{eqnarray}
\begin{aligned}
\hat{\boldsymbol x}_{n}(t+1) &= \boldsymbol g_{\mathbf{x}}(\hat{\boldsymbol r}_{n}(t), v_{n}^{\mathsf r}(t)\mathbf I) \\&=\beta_n\phi\left(\hat{\boldsymbol r}_{n}\left(t\right)\right)\left(\beta_n+ v_{n}^{\mathsf r}\left(t\right)\right)^{-1}\hat{\boldsymbol r}_{n}\left(t\right),
\label{eq:xhat}
\end{aligned}
\end{eqnarray}
where
\begin{eqnarray}
\begin{aligned}
\phi\left(\hat{\boldsymbol r}_{n}\left(t\right)\right)=\frac{1}{1+\frac{1-\varepsilon}{\varepsilon}\exp\left(-M\psi_n(t)\right)},
\label{eq:thet}
\end{aligned}
\end{eqnarray}
\begin{eqnarray}
\begin{aligned}
\psi_n(t)=&\left(\frac{1}{v_{n}^{\mathsf r}(t)}-\frac{1}{\beta_n+ v_{n}^{\mathsf r}(t)}\right)\frac{\hat{\boldsymbol r}_{n}^H(t)\hat{\boldsymbol r}_{n}(t)}{M}\\&-\log\left(1+\frac{\beta_n}{v_{n}^{\mathsf r}(t)}\right).
\label{eq:xhat2}
\end{aligned}
\end{eqnarray}
The variance is
\begin{eqnarray}
\begin{aligned}
& v_{n}^{\mathsf{x}}(t+1) \\ & = \frac{1-\varepsilon}{\varepsilon}\beta_n^2\phi^2(\hat{\boldsymbol r}_{n}(t))\exp\left(-M\psi_n(t)\right)\frac{\hat{\boldsymbol r}_{n}^H(t)\hat{\boldsymbol r}_{n}(t)}{M(\beta_n+ v_{n}^{\mathsf r}(t))^{2}}\\&\quad +\beta_n v_{n}^{\mathsf r}(t)\phi(\hat{\boldsymbol r}_{n}(t))\left(\beta_n+ v_{n}^{\mathsf r}(t)\right)^{-1}
\label{eq:varx}
\end{aligned}
\end{eqnarray}

{\it{Proof:}} Please refer to Appendix \ref{appendix:p2}.

According to (\ref{eq:active}) in Appendix \ref{appendix:p2}, $\phi(\hat{\boldsymbol r}_{n}(t))$ describes the estimated probability that device $n$ is active. Examining the above non-linear functional form of the MMSE denoiser (\ref{eq:xhat})-(\ref{eq:xhat2}), it is worth noting that if device $n$ is active, $\phi(\hat{\boldsymbol r}_{n}(t))$ tends to $1$. Otherwise, it tends to $0$. As a result, the algorithm adopts a threshold strategy for activity detection, and the proposed activity detector and channel estimator are as follows.

\textbf{Definition 1}: For each device $n$, after $t$ iterations, the device activity detector is defined as
\begin{eqnarray}
\begin{aligned}
\hat{\alpha}_{n,t} = \left\{ \begin{array}{l}
1,\quad \hfill \phi(\hat{\boldsymbol r}_{n}(t)) > \varepsilon\\
0,\quad \hfill \phi(\hat{\boldsymbol r}_{n}(t)) \leq \varepsilon
\end{array}. \right.
\end{aligned}
\end{eqnarray}

From (\ref{eq:active}), the estimated active probability of device $n$ is $\phi(\hat{\boldsymbol r}_{n}(t))$. When $\phi(\hat{\boldsymbol r}_{n}(t))$ is larger than the prior activity probability $\varepsilon$, the device $n$ is considered to be active. Otherwise, it is inactive. For active device $k$, its channel and signal are estimated as:
\begin{eqnarray}
\begin{aligned}
\hat{\boldsymbol h}_{k,t} = \hat{\boldsymbol x}_{k}(t),\quad \hat{\boldsymbol d}_{k,t} = [\hat{a}_{L_p+1,k}(t),\ldots,\hat{a}_{L,k}(t)].
\label{eq:esHA}
\end{aligned}
\end{eqnarray}

\begin{table}[ht]
	\vspace{-0.3cm}
	\centering
	\resizebox{\linewidth}{!}{
		\begin{tabular}{l}
			\hline  
			\textbf{Algorithm 1:} The proposed BiGAMP algorithm\\
			\hline
			Give the system output $\mathbf Y$ and estimation functions $\bm g_{\mathbf z}$, $\bm g_{\mathbf x}$, and $g_{\mathsf a}$. \\For $t=1,\ldots, T_{\text{max}}$, generate the estimates $\hat{\mathbf X}(t)$, $\hat{\mathbf A}(t)$, and $\hat{\mathbf Z}(t)$ by\\ the following recursion: \\
			1: Initialization: For each $l,n,m$, set $\hat {\bm s}_l(0)=0$, $\hat { a}_{ln}(1)=c_{ln}(l\leq L_p)$,\\ \quad $\hat { a}_{ln}(1)=0(l> L_p)$, $\hat {\bm x}_n(1)=0$,   $ v^{\mathsf a}_{ln}(1)=1$ and $\bm v^{\mathbf x}_n(1)=\beta_n \mathbf I$. \\
			2: \textbf{Repeat} \\
			3: Update the estimate $\bar{\bm p}_l(t)$ of the matrix product $[\mathbf {AX}]_{l,:}$ and  the\\ \quad corresponding   covariance  matrix $\bar{\bm v}^{\mathsf p}_l(t)$ by (\ref{eq:barp}).\\
			4: Apply Onsager correction to compute the corrected estimate $\hat{\bm p}_l(t)$  and\\ \quad covariance matrix ${\bm v}^{\mathsf p}_l(t)$ by (\ref{eq:vphat}).\\
			5:  Update the approximate posterior mean $\hat{\bm z}_l(t)$ and covariance  matrix \\ \quad${\bm v}^{\mathsf z}_l(t)$ by (\ref{eq:pvar1}).
			\\
			6: Update the scaled residual $\hat{\bm s}_l(t)$ and the set of inverse-residual-covariance\\ \quad  ${\bm v}^{\mathsf s}_l(t)$ by (\ref{eq:s}).\\
			7: Update $\hat{q}_{ln}(t)$ and ${ v}^{\mathsf q}_{ln}(t)$by (\ref{eq:Eq}) and (\ref{eq:qvar}). \\
			8: Update $\hat{\bm r}_{n}(t)$ and $\bm{ v}^{\mathsf r}_{n}(t)$ by (\ref{eq:Er}) and (\ref{eq:xva}). \\
			9: Compute the estimate $\hat{a}_{ln}(t+1)$  and variance $v_{ln}^{\mathsf a}(t+1)$  of $\mathsf a_{nl}$ by  (\ref{eq:avard}).
			\\
			10: Compute the estimate $\hat{\bm x}_{n}(t+1)$, $\bm v_{n}^{\mathbf x}(t+1)$ and $\phi(\hat{\boldsymbol r}_{n}(t))$  by  (\ref{eq:xhat})-(\ref{eq:varx}).
			\\
			11: \textbf{Until}
			 \qquad $\|\bar{\mathbf P}(t+1)-\bar{\mathbf P}(t)\|_F^2\le\kappa\|\bar{\mathbf P}(t)\|_F^2$\\
			12: \textbf{Return}
			$\hat{\boldsymbol x}_{n}(t)$,	$\hat{a}_{ln}(t)$,$v_{ln}^{\mathsf q}(t)$,  $ v_{n}^{\mathsf r}(t)$, and $\phi(\hat{\boldsymbol r}_{n}(t))$.\\
			\hline
		\end{tabular}
	}
\end{table}

We summarize the proposed algorithm in Algorithm 1\footnote{Note the computations involving variances in Algorithm 1 require considering {\it Assumption 1} to be simplified. Adaptive damping which is not included in Algorithm 1 is employed to ensure the convergence of the proposed BiGAMP algorithm. The details of damping are similar to \cite[Section IV]{RN157}.  Due to space limitations, we will no longer discuss this issue, and interested readers can refer to \cite{RN157}.}, where $T_{\text{max}} $ is the maximum number of iterations. According to (\ref{eq:barp1}), define $\bar{\mathbf P}(t)=\hat{\mathbf A}(t)\hat{\mathbf X}(t)=[\bm{\bar p}_1(t),\ldots,\bm{\bar p}_L(t)]^T$. Algorithm 1 stops when the difference between the updated $\bar{\mathbf P}(t+1)$ and $\bar{\mathbf P}(t)$ is small enough. Given $\kappa=10^{-4}$, the stopping criterion is  $\|\bar{\mathbf P}(t+1)-\bar{\mathbf P}(t)\|_F^2\le\kappa\|\bar{\mathbf P}(t)\|_F^2$. In the following, where no ambiguity arises, {\it the BiGAMP algorithm} always means Algorithm 1. The complexity of Algorithm 1 depends on the multiplication of the channel matrix and the signal matrix, i.e., $\hat{\mathbf A}(t)\hat{\mathbf X}(t)$ in `3' of Algorithm 1. Since $\hat{\mathbf A}(t) \in \mathbb{C}^{L\times N}$ and  $\hat{\mathbf X}(t) \in \mathbb{C}^{N\times M}$, the complexity scales as $ {O}(LNM) $.

\section{Performance for BiGAMP Algorithm}
\label{Per}
In this section, we first construct the SE to describe the convergence of the algorithm, then analyze the theoretical performance of the proposed algorithm for DAD-CE-SD, which includes the error probability of DAD, the corresponding MSE of CE, and the SER of SD.

\subsection{State Evolution}
\label{SE}
The estimates $\hat{\mathbf A}(t)$ and $\hat{\mathbf X}(t)$ are obtained from the observation $\mathbf Y$. $\hat{\mathbf P}(t)=[\bm{\hat p}_1(t),\ldots,\bm{\hat p}_L(t)]^T$is the estimate of $\mathbf Z$ after applying Onsager correction to decouple the errors of $\mathbf Y-\bar{\mathbf P}(t)$. For the proposed BiGAMP algorithm, according to \cite{2010arXiv1010.5141R,7457269}, we try to track the evolution of the MSE as its iteration.  Therefore, we define
\begin{eqnarray}
	\begin{aligned}
	\bm{\Gamma}(t) \triangleq  \mathbb{E}[(\bm y_l-\hat{\bm p}_l(t))(\bm y_l-\hat{\bm p}_l(t))^H].
	\label{xtau}
	\end{aligned}
\end{eqnarray}
Note $\mathbb{E}[\cdot]$ is also the mean for all $l$.
It is evident that $\bm{\Gamma}(t)$ characterizes the convergence performance of  Algorithm 1.

{\it Assumption 2:} To facilitate the analysis, we simplify the variance estimations as follows (omitting iteration  $t$):
\begin{eqnarray}
\begin{aligned}
{v}_{l}^{\mathsf{p}}&\approx {v}^{\mathsf{p}}\triangleq \frac{1}{L}\sum_{l=1}^{L}{v}_{l}^{\mathsf{p}},\\\quad {v}_{n}^{\mathsf{r}}&\approx {v}^{\mathsf{r}}\triangleq \frac{1}{N}\sum_{n=1}^{N}{v}_{n}^{\mathsf{r}},\\\quad
{v}_{ln}^{\mathsf{q}}&\approx {v}^{\mathsf{q}}\triangleq \frac{1}{LN}\sum_{l=1}^{L}\sum_{n=1}^{N}{v}_{ln}^{\mathsf{q}}.
\label{eq:simply}
\end{aligned}
\end{eqnarray}
{\it Assumption 2} holds in the asymptotic regime.

\textbf{Theorem 1}:
In the asymptotic regime, it can be proved that
\begin{eqnarray}
\begin{aligned}
\bm{\Gamma}(t)=\tau(t) \mathbf I =(v^{\mathsf p}(t)+\sigma^2)\mathbf I,
\label{tauxa}
\end{aligned}
\end{eqnarray}
where  $\tau(t) $ is called ``State Evolution" (SE), and it updates as the recursion value $v^{\mathsf p}(t)$. The algorithm is convergent under the condition
\begin{eqnarray}
\begin{aligned}
L>c_1K, \quad and \quad
M>c_2K,
\label{eq:cond}
\end{aligned}
\end{eqnarray}
where $\frac{1}{4}<c_1,c_2<2$ are constants constrained by (\ref{eq:vr1}) in Appendix \ref{appendix:t1}.

{\it Proof: } Please refer to Appendix \ref{appendix:t1}.

According to Theorem 1, SE updates as  $v^{\mathsf p}(t)$ which hinges upon $v^{\mathsf r}(t)$ and $v^{\mathsf q}(t)$ in Appendix \ref{appendix:t1}. $v^{\mathsf r}(t)$ and $v^{\mathsf q}(t)$ characterize the estimation error of $\mathbf X$ and $\mathbf A$. To guarantee SE converges, $v^{\mathsf r}(t)$ and $v^{\mathsf q}(t)$ must be convergent. Therefore, the behavior of the BiGAMP algorithm can be described by SE. Meanwhile, to ensure the convergence of the BiGAMP algorithm, (\ref{eq:cond}) gives the relationship among the number of active devices $K$, pilot length $L$, and the number of antennas $M$.

\subsection{Error Probability of Device Activity Detection}
Now, we analyze the error probability of DAD according to the detector in Definition 1. The error probability of device $n$ after the $t$th iteration is defined as
\begin{eqnarray}
\begin{aligned}
P_{n,t}^e(M) =& P(\alpha_n = 0)P(\hat{\alpha}_{n,t} = 1|\alpha_n = 0)\\&+ P(\alpha_n = 1)P(\hat{\alpha}_{n,t} = 0|\alpha_n = 1),
\end{aligned}
\end{eqnarray}
which is proved to be a function of $v_n^{\mathsf r}(t)$ and the number of BS antennas $M$.

\textbf{Theorem 2:} For device $n$, the error probability of DAD after $t$ iterations is expressed as
\begin{eqnarray}
P_{n,t}^e(M)
=(1-\varepsilon)\frac{\bar{\Gamma}(M,Mb_{n,t})}{\Gamma(M)} + \varepsilon\frac{\underline{\Gamma}(M,Mc_{n,t})}{\Gamma(M)},
\end{eqnarray}
where $\bar{\Gamma}(\cdot),\Gamma(\cdot)$ and $\underline{\Gamma}(\cdot)$ are the upper incomplete Gamma function, the Gamma function, and the lower incomplete Gamma function, respectively.\footnote{For the Gamma function $\Gamma(a)=\int_0^{\infty}e^{-t}t^{a-1}dt$, the incomplete gamma functions are obtained by decomposing it into an integral from $0$ to $x$ and another from $x$ to $\infty$, i.e., $\underline\Gamma(a)=\int_0^{x}e^{-t}t^{a-1}dt$ and $\bar\Gamma(a)=\int_x^{\infty}e^{-t}t^{a-1}dt$.} With  the path-loss and shadowing component $\beta_n$, it has
\begin{subequations}
	\begin{align}
	b_{n,t}=&\frac{\beta_n + v_n^{\mathsf r}(t)}{\beta_n}\log \frac{\beta_n+v_{n}^{\mathsf r}(t)}{ v_{n}^{\mathsf r}(t)},\\ c_{n,t}=&\frac{v_n^{\mathsf r}}{\beta_n}\log \frac{\beta_n+v_{n}^{\mathsf r}(t)}{ v_{n}^{\mathsf r}(t)}.
	\end{align}
\end{subequations}

{\it Proof:} Please refer to Appendix \ref{appendix:t2}

Since $ u/(1+u)\leq \log(1+u)\leq u$ for $ u\in [0,\infty)$, we have $(\frac{v_n^{\mathsf r}}{\beta_n})\log \frac{\beta_n+v_{n}^{\mathsf r}} { v_{n}^{\mathsf r}}\leq 1$ and $(\frac{\beta_n + v_n^{\mathsf r}}{\beta_n})\log \frac{\beta_n+v_{n}^{\mathsf r}} { v_{n}^{\mathsf r}}\geq 1$, i.e.,  $b_{n,t}\geq 1$ and $c_{n,t}\leq 1$.
According to \cite{Gautschi98theincomplete} and Appendix E in \cite{RN164}, for $b_{n,t}\geq 1$ and $c_{n,t}\leq 1$, $ \frac{\bar{\Gamma}(M,Mb_{n,t})}{\Gamma(M)}\to 0$ and $\frac{\underline{\Gamma}(M,Mc_{n,t})}{\Gamma(M)}\to 0$ as  $M \to \infty$. Hence, there is $P_{n,t}^e\to 0$ as $M\to \infty$, which means the detection error probability goes to zero as $M\to \infty$ in the asymptotic regime.

\subsection{Mean Square Error of Channel Estimation}
When device $k\in\mathcal{K}$ is detected as active,  the estimated channel $\hat{\bm h}_{k,t}$ is defined in  (\ref{eq:esHA}). Define the difference between the actual channel ${\bm h}_{k}$ and the estimated $\hat{\bm h}_{k,t}$ as $\Delta{\bm h}_{k,t}\triangleq {\bm h}_{k}-\hat{\bm h}_{k,t}$. Then we can give the following theorem.

\textbf{Theorem 3:} For active device $k\in\mathcal{K}$, the MSE of CE is given by
\begin{eqnarray}
\begin{aligned}
Cov(\Delta{\bm h}_{k,t},\Delta{\bm h}_{k,t})= v_{k,t}^{\Delta\mathsf h}(M)\mathbf I,
\end{aligned}
\end{eqnarray}
where
\begin{eqnarray}
\begin{aligned}
v_{k,t}^{\Delta \mathsf h}(M) =\frac{1}{M}\mathbb{E}&\left[{ \phi^2_{k,t}\left(\frac{\beta_k\left(\bm h_k + \bm w_k^{\mathsf r}(t)\right)}{\beta_k+ v_{k}^{\mathsf r}(t)}-\bm h_k\right)^H}\right.\\&\quad\left.{\left(\frac{\beta_k\left(\bm h_k + \bm w_k^{\mathsf r}\left(t\right)\right)}{\beta_k+ v_{k}^{\mathsf r}(t)}-\bm h_k\right)}\right],
\label{eq:cherror}
\end{aligned}
\end{eqnarray}
and $\bm w_k^{\mathsf r}(t)$ is generated by $\mathcal{CN}(\bm 0, v_k^{\mathsf r}(t)\mathbf I)$ according to Appendix \ref{appendix:p2}. In the asymptotic regime, $v_{k,t}^{\Delta \mathsf h}(M)$ converges to
\begin{eqnarray}
\lim_{M\to \infty} v_{k,t}^{\Delta\mathsf h}(M) =\frac{\beta_k v_{k}^{\mathsf r}(t)}{\beta_k+ v_{k}^{\mathsf r}(t)}.
\label{eq:th}
\end{eqnarray}

 {\it Proof:} Please refer to Appendix \ref{appendix:t3}.

Theorem 3 shows that the MSE of CE is related to $v_k^{\mathsf r}(t)$. According to Section \ref{SE}, $v_k^{\mathsf r}(t)$ should converge for the BiGAMP algorithm to work, so that $v_{k,t}^{\Delta\mathsf h}$ converges to the fixed point when $M$ is large enough. Note that the residual noise in (\ref{eq:errorh}) is considered uncorrelated across the antennas since each active device’s channels across the multiple receive antennas at the BS are considered uncorrelated.
\subsection{Symbol Error Rate of Signal Detection}
\label{SER}

\begin{figure*}[!t]
	\vspace*{-8pt}
	\normalsize	
	\centering
	\subcaptionbox{Error probability of DAD\label{Fig3a}}{\includegraphics[width = 0.3\textwidth]{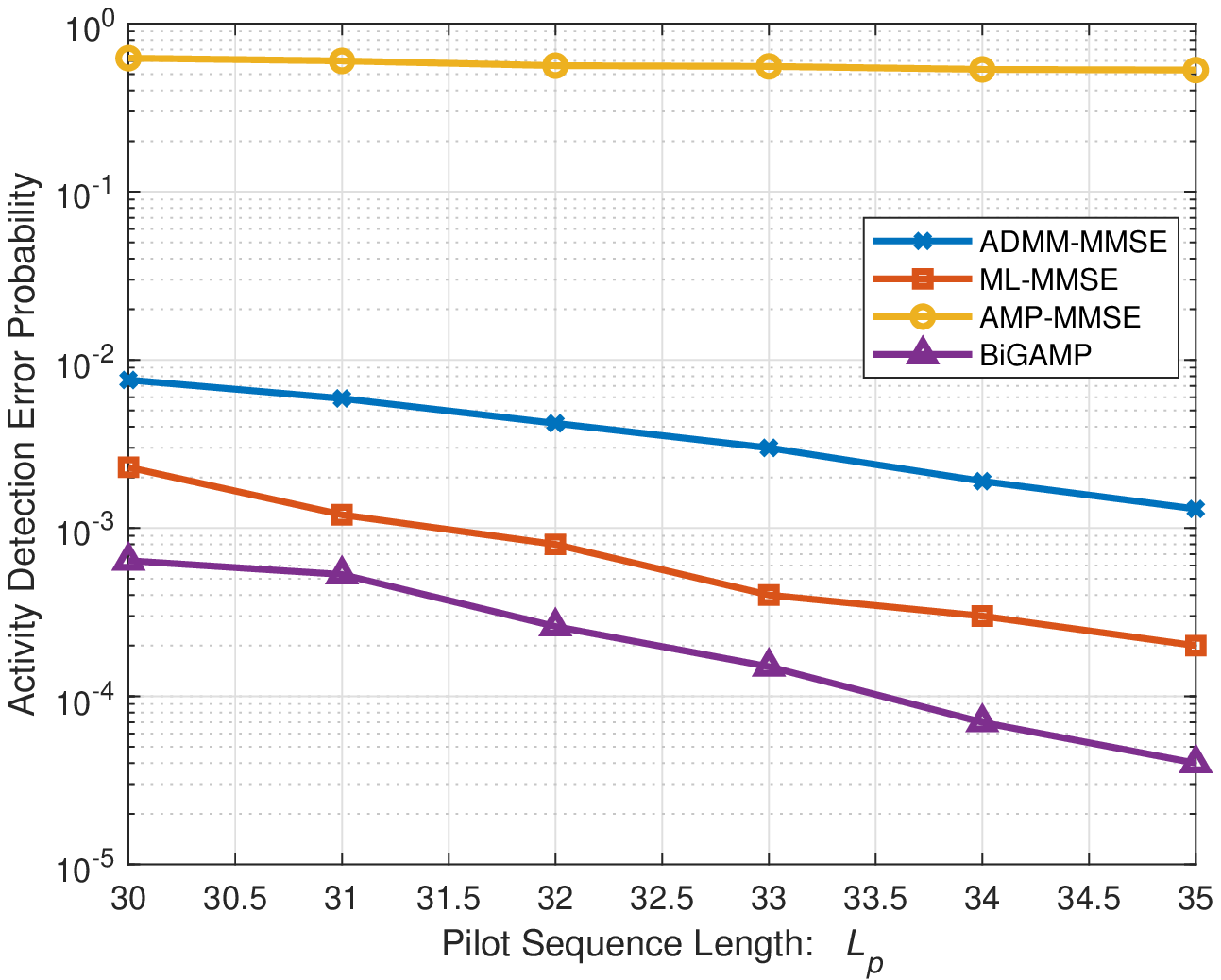}}
	\subcaptionbox{MSE of CE\label{Fig3b}}{\includegraphics[width = 0.3\textwidth]{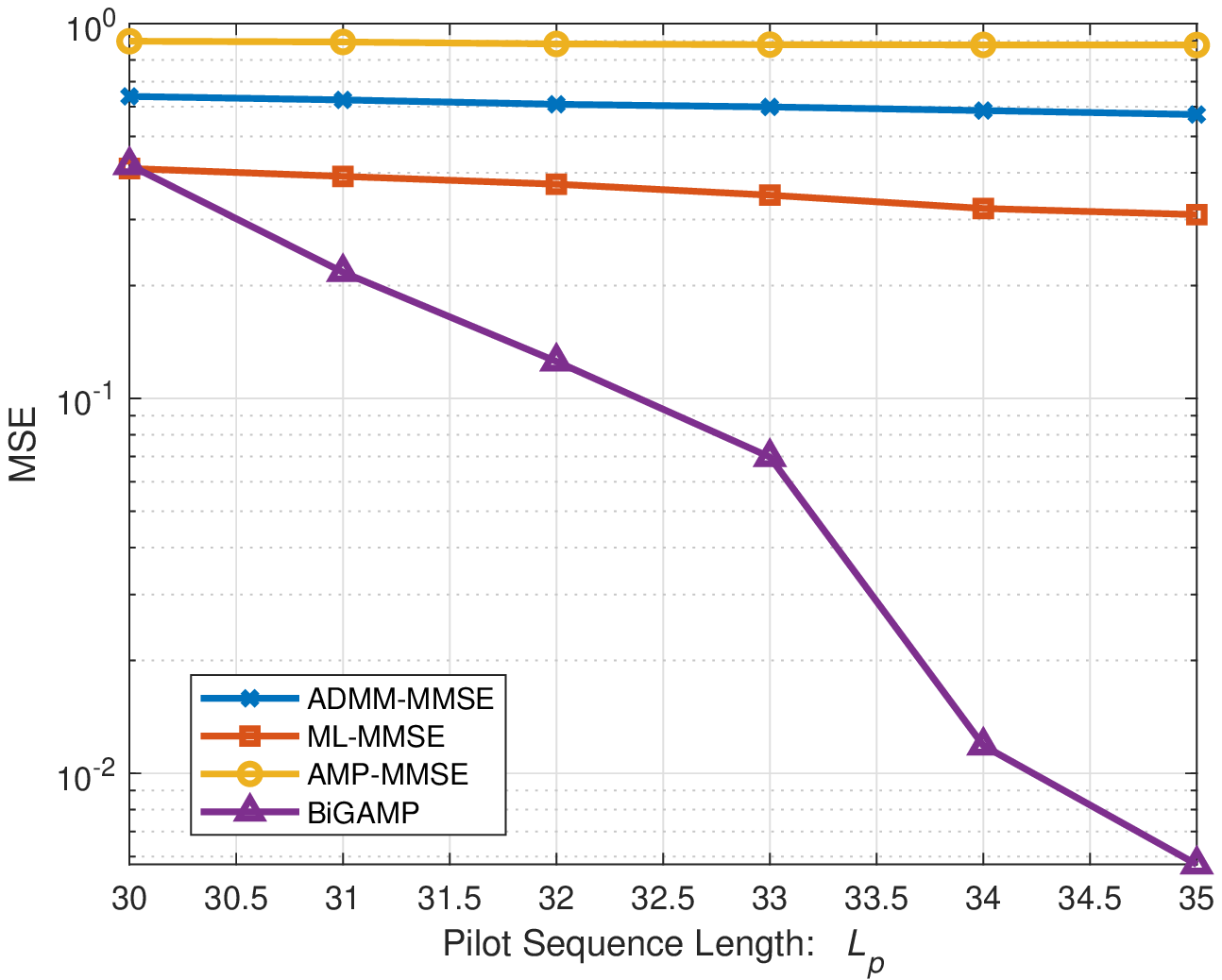}}
	\subcaptionbox{SER of SD\label{Fig3c}}{\includegraphics[width = 0.3\textwidth]{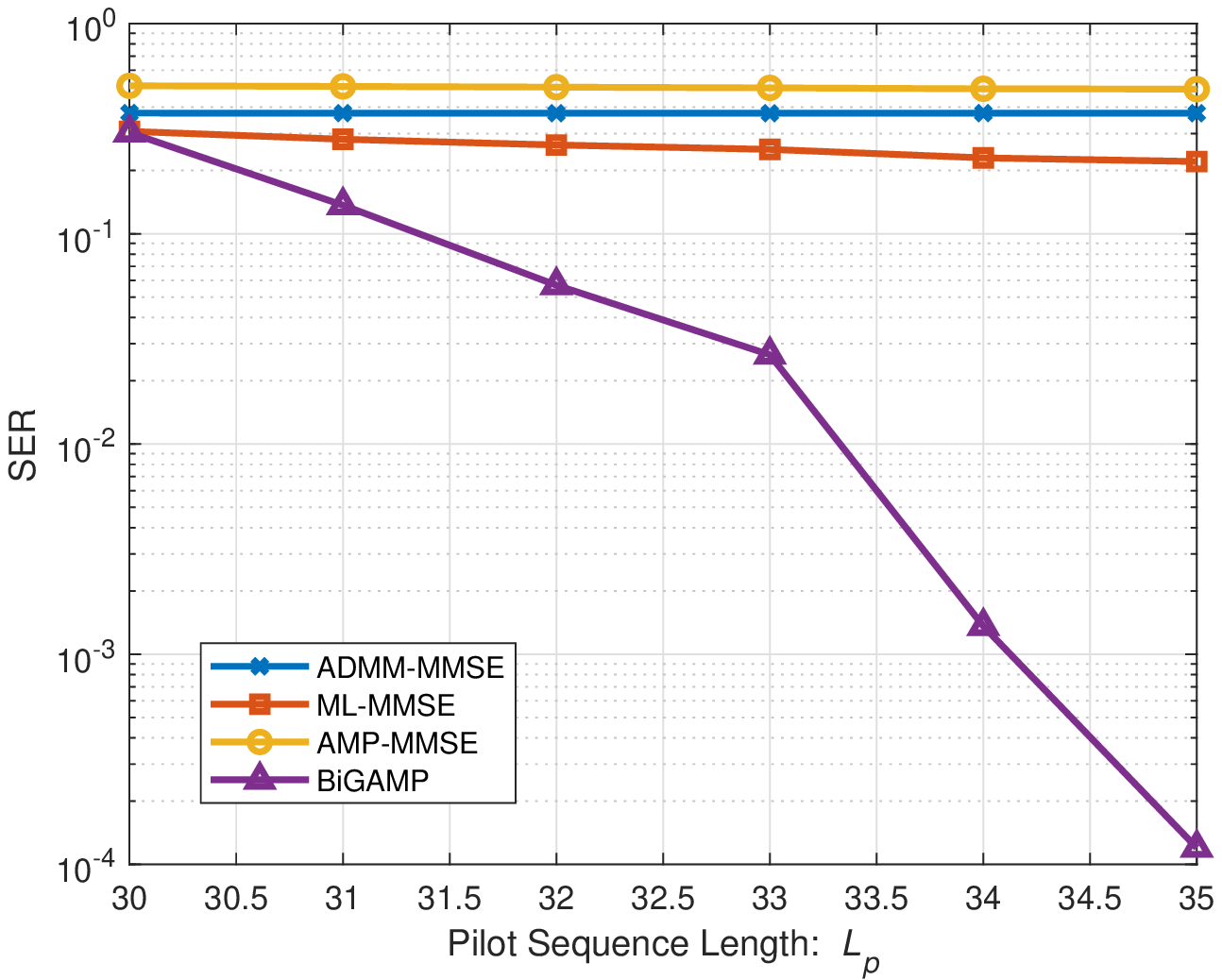}}
	\caption{\scriptsize{There is  $N=1000$, $\varepsilon =0.05$, $L_d=100$. \subref{Fig3a}, \subref{Fig3b}, and \subref{Fig3c} are error probability of DAD, MSE of CE, and SER of SD, respectively, versus the length of pilot $L_p$ with $M=64$. }}
	\label{fig:3}
	\vspace*{-8pt}
\end{figure*}
\begin{figure*}[!t]
	\normalsize	
	\centering
	\subcaptionbox{Error probability of DAD\label{Fig4a}}{\includegraphics[width = 0.3\textwidth]{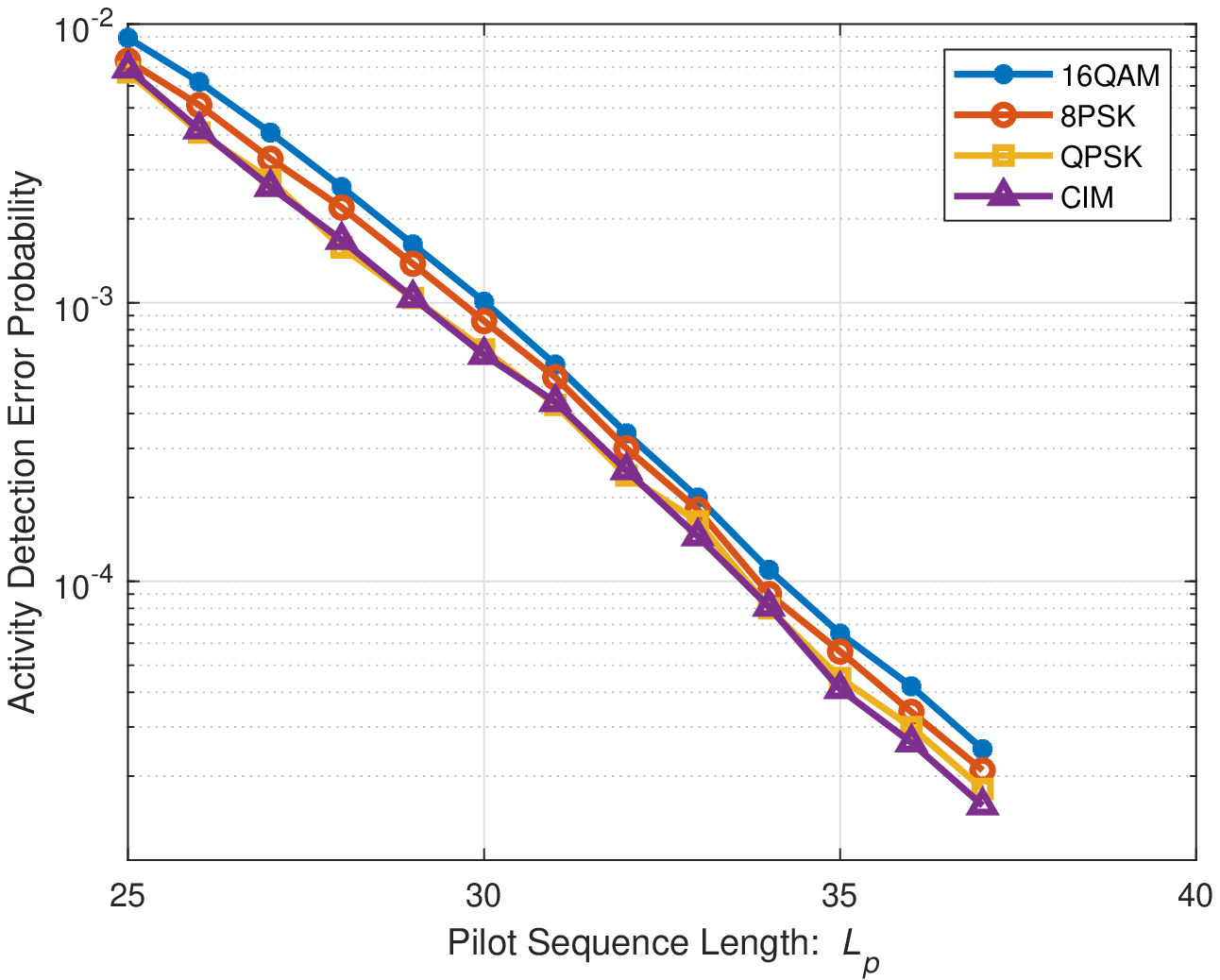}}
	\subcaptionbox{MSE of CE\label{Fig4b}}{\includegraphics[width = 0.3\textwidth]{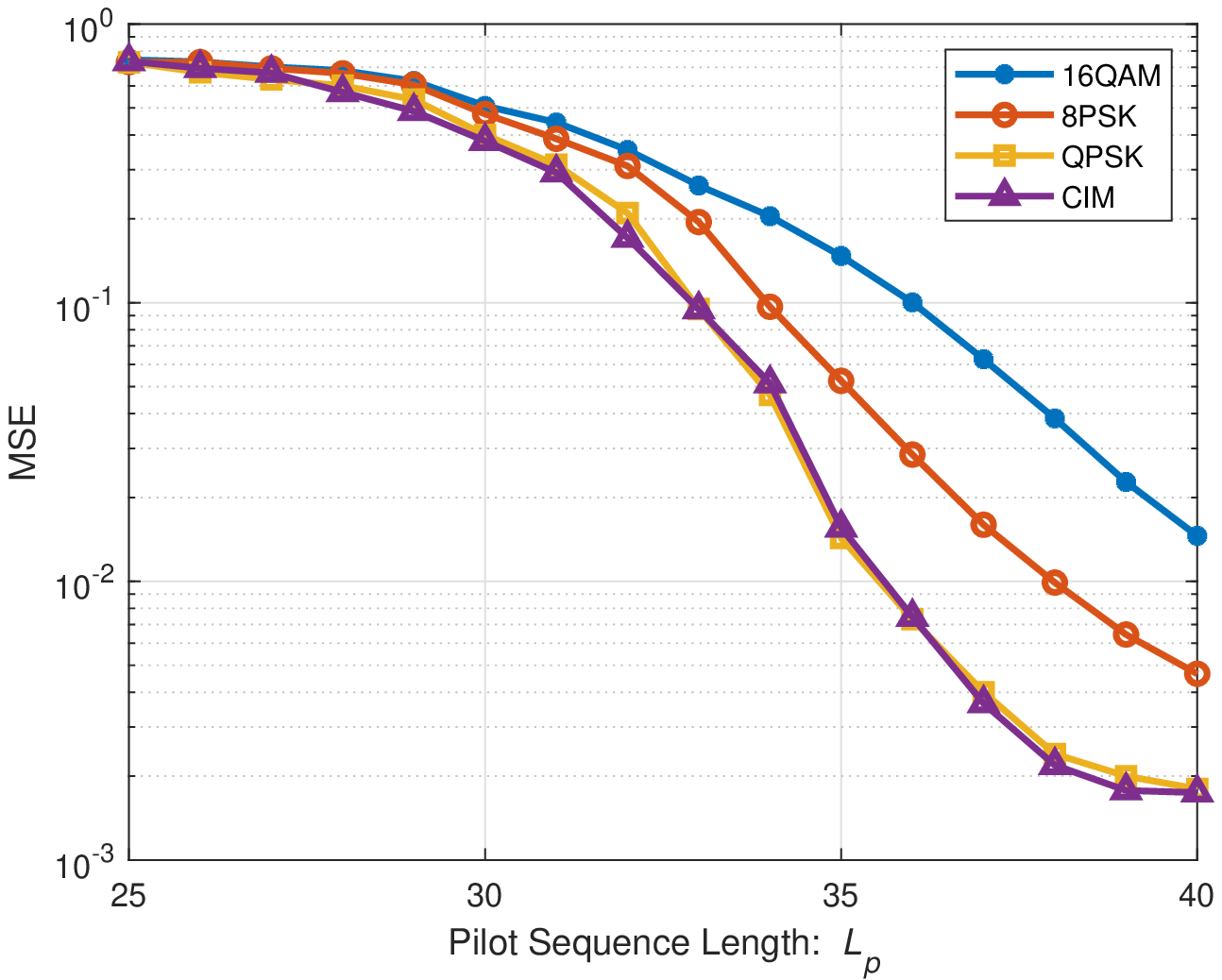}}
	\subcaptionbox{SER of SD\label{Fig4c}}{\includegraphics[width = 0.3\textwidth]{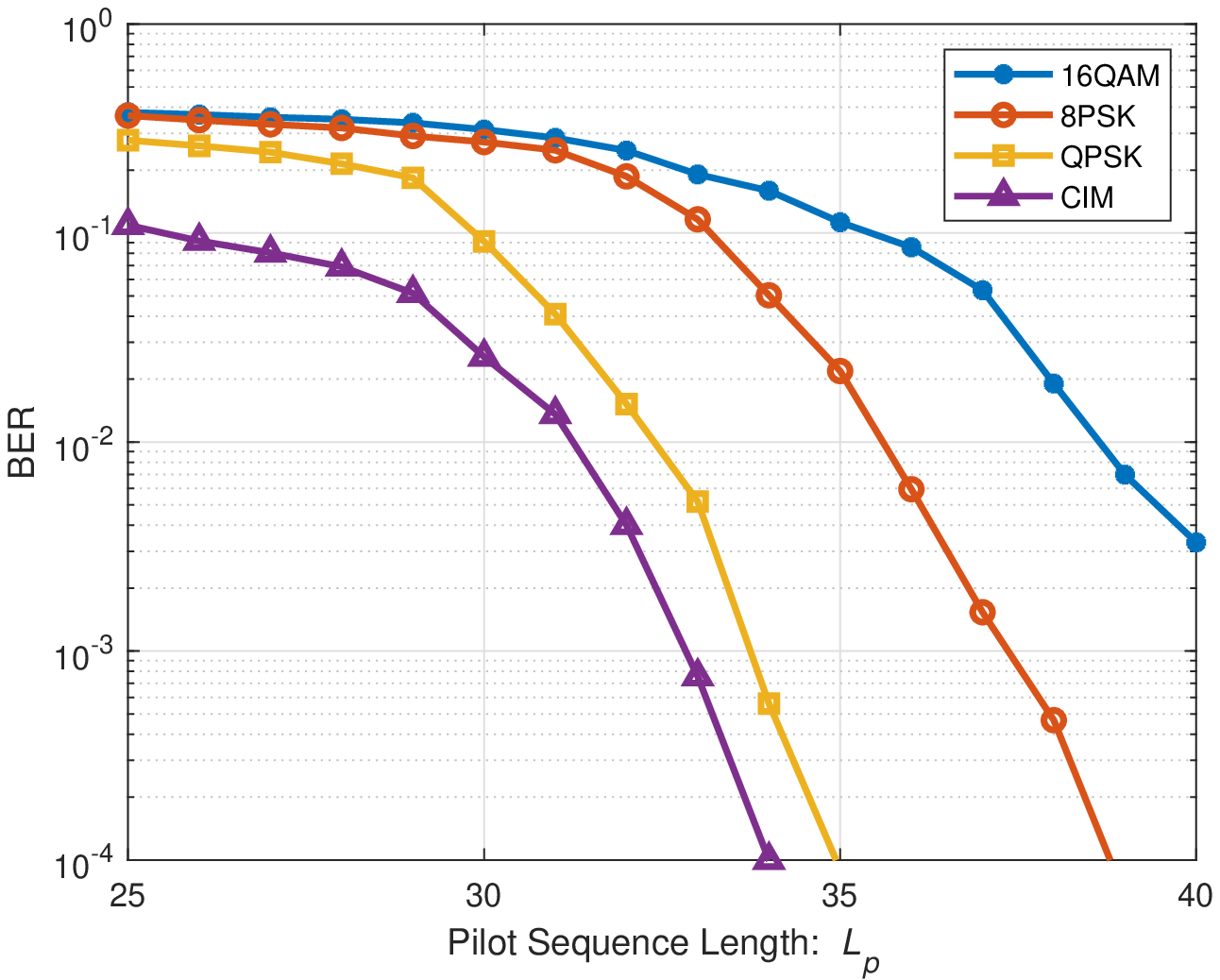}}
	\caption{\scriptsize{There is  $N=1000$, $\varepsilon =0.05$, $L_d=128$. \subref{Fig4a}, \subref{Fig4b}, and \subref{Fig4c} are error probability of DAD, MSE of CE, and SER of SD, respectively, versus the length of pilot $L_p$ with $M=64$.}}
	\label{fig:4}
	\vspace*{-8pt}
\end{figure*}

For any active device $k\in \mathcal{K}$, the estimated $\hat {\bm d}_{k,t}$ is as defined in  (\ref{eq:esHA}). For simplicity, we omit $t$ in the following. Assume that the system adopts a Gaussian codebook $\mathcal{D}\subset \mathbb{C}^{J\times1}$ and $\mathcal{D}=|D|$, where $J$ is the length of codewords. There is $L_d = N_{s}\times J$, where $N_{s}$ is the number of codewords. With $\bm d_k^{n_{s}}\in \mathcal{D}$, the transmitted symbols are $\bm d_k=[(\bm d_k^1)^T,\ldots, (\bm d_k^{N_{s}})^T]^T$. For given estimate $\hat{\bm d}_k=[(\hat{\bm d}_k^1)^T,\ldots, (\hat{\bm d}_k^{N_{s}})^T]^T$, the $n_{s}$th detected codeword for device $k$ could be expressed as
	\begin{eqnarray}
	\begin{aligned}
	\bm{d}_k^{'n_{s}} = \arg \min_{\bm d\in \mathcal{D}}\|\hat{\bm d}_k^{n_{s}}-\bm d\|_2.
	\end{aligned}
	\end{eqnarray}
When the detected symbol  $\bm{d}_k^{'n_{s}}\neq \bm{d}_k^{n_{s}}$, the result of SD is wrong. The SER is defined as
	\begin{eqnarray}
	\begin{aligned}
	P_{d}^e&= \mathbb{E}[\frac{1}{K'N_{s}}\sum_{k=1}^{K'}\sum_{n_{s}=1}^{N_{s}}1\{\bm{d}_k^{'n_{s}}\neq \bm{d}_k^{n_{s}}\}] \\&=
	\mathbb{P}(\bm{d}_k^{'n_{s}}\neq \bm{d}_k^{n_{s}}),
	\end{aligned}\label{eq:Pe}
	\end{eqnarray}
where $K'$ is the number of active devices detected. According to the above definition, we give Theorem 4.

\textbf{Theorem 4}: %
The SER of signal detection is
\begin{eqnarray}
\begin{aligned}
P_{d}^e\leq\exp\left({-\rho\ln\left(D-1\right)-J\rho\ln \left(1+\frac{1}{Lv^{\mathsf a}(t)(1+\rho)}\right)}\right).\nonumber
\end{aligned}
\end{eqnarray}

{\it Proof:} Please refer to Appendix \ref{appendix:t4}.

The above SER is an upper bound based on the Gallager-type upper bound and $\rho\in(0,1)$ represents Gallager's $\rho$-trick.
The effect of $L$ on SER is mainly by affecting the signal power. But the signal power also affects $v^{\mathsf a}(t)$ in the simulation. According to (\ref{eq:avard}), we have $Lv^{\mathsf a}(t)  = 1-\frac{1}{1+Lv^{\mathsf q}(t-1)}$. Therefore, with fixed $D$ and $J$, $P_{d}^e$ increases as $Lv^{\mathsf q}(t)$ increases.

\section{Numerical Results}
\label{Num}
In this section, we provide numerical results to verify the performance of the proposed algorithm. In the simulation, the signal-to-noise ratio (SNR) is $10$ dB. In addition, we assume that devices are static or immobile in this cellular, so the path-loss and shadowing component $\beta_1=\ldots=\beta_N=\bar\beta=1$.
For the Gaussian codebook, we set $\rho=1/2$, $J=5$, and $D=|\mathcal{D}|=64$.
Moreover, all numerical results are obtained by averaging over $1000$ simulation realizations.

\begin{figure*}[!t]
	\normalsize	
	\centering
	\subcaptionbox{Error probability of DAD\label{Fig5a}}{\includegraphics[width = 0.3\textwidth]{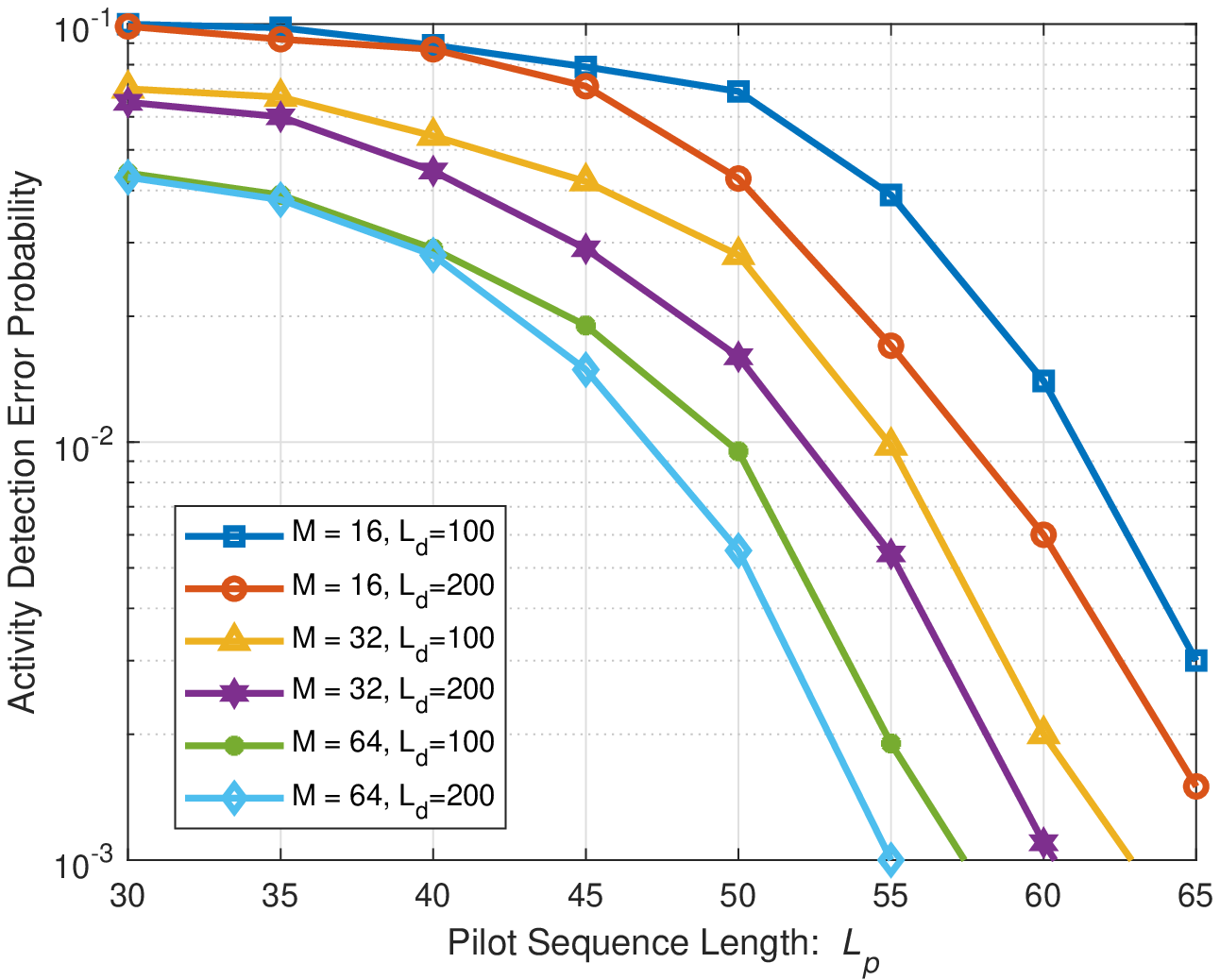}}
	\subcaptionbox{MSE of CE\label{Fig5b}}{\includegraphics[width = 0.3\textwidth]{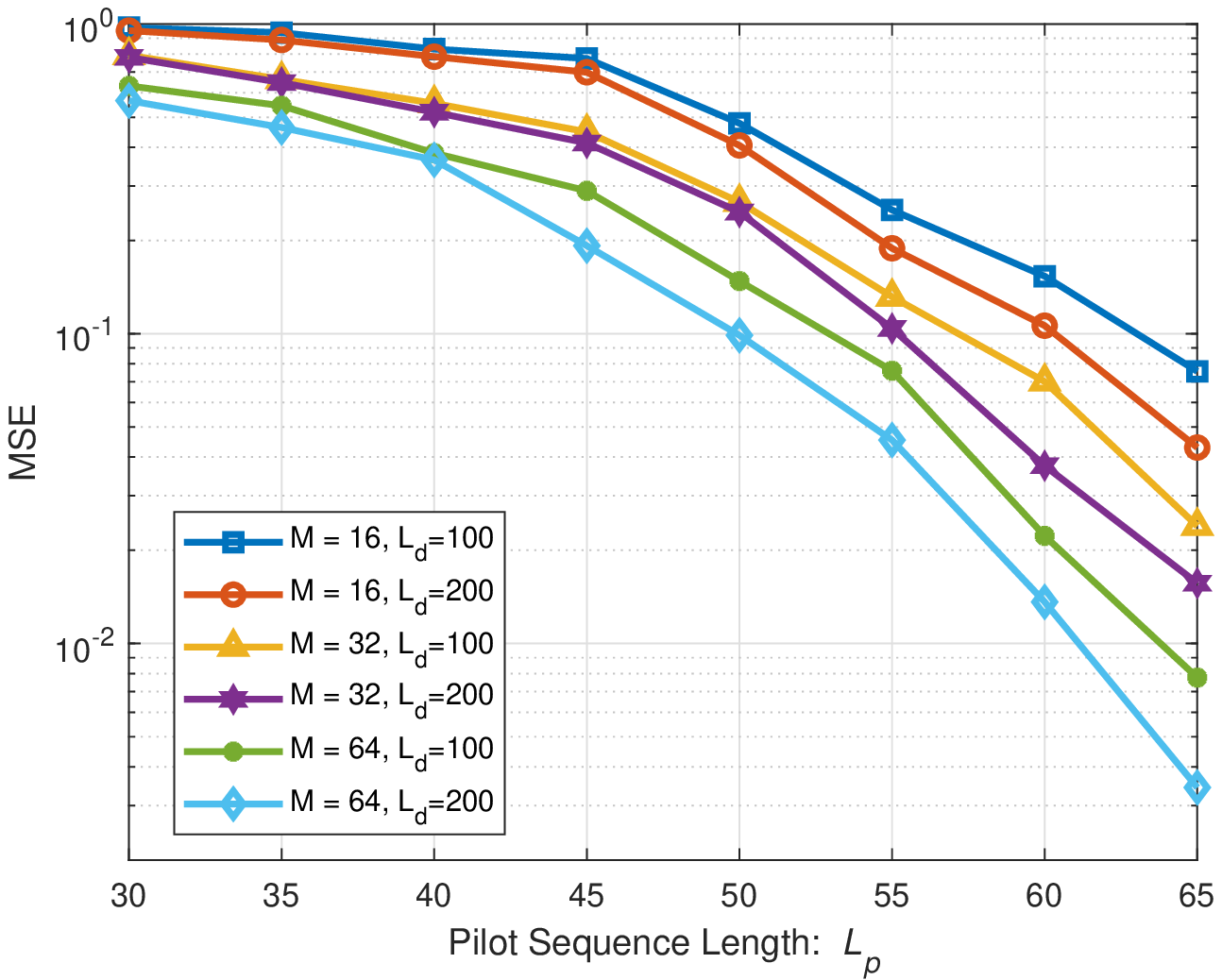}}
	\subcaptionbox{SER of SD\label{Fig5c}}{\includegraphics[width = 0.3\textwidth]{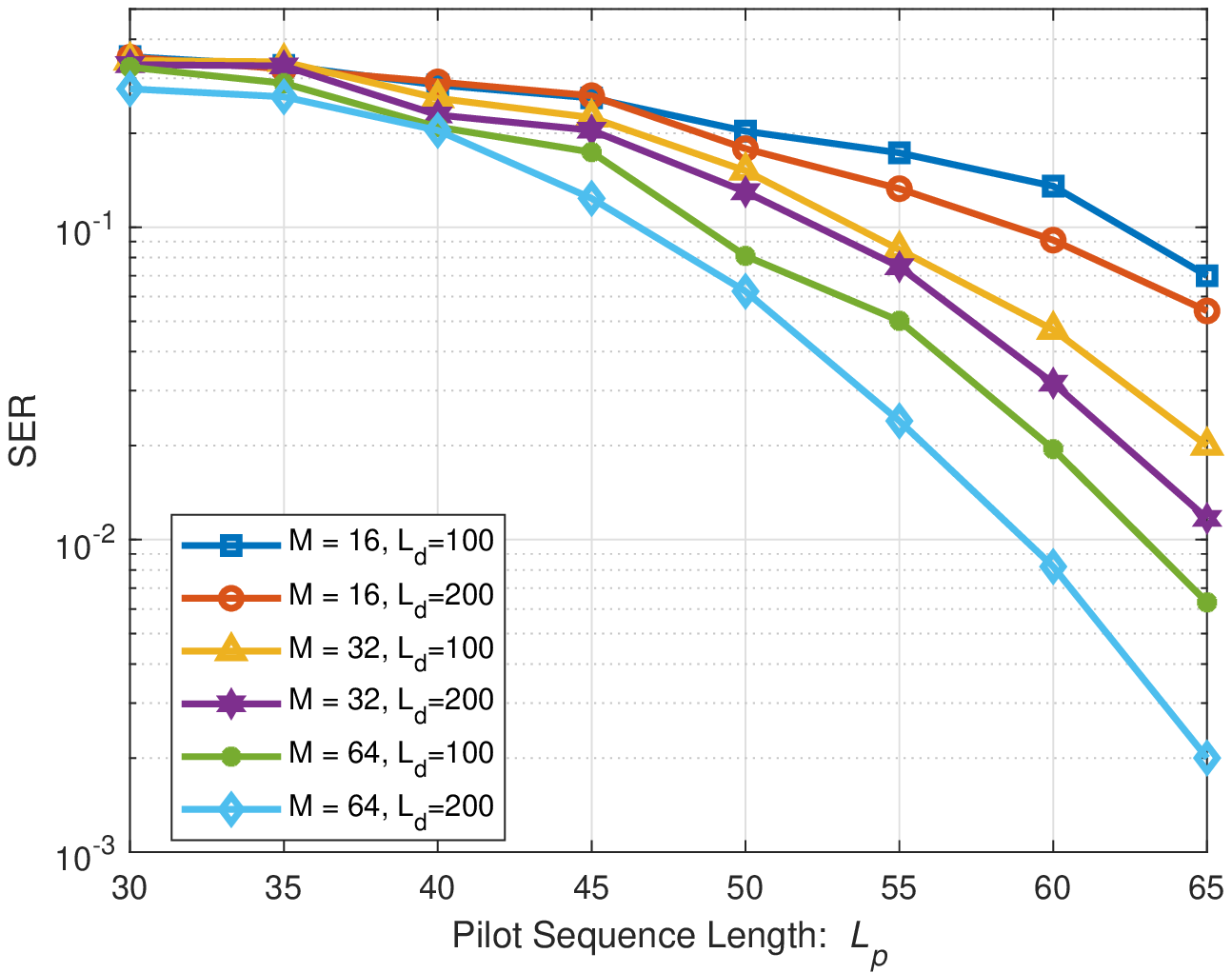}}
	\caption{\scriptsize{There is  $N=1000$ and $\varepsilon =0.05$. \subref{Fig5a}, \subref{Fig5b}, and \subref{Fig5c} are error probability of DAD, MSE of CE, and SER of SD, respectively, versus the length of pilot $L_p$ with different $M$ and $L_d$. }}
	\label{fig:5}
\end{figure*}
\begin{figure*}[!t]
	\vspace*{-8pt}
	\normalsize	
	\centering
	\subcaptionbox{Error probability of DAD\label{Fig6a}}{\includegraphics[width = 0.3\textwidth]{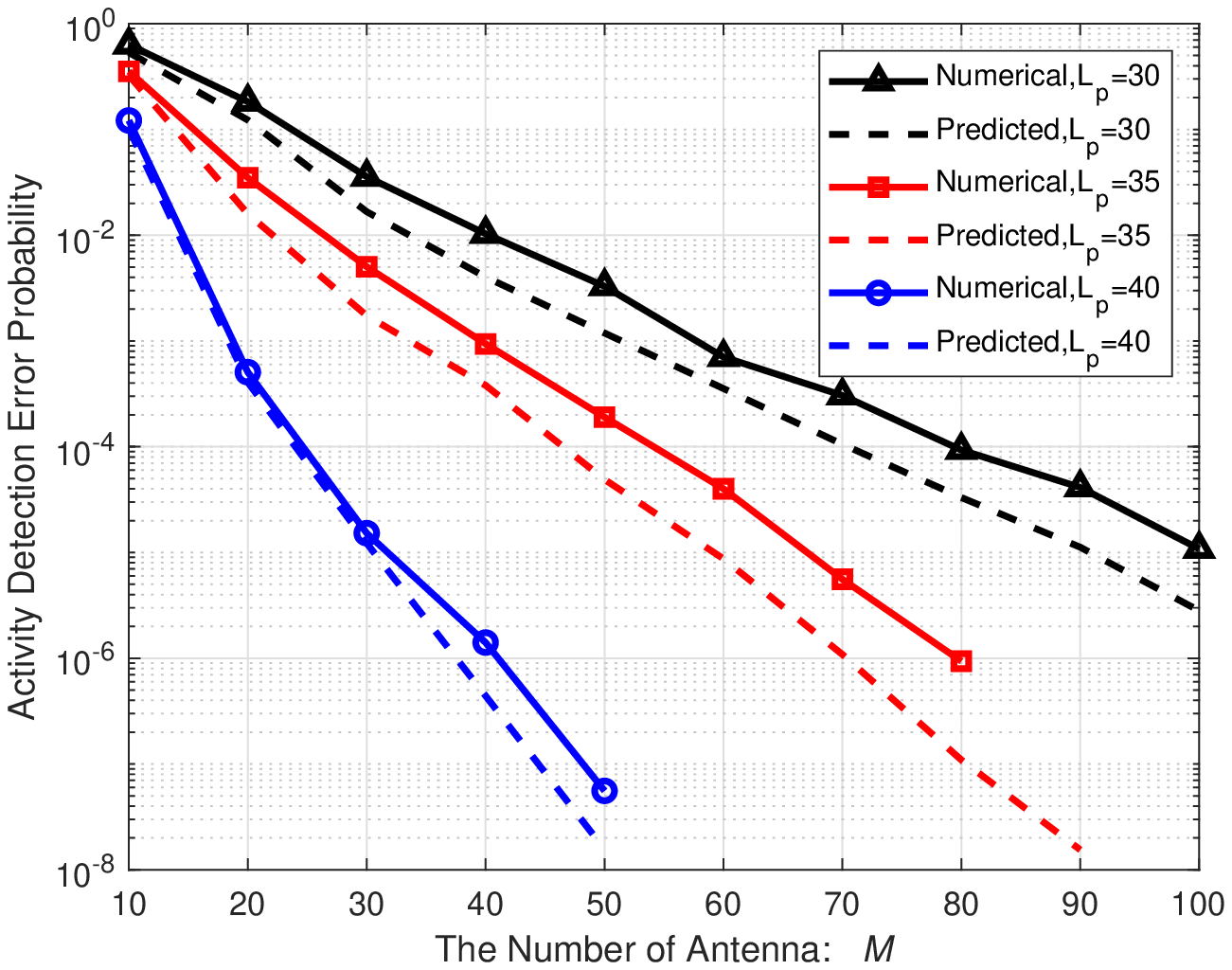}}
	\subcaptionbox{MSE of CE\label{Fig6b}}{\includegraphics[width = 0.3\textwidth]{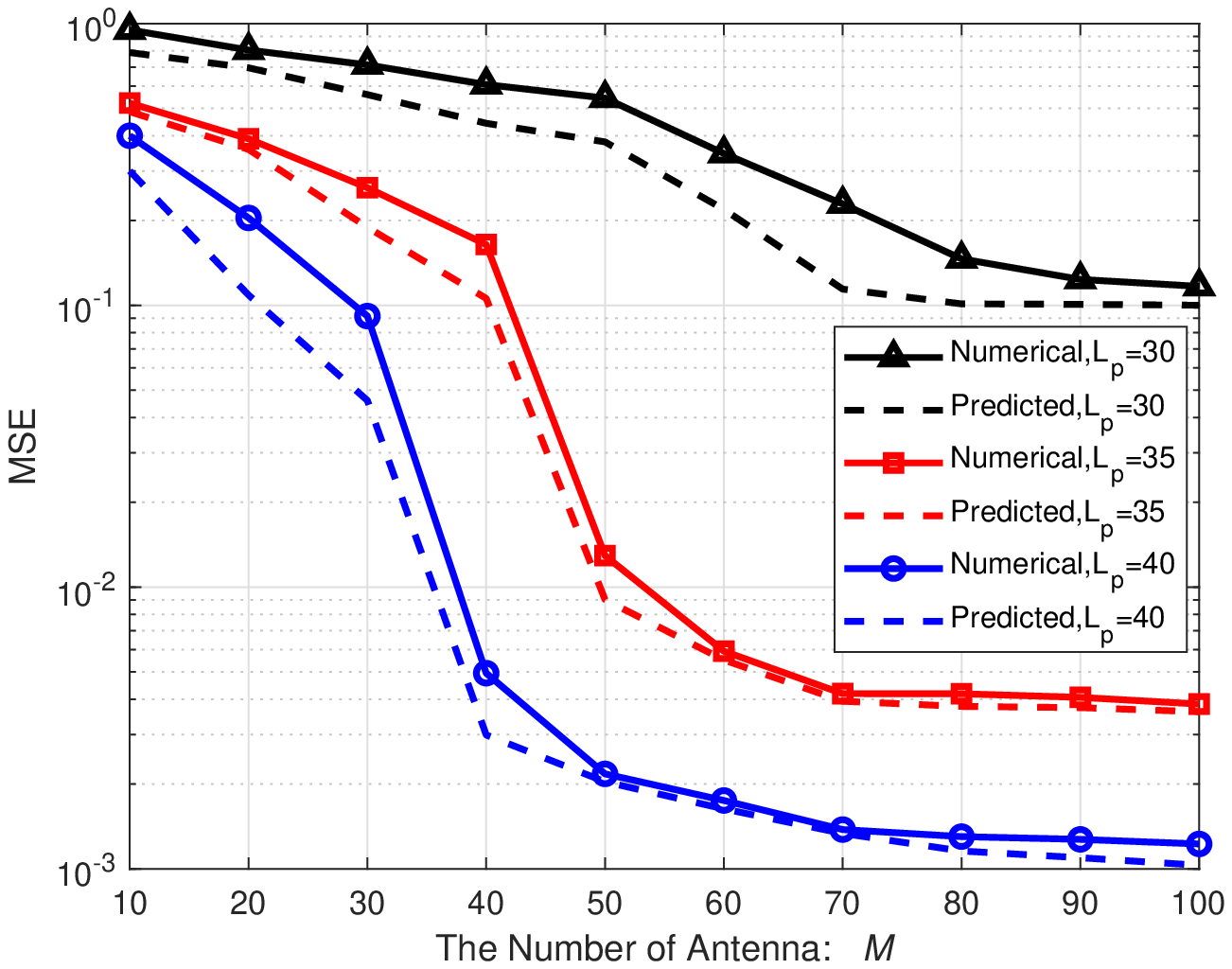}}
	\subcaptionbox{SER of SD\label{Fig6c}}{\includegraphics[width = 0.3\textwidth]{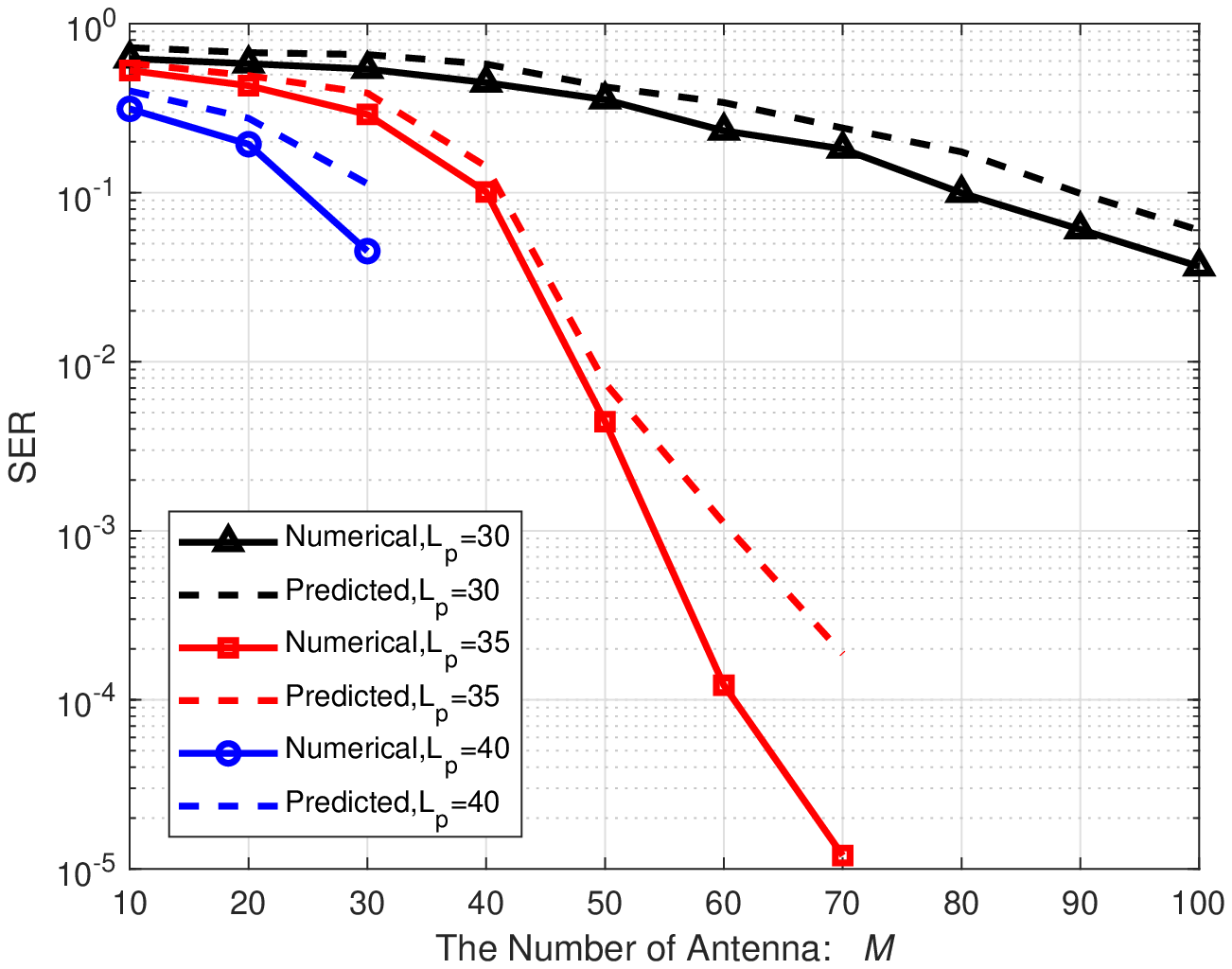}}
	\caption{\scriptsize{There is  $N=1000$, $\varepsilon =0.05$, $L_d=100$.  \subref{Fig6a}, \subref{Fig6b} and \subref{Fig6c} are numerical results and the predictions  versus $M$ with $L_p=30$, $L_p=35$, and $L_p=40$, respectively.}}
	\vspace*{-8pt}
\end{figure*}
\subsection{The DAD-CE-SD Performance}
\label{Numceasd}
First, we choose three extensively studied methods that perform well in DAD-CE-SD as baselines. ML-MMSE is an optimization-based method that uses the coordinate descent method for the ML estimation in \cite{9374476} to detect device activities. Then, it uses the standard MMSE to estimate channels and detect signals of the devices that have been detected to be active. The complexity of ML is ${O}(NL_p^2)$, plus the complexity of MMSE, i.e., $\max\{ {O}(L_p^2K), {O}(L_pKM), {O}(L_p^3)\}$ plus $\max\{ {O}(M^2K), {O}(L_dKM),{O}(M^3)\}$. ADMM is also one optimization-based method to solve group LASSO which conducts CE with the block coordinate descent algorithm \cite{RN281, RN148}. Then MMSE is used to estimate signals. The complexity of ADMM-MMSE is $ {O}(L_pNM)$ plus $\max\{ {O}(M^2K), {O}(L_dKM), {O}(M^3)\}$. AMP is an approximate message passing algorithm based on MMSE, which is used to detect activities and estimate channels and signals using MMSE estimation\cite{RN164, RN166}. The complexity of AMP-MMSE is the same as ADMM-MMSE.

Figure \ref{Fig3a} shows the error probability of DAD.  The proposed algorithm  performs better than ADMM-MMSE, ML-MMSE, and AMP-MMSE when the pilot length is limited. Figures \ref{Fig3b} and \ref{Fig3c} illustrate the MSE of CE and the SER of SD, respectively. It can be observed that the proposed algorithm outperforms others. Note that the SER and MSE are only measured when active devices are detected correctly, which is based on the following two reasons: a) to avoid the situation that CE and SD heavily rely on the performance of DAD; b) to eliminate the effects of devices that are mistaken for active. In this simulation, the setup $L_p/N$ is smaller than $\varepsilon$. Thus the proposed algorithm has advantages in a short pilot length, which can significantly save pilot overhead.

Considering the existing communication system, Figs. \ref{Fig4a}, \ref{Fig4b}, and \ref{Fig4c} give the numerical results when the signal modulations are QPSK\footnote{More details about applying the proposed algorithm in communication systems with QPSK modulation can be found in our work \cite{9940392}.}, 8PSK, 16QAM, and code index modulation (CIM)\footnote{The CIM is based on direct sequence-spread spectrum modulation. In this paper, the CIM is referenced to \cite{RN442}, where the bit stream is divided into modulated subblocks of length $2$ bits and mapped subblocks of length $6$ bits. The combination of 2 bits in each modulated subblock is modulated  into a constellation symbol by QPSK. The combination of $6$ bits in each mapped subblock is mapped as a spreading code to spread the QPSK symbol and each spreading code is a $2^6$ orthogonal Walsh code. Since the modulated subblock of CIM adopts QPSK, the statistics of symbols in CIM are the same as the statistics of constellation symbols in QPSK.}, respectively.
Because CIM is a kind of direct sequence spread spectrum modulation and the bits are embedded in the spreading code, the bit error rate (BER) is used instead of SER in Fig. 4(c) to show the error probability of SD.
The results show that the modulation method has little effect on the performance of the DAD. Since the estimated activity probability is determined by the channels across antennas according to (\ref{eq:thet}), the effect is small enough if $M$ is large enough. The performance of MSE and BER differs due to the statistical characteristics of the codewords. With the same SNR, it is observed that increasing the spectral efficiency will result in a rise in BER for QPSK, 8PSK, and 16QPSK in Fig. \ref{Fig4c}. The MSE of QPSK is close to that of CIM  from Fig. \ref{Fig4b} since the statistics of constellation symbols in CIM are the same as that of QPSK. However, from Fig. \ref{Fig4c}, the BER of CIM is smaller than that of QPSK because CIM applies sequence-spread spectrum technology and embeds bits in spreading codes. By applying spreading codes, the coding gain is enhanced, the system is immunized against errors, and the BER is further decreased according to \cite{RN442}.
The numerical results show that the proposed algorithm also applies to discrete codewords in existing communication systems.

Figures \ref{Fig5a}, \ref{Fig5b}, and \ref{Fig5c} describe the error probability of DAD, MSE of CE, and SER of SD when channels are correlated between the elements in $\bm h_n$. The correlated channels are modeled as \cite{RN105}. Figure \ref{Fig3c} shows that the error becomes smaller when the number of antennas is higher, the $L_d$ is longer, and the $L_p$ is longer. But compared with channels uncorrelated between the elements in $\bm h_n$, the correlated channels are addressed with longer $L_p$ to obtain acceptable results. Figure \ref{Fig5a} shows that if $M=64,L_d = 200$, $L_p\ge K=50$ can make DAD less than $5\times10^{-3}$. But $L_p\ge 60$ is needed to make MSE and SER less than $10^{-2}$ as $M=64,L_d = 200$ according to Figs. \ref{Fig5b} and \ref{Fig5c}. Thus, the proposed algorithm is applicable for the communication system with correlated channels, but the communication system needs to take on higher overheads to obtain satisfactory results.

\subsection{Analysis of Theoretical Performance}
In this section, we try to use the numerical results to verify the predicted performance in Section \ref{Per}. Figure \ref{Fig6a} illustrates the error probability of DAD and the predicted error probability by Theorem 2 versus antenna $M$ with different $L_p$. It is observed that the error probability decreases as $M$ increases, and the predictions of Theorem 2 characterize the results of numerical simulations. In addition, the reduction is more significant when $L_p$ is larger. Specifically, when $L_p=30$, $M$ is about $100$ to drive the error probability below $10^{-5}$; when $L_p=35$, $M\approx 67$ is needed; when $L_p=40$, just $M\approx 40$ is enough. Figure \ref{Fig6b} illustrates the MSE of CE and the predicted MSE of CE by (\ref{eq:cherror}) in Theorem 3 versus antenna $M$ with different $L_p$. It is observed that the MSE obtained numerically from the proposed algorithm is close to that predicted by Theorem 3. Although MSE decreases as $M$ and $L_p$ increase, the reduction is small when $M\ge 90$, $M\ge 80$, and $M\ge 70$ for $L_p=30$, $L_p=35$, and $L_p=40$, respectively. This is because the MSE converges to the point of  (\ref{eq:th}) in Theorem 3 when $v^{\mathsf r}$ converges. Figure \ref{Fig6c} illustrates the SER of SD and their predictions by Theorem 4 versus $M$ with different $L_p$. The numerical results match the predictions for different $L_p$. In addition, it is observed that SER decreases as $M$ increases, and SER reduces faster as $L_p$ increases. Note that we ignore some predicted values below $10^{-15}$.

\begin{figure*}[!t]
	\normalsize	
	\centering
	\subcaptionbox{Error probability of DAD\label{Fig7a}}{\includegraphics[width = 0.32\textwidth]{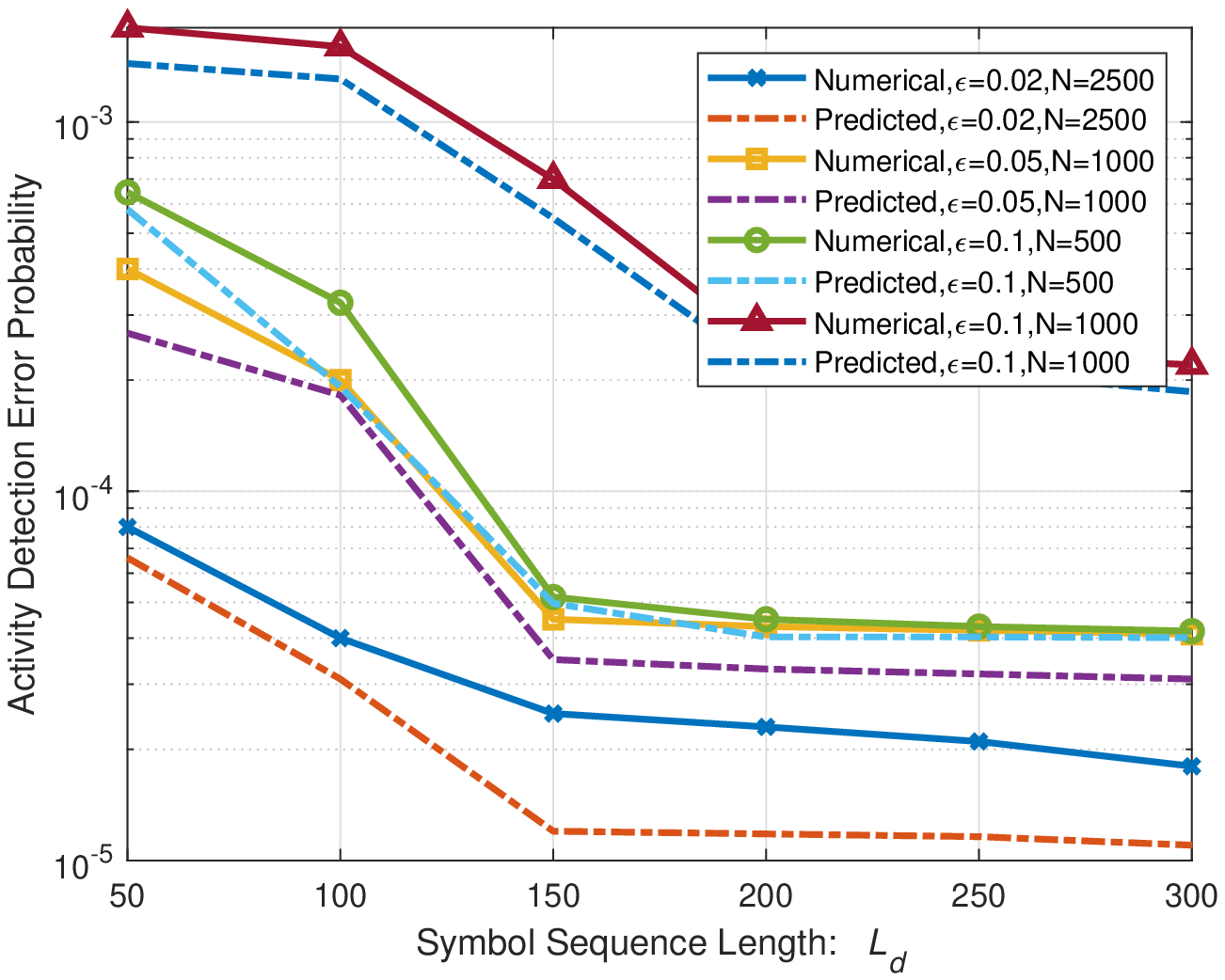}}
	\subcaptionbox{MSE of CE\label{Fig7b}}{\includegraphics[width = 0.32\textwidth]{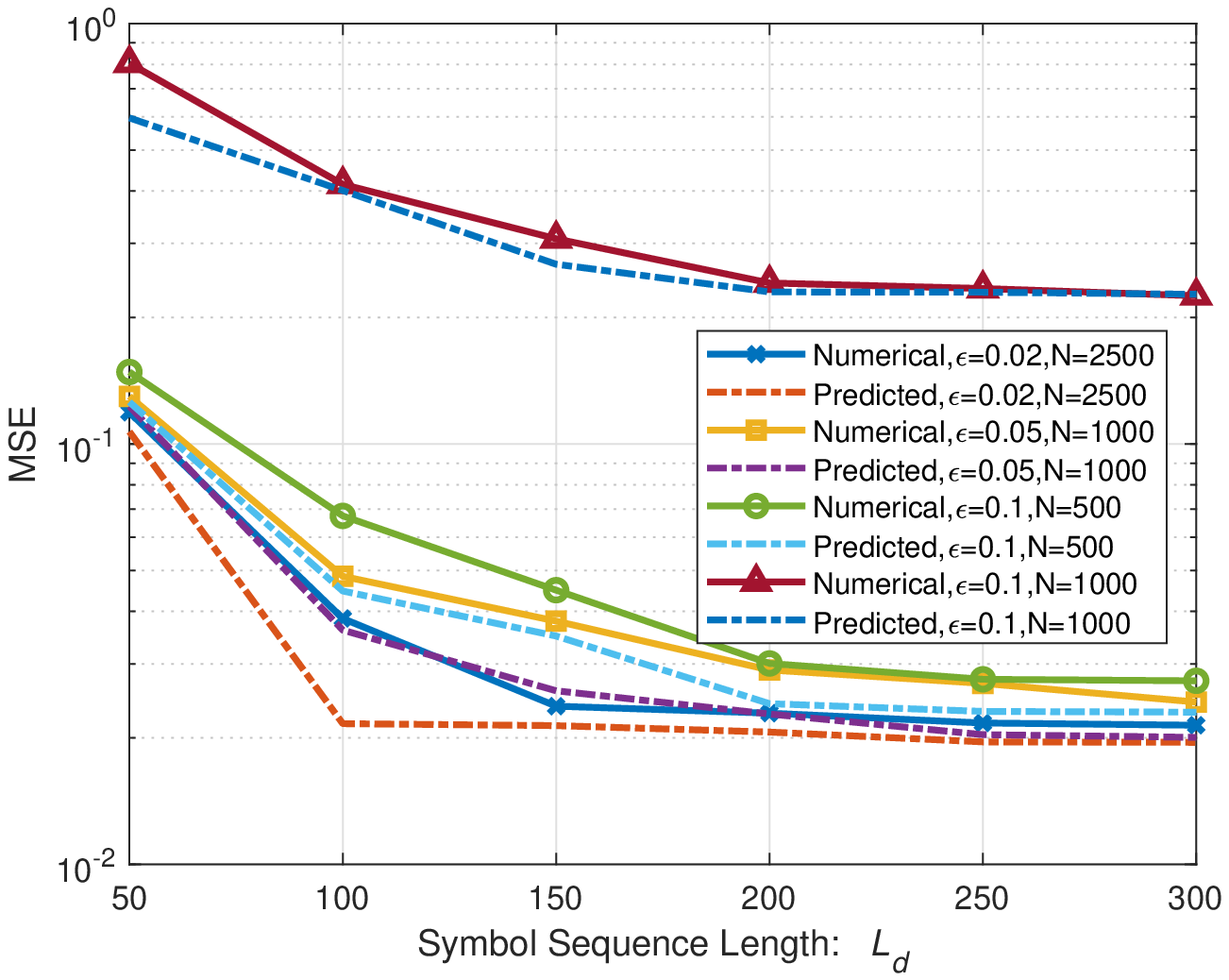}}
	\subcaptionbox{SER of SD\label{Fig7c}}{\includegraphics[width = 0.32\textwidth]{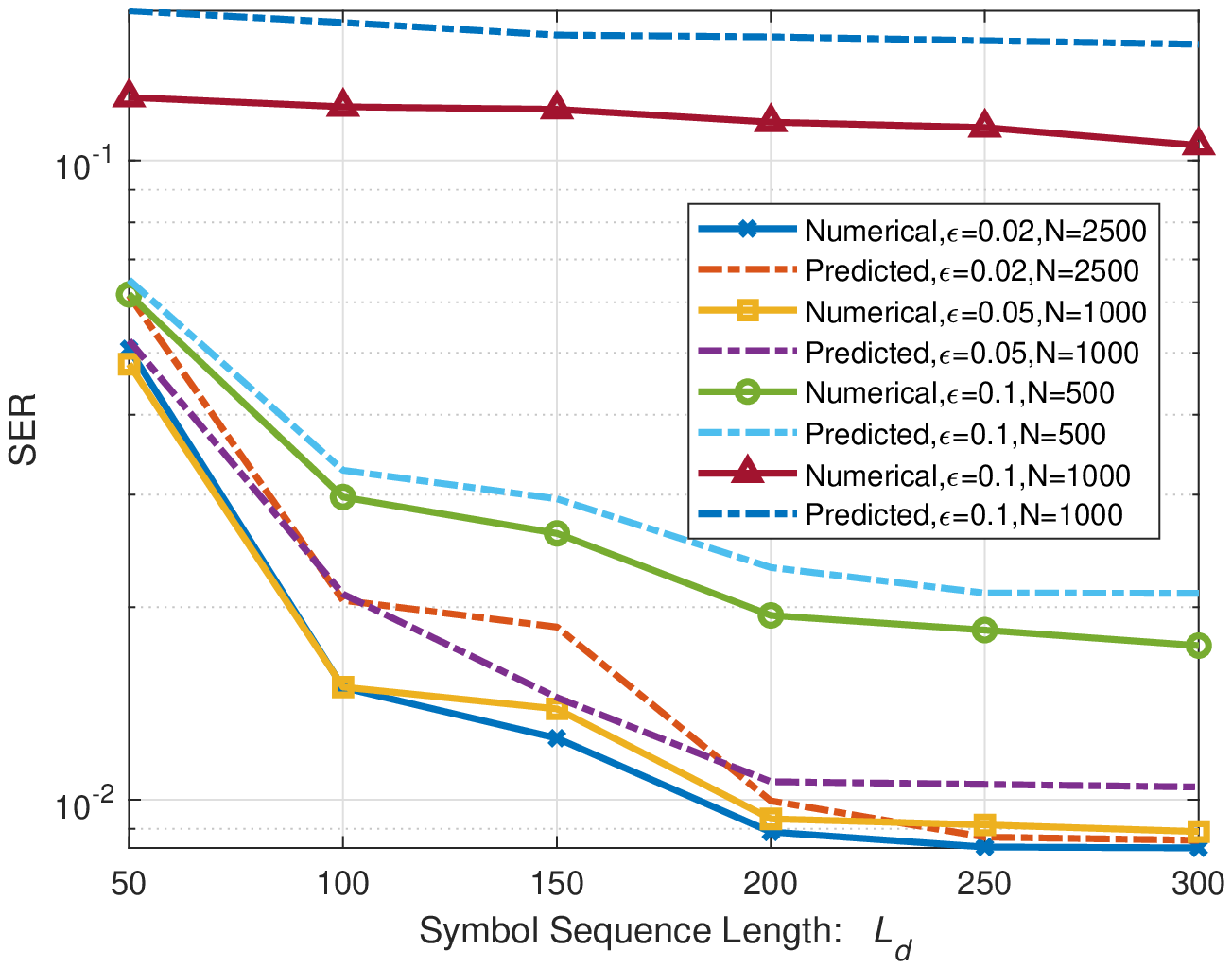}}
	\caption{\scriptsize{\subref{Fig7a}, \subref{Fig7b}, and \subref{Fig7c} are numerical results and the predictions for error probability of DAD, MSE of CE, and SER of SD, respectively, versus the symbol length $L_d$ with $L_p/K=0.66$ and $M=64$.}}
	\label{fig:7}
	\vspace*{-8pt}
\end{figure*}

Figures \ref{Fig7a}, \ref{Fig7b}, and \ref{Fig7c} show the numerical results and predictions for the error probability of DAD, MSE of CE, and SER of SD versus the symbol length $L_d$.  The results show that the longer the $L_d$ is, the lower the error probability, MSE, and SER are. But the performance improves very little when $L_d>100$. In addition,  according to Theorem 1, the proposed algorithm mainly relies on the relationship of $K$, $L$, and $M$. It is observed that even if the algorithm performs better as the $\varepsilon$ decreases, if $N\times\varepsilon=K$ is the same, the performance improvement of the algorithm is very limited, especially for MSE of CE and SER of SD.

\subsection{State Evolution}

\begin{figure}	
	\centering
	\begin{center}
		\includegraphics[width = 0.45\textwidth]{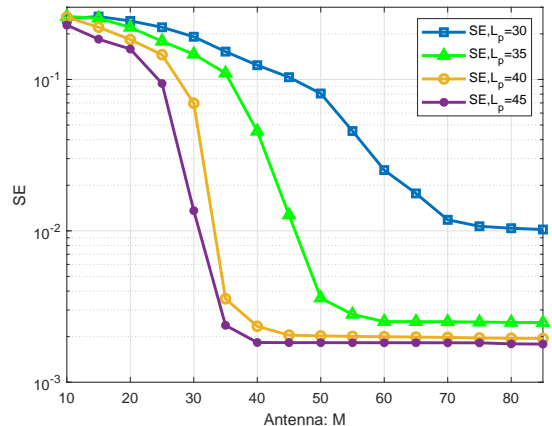}
	\end{center}
	\caption{\scriptsize{There is $N=1000$, $\varepsilon =0.05$, $L_d=100$. This figure shows the SE given by Theorem 1 versus $M$ with $L_p=30$, $L_p=35$, $L_p=40$, and $L_p=45$, respectively.}}
	\label{fig:8}
	\vspace{-1em}
\end{figure}

Figure \ref{fig:8} describes the SE in Theorem 1 versus $M$ with $L_p=30, L_p=35, L_p=40$, and $ L_p=45$, respectively. It shows that the SE decreases as $M$ increases, which means that the BiGAMP tends to obtain a more precise estimate of $\mathbf{AX}$. At the same time, the results show that the SE reduces rapidly when $L_p$ becomes larger. When $M=40$, $\tau$ approaches the convergence for $L_p=45$. However, it comes up to the convergence when $M=45$ and $M=60$ for $L_p=40$ and $L_p=35$, respectively. For $L_p=30$, SE converges until $M=80$.

\section{Conclusion}
\label{Con}
The joint DAD-CE-SD is a crucial issue for massive wireless connectivity applications. This paper proposes a BiGAMP algorithm to solve the joint DAD-CE-SD problem, which can take full advantage of the statistics of channels and signals, and helps to estimate channels and detect signals more accurately. The SE is adopted to describe the convergence performance and obtain the convergence conditions of the proposed algorithm. Meanwhile, we analyze the theoretical performance of DAD-CE-SD, which can be applied to predict the DAD-CE-SD's performance theoretically. Finally, the numerical results show that the proposed algorithm performs well for the DAD-CE-SD problem with fewer pilots, which is essential to support massive IoT scenarios.

\appendices

\section{ Derivation of ${\Delta}_{l\to n}^{\mathbf{x}}(\boldsymbol{x}_n,t)$}
\label{x}
Since $\mathbf z_l=\sum_{k=1}^N \mathsf a_{lk}\mathbf x_k\in \mathbb{C}^{M\times1}$, the mean and covariance matrix of $\mathbf z_l$ under the condition of $\mathbf x_l = \bm x_l$ are $\mathbb{E}[\mathbf{z}_l|\mathbf{x}_n=\boldsymbol{x}_n]=\hat{a}_{l,ln}(t)\boldsymbol{x}_n+{\hat {\boldsymbol p}_{l,n}(t)}$ and $\mathsf{var}\left[\mathbf{z}_l|\mathbf{x}_n=\boldsymbol{x}_n\right]=v_{l,ln}^{\mathsf{a}}(t)\boldsymbol{x}_n\boldsymbol{x}_n^H+\boldsymbol{v}_{l,n}^\mathsf{p}(t)$, respectively, where ${\hat {\boldsymbol p}_{l,n}(t)}=\sum_{k=1,k\ne n}^{N}\hat{a}_{l,lk}(t)\hat{\boldsymbol{x}}_{l,k}(t)$ and $\boldsymbol{v}_{l,n}^{\mathsf{p}}(t)=\sum_{k=1,k\ne n}^{N}|\hat{a}_{l,lk}(t)|^2{\boldsymbol{v}}_{l,k}^\mathbf{x}(t)+v_{l,lk}^{\mathsf{a}}(t)\hat{\boldsymbol{x}}_{l,k}(t)\hat{\boldsymbol{x}}_{l,k}^H(t)+v_{l,lk}^{\mathsf{a}}(t){\boldsymbol{v}}_{l,k}^\mathbf{x}(t)$.  According to the CLT, the distribution of the random variable $\mathbf z_l$ conditioned on $\mathbf{x}_n=\boldsymbol{x}_n$ can be characterized by a complex Gaussian distribution with a conditional mean and covariance matrix. Thus, the message ${I}_{l\to n}^{\mathbf{x}}(\boldsymbol{x}_n,t)$ is approximated as (\ref{eq:xmton1}),
\begin{figure*}[!t]
	\vspace*{-8pt}
	\normalsize
\begin{align}
I_{l\to n}^{\mathbf{x}}(\boldsymbol{x}_n,t)=&const\cdot \int_{\bm{a}_{l},\{{\boldsymbol{x}_r}\}_{r\ne n}} p_{{\mathbf{y}}_l|{\mathbf z_l}}({\boldsymbol{y}}_l|\boldsymbol{z}_l)\prod_{r=1,r\ne n}^{N} I_{l\leftarrow r}^{\mathbf{x}}(\boldsymbol{x}_r,t)\prod_{k=1}^N I_{l\leftarrow lk}^{\mathsf{a}}({a}_{lk},t)\nonumber\\\approx& const\cdot \int_{\boldsymbol{z}_l} p_{{\mathbf{y}}_l|{\mathbf{z}}_l}({\boldsymbol{y}}_l|\boldsymbol{z}_l)\mathcal{CN}\left(\boldsymbol{z}_l;\mathbb{E}[\mathbf{z}_l|\mathbf{x}_n=\boldsymbol{x}_n],\mathsf{var}[\mathbf{z}_l|\mathbf{x}_n=\boldsymbol{x}_n]\right)\nonumber \\
=&\exp\left({H_l\left({\hat{a}_{l,ln}(t)\boldsymbol{x}_n+{\hat {\boldsymbol p}_{l,n}(t)},v_{l,ln}^{\mathsf{a}}(t)\boldsymbol{x}_n\boldsymbol{x}_n^H+\boldsymbol{v}_{l,n}^\mathsf{p}(t);\boldsymbol{y}_l}\right)+ const}\right).\label{eq:xmton1}\\
\approx& const \cdot \exp(H_l)\cdot
\exp(2\mathsf{Re}[(\boldsymbol{x}_n-\hat{\boldsymbol{x}}_{n}(t))^H({\hat a}_{l,ln}^{*}(t)\nabla_{\boldsymbol u^{*}}H_l+v_{l,ln}^{\mathsf{a}}(t)\nabla_{\boldsymbol \Sigma}H_l\hat{\boldsymbol{x}}_{n}(t) +O({1/N^{3/2}}))]\nonumber\\ &+\mathsf{Re}[(\boldsymbol{x}_n-\hat{\boldsymbol{x}}_{n}(t))^H(\nabla_{\boldsymbol u^{*}\boldsymbol u}H_l|\hat{a}_{l,ln}(t)|^2 +v_{l,ln}^{\mathsf{a}}(t)\nabla_{\boldsymbol \Sigma}H_l +O({1/N^{3/2}}))(\boldsymbol{x}_n-\hat{\boldsymbol{x}}_{n}(t))]),\label{eq:xmn1}\\ \approx &const \cdot \exp(
2\mathsf{Re}[(\boldsymbol{x}_n-\hat{\boldsymbol{x}}_{n}(t))^H({\hat a}_{l,ln}^{*}(t)\nabla_{\boldsymbol u^{*}}H_l+v_{ln}^{\mathsf{a}}(t)\nabla_{\boldsymbol \Sigma}H_l\hat{\boldsymbol{x}}_{n}(t) )]\nonumber\\ &  +\mathsf{Re}[(\boldsymbol{x}_n-\hat{\boldsymbol{x}}_{n}(t))^H(\nabla_{\boldsymbol u^{*}\boldsymbol u}H_l|\hat{a}_{ln}(t)|^2 +v_{ln}^{\mathsf{a}}(t)\nabla_{\boldsymbol \Sigma}H_l )(\boldsymbol{x}_n-\hat{\boldsymbol{x}}_{n}(t))]).\label{eq:xmton2}
\end{align}
\hrulefill
	\vspace*{-8pt}
\end{figure*}
where
$
H_l(\boldsymbol u,\boldsymbol \Sigma;\boldsymbol y_l)\triangleq  \log \int_{\mathbf z_l} p_{\mathbf{y}_l|\mathbf{z}_l}(\boldsymbol y_l|\boldsymbol z_l)\mathcal{CN}(\boldsymbol z_l;\boldsymbol u,\boldsymbol \Sigma).
$ Plugging (\ref{eq:pmvar}) into $H_l$ term in (\ref{eq:xmton1}), there is
\begin{eqnarray}
\begin{aligned}
&H_l(\hat{a}_{l,ln}(t)\boldsymbol{x}_n+{\hat {\boldsymbol p}_{l,n}(t)},v_{l,ln}^{\mathsf{a}}(t)\boldsymbol{x}_n\boldsymbol{x}_n^H+\boldsymbol{v}_{l,n}^\mathsf{p}(t);\boldsymbol{y}_l)\\
&=H_l\left({\hat{a}_{l,ln}(t)(\boldsymbol{x}_n-\hat{\boldsymbol{x}}_{n}(t))+{\hat {\boldsymbol p}_{l}(t)}+O\left({1/N}\right),}\right.\\& \left.{ v_{l,ln}^{\mathsf{a}}(t)(\boldsymbol{x}_n\boldsymbol{x}_n^H-\hat{\boldsymbol{x}}_{n}(t)\hat{\boldsymbol{x}}_{n}^H(t))+\boldsymbol{v}_{l}^\mathsf{p}(t)+O\left({1/N}\right);\boldsymbol{y}_l}\right)
\label{eq:xltn}
\end{aligned}
\end{eqnarray}
Expanding (\ref{eq:xltn}) with the Taylor series in $\boldsymbol{x}_n$ at the point $\hat{\boldsymbol{x}}_{n}(t)$, then (\ref{eq:xmton1}) is written as (\ref{eq:xmn1}),
where $H_l$ is a simplified representation of $H_l({\hat {\boldsymbol p}_{l}(t)},\boldsymbol{v}_{l}^\mathsf{p}(t);\boldsymbol{y}_l)$ and $\nabla_{\boldsymbol u^{*}\bm u}H_l\triangleq \nabla_{\boldsymbol u^{*}}(\nabla_{\boldsymbol u}H_l)$. $\nabla_{\boldsymbol u^{*}}H_l$ and $\nabla_{\boldsymbol \Sigma}H_l$ are the derivations of $H_l$ with respect to conjugate $\bm u^*$ of the first parameter (under plural conditions) and  the second parameter $\bm \Sigma$, respectively.
As $N\to \infty$, the higher-order terms $O(1/N^{3/2})$ and $O(1/N)$ inside $H_l$ vanish. Replacing $|\hat{a}_{l,ln}(t)|^2$ by $|\hat{a}_{ln}(t)|^2$ and $v_{l,ln}^{\mathsf{a}}(t)$ by $v_{ln}^{\mathsf{a}}(t)$ since their error is $O(1/N^{3/2})$, ${I}_{l\to n}^{\mathbf{x}}(\boldsymbol{x}_n,t)$ is approximated as (\ref{eq:xmton2}).
Appendix A in \cite{RN157} proved  that
\begin{equation}
\begin{aligned}
\hat{\boldsymbol s}_l(t)=\nabla_{\boldsymbol u^{*}}H_l({\hat {\boldsymbol p}_{l}(t)},\boldsymbol{v}_{l}^\mathsf{p}(t);\boldsymbol{y}_l) = \boldsymbol{v}_{l}^\mathsf{p}(t)^{-1}(\hat{\boldsymbol{z}}_l(t)-\hat {\boldsymbol p}_{l}(t)),
\label{eq:Esa}
\end{aligned}
\end{equation}
\begin{equation}
\begin{aligned}
\boldsymbol{v}_{l}^\mathsf{s}(t)&=-\nabla_{\boldsymbol u^{*}\boldsymbol u}H_l({\hat {\boldsymbol p}_{l}(t)},\boldsymbol{v}_{l}^\mathsf{p}(t);\boldsymbol{y}_l)\\&= \boldsymbol{v}_{l}^\mathsf{p}(t)^{-1}(\mathbf I-\boldsymbol{v}_{l}^\mathsf{z}(t)\boldsymbol{v}_{l}^\mathsf{p}(t)^{-1}).
\label{eq:varsa}
\end{aligned}
\end{equation}
At the same time, $\nabla_{\boldsymbol u^{*}}H_l$,$\nabla_{\boldsymbol u^{*}\boldsymbol u}H_l$, and $\nabla_{\boldsymbol \Sigma}H_l$ satisfy the relationship
\begin{equation}
\nabla_{\boldsymbol \Sigma}H_l=\nabla_{\boldsymbol u^{*}}H_l(\nabla_{\boldsymbol u}H_l)^T+\nabla_{\boldsymbol u^{*}\boldsymbol u}H_l.
\label{eq:Es-vars}
\end{equation}
Plug (\ref{eq:Esa})-(\ref{eq:Es-vars}) into (\ref{eq:xmton2}), then
\begin{eqnarray}
\begin{aligned}
&{\textsc{I}}_{l\to n}^{\mathbf{x}}(\boldsymbol{x}_n,t)
 \approx const \\&\cdot \exp( \mathsf{Re}[2({\hat a}_{l,ln}^{*}(t)\hat{\boldsymbol s}_l(t) +\hat{\boldsymbol{x}}_{n}(t)\boldsymbol{v}_{l}^\mathsf{s}(t)|\hat{a}_{ln}(t)|^2)\boldsymbol{x}_n^H\\& +{\boldsymbol{x}_n^H}(v_{ln}^{\mathsf{a}}(t)(\hat{\boldsymbol s}_l(t)\hat{\boldsymbol s}_l^H(t) -\boldsymbol{v}_{l}^\mathsf{s}(t))-|\hat{a}_{ln}(t)|^2\boldsymbol{v}_{l}^\mathsf{s}(t))\boldsymbol{x}_n]).
\end{aligned}
\end{eqnarray}

\section{ Proof of (\ref{eq:vphat})}
\label{simplify}
Plug (\ref{eq:xn}) and (\ref{eq:xn2}) into (\ref{eq:pmvara}).
As $M\to \infty$, $\mathsf{Re}\left[\hat{\bm x}_{k}(t-1)^H\hat{\bm s}_{l}(t-1)v_{lk}^{\mathsf a}(t)/M\right] \hat{\boldsymbol x}_{k}(t)\to 0$, which will lose the massages to correct the $\hat{a}_{l,lk}(t)$. Hence, we use $\mathsf{Re}\left[\hat{\boldsymbol x}_{k}^*(t)\odot \hat {\boldsymbol s}_{l}(t-1)v_{lk}^{\mathsf a}(t)\right]\odot \hat{\boldsymbol x}_{k}(t)$ in place of $\mathsf{Re}\left[{\hat{\bm x}_{k}(t-1)^H}\right.\hat{\bm s}_{l}(t-1)\left.{v_{lk}^{\mathsf q}(t)/M}\right] \hat{\boldsymbol x}_{k}(t)$ and get (\ref{eq:p1}). Then replacing the $\hat{a}_{lk}^*(t-1)$ with $\hat{a}_{lk}^*(t)$ and neglecting terms $O({1/\sqrt{N}})$, ${\hat {\boldsymbol p}_{l}(t)}$ can be approximated as (\ref{eq:phat}).
$(a)$ denotes equal in probability when the real and imaginary parts are identically distributed. Similarly, plug (\ref{eq:xn}), (\ref{eq:xn2}), $\boldsymbol v_{n}^{\mathbf{x}}(t)=\boldsymbol v_{l,n}^{\mathbf{x}}(t)+ O(1/\sqrt N)$, and $v_{ln}^{\mathsf a}(t)=v_{l,ln}^{\mathsf a}(t)+O\left(1/N^{2/3}\right)$ into (\ref{eq:pmvarb}) with retaining only the $O(1)$ terms, $\boldsymbol{v}_{n}^{\mathsf{p}}(t)$ is approximated as (\ref{eq:vpha}).
\begin{figure*}[!t]
	\vspace*{-0.8cm}
	\normalsize
\begin{align}
\hat {\boldsymbol p}_{l}(t)&
 = \sum_{k=1}^{N}\hat{a}_{lk}(t)\hat{\boldsymbol x}_{k}(t)
-2\mathsf{Re}\left[\frac{1}{M}\hat{\bm x}_{k}^H(t-1)\hat{\bm s}_{l}(t-1)v_{lk}^{\mathsf a}(t)\right] \hat{\boldsymbol x}_{k}(t) - 2\mathsf{Re}\left[\left(\hat{a}_{lk}^*(t-1) \hat{\boldsymbol s}_l(t-1)\right)^H\boldsymbol v_{k}^{\mathbf x}(t)\right]^H\hat{a}_{lk}(t)+O(\frac{1}{\sqrt{N}})\nonumber
\nonumber\\&\approx \sum_{k=1}^{N}\hat{a}_{lk}(t)\hat{\boldsymbol x}_{k}(t)-
2\mathsf{Re}\left[\hat{\boldsymbol x}_{k}^*(t)\odot \hat {\boldsymbol s}_{l}(t-1)v_{lk}^{\mathsf a}(t)\right] \odot \hat{\boldsymbol x}_{k}(t)  - 2\mathsf{Re}\left[\left(\hat{a}_{lk}^*(t) \hat{\boldsymbol s}_l(t-1)\right)^H\boldsymbol v_{k}^{\mathbf x}(t)\right]^H\hat{a}_{lk}(t)\label{eq:p1}
\\ &\overset{(a)}{\approx} \sum_{k=1}^{N}\hat{a}_{lk}(t)\hat{\boldsymbol x}_{k}(t)  -\left( \sum_{k=1}^{N}|\hat{a}_{lk}(t)|^2\boldsymbol v_{k}^{\mathbf x}(t)+v_{lk}^{\mathsf a}(t)\hat{\boldsymbol x}_{k}(t)\hat{\boldsymbol x}_{k}^H(t)\right)\hat{\boldsymbol s}_l(t-1).\label{eq:phat}
\end{align}
\vspace*{-8pt}
\begin{eqnarray}
\begin{aligned}
\boldsymbol{v}_{n}^{\mathsf{p}}(t)\approx \sum_{k=1}^{N}|\hat{a}_{lk}(t)|^2 \boldsymbol v_{k}^{\mathbf x}(t)+v_{lk}^{\mathsf a}(t)\hat{\boldsymbol x}_{k}(t)\hat{\boldsymbol x}_{k}^H(t)+v_{lk}^{\mathsf a}(t)\boldsymbol v_{k}^{\mathbf x}(t)+O\left({1/\sqrt{N}}\right)\approx\bar {\boldsymbol v}_{l}^{\mathsf p}+\sum_{k=1}^{N}v_{lk}^{\mathsf a}(t)\boldsymbol v_{k}^{\mathbf x}(t).	
\end{aligned}\label{eq:vpha}
\end{eqnarray}
\vspace*{-8pt}
\end{figure*}

\section{ Proof of Proposition 1}
\label{appendix:p2}

To simplify the notation, we omit iteration $t$. Define a random vector $\mathbf r_n\triangleq \mathbf x_n + \mathbf w^{\mathsf r}$, where $\mathbf w^{\mathsf r}$ is a random vector following $\mathcal{CN}\left(\bm 0,  v_n^{\mathsf r}\mathbf I\right)$. Thus, $\mathbf r_n\sim\mathcal{CN}(\bm 0, \left(\beta_n+ v_n^{\mathsf r}\right)\mathbf I)$ if device $n$ is active; otherwise, $\mathbf r_n\sim\mathcal{CN}(\bm 0, v_n^{\mathsf r}\mathbf I)$. According to (\ref{eq:chap26}), (\ref{eq:Ex}),  and {\it Assumption 1}, we have
\begin{eqnarray}
\begin{aligned}
\hat{\boldsymbol x}_{n} =\mathbb{E}[\mathbf x_n|\mathbf r_n =\hat{\bm r}_n] =  \frac{\varepsilon}{C_{\bm x}}\int_{\boldsymbol x}\boldsymbol xp_{\mathbf h_n}(\bm x)\mathcal{CN}(\boldsymbol x; \hat{\boldsymbol r}_{n}(t),v_{n}^{\mathsf r}(t)\mathbf I),\nonumber
\end{aligned}
\end{eqnarray}
where $C_{\bm x}$ can be interpreted as $p(\mathbf r_n = \hat{\bm r}_n)$.  With
\begin{eqnarray}
\begin{aligned}
p_{\mathbf h_n}&(\bm x)\mathcal{CN}(\boldsymbol x; \hat{\boldsymbol r}_{n},v_{n}^{\mathsf r}\mathbf I)=\frac{\exp\left(-(\beta_n + v_{n}^{\mathsf r})^{-1}\hat{\boldsymbol r}_{n}^H\hat{\boldsymbol r}_{n}	\right)}{|\pi ( v_{n}^{\mathsf r}+\beta_n)|^M}\\&\cdot\mathcal{CN}(\bm x;\beta_n(\beta_n +v_{n}^{\mathsf r})^{-1}\hat{\boldsymbol r}_{n},\beta_n v_{n}^{\mathsf r}(t)\left(\beta_n+ v_{n}^{\mathsf r}(t)\right)^{-1}\mathbf I),
\end{aligned}
\end{eqnarray}
we have
\begin{eqnarray}
\begin{aligned}
\hat{\boldsymbol x}_{n}& =\phi(\hat{\boldsymbol r}_{n})\beta_n(\beta_n +v_{n}^{\mathsf r})^{-1}\hat{\boldsymbol r}_{n},
\label{eq:Ex2}
\end{aligned}
\end{eqnarray}
where
\begin{eqnarray}
\begin{aligned}
\phi(\hat{\boldsymbol r}_{n})&=\frac{p(\mathbf r_n=\hat{\bm r}_n,\alpha_n =1)}{p(\mathbf r_n=\hat{\bm r}_n)}\\&=\frac{\varepsilon}{C_{\bm x}}\frac{\exp(-(\beta_n + v_{n}^{\mathsf r})^{-1}\hat{\boldsymbol r}_{n}^H\hat{\boldsymbol r}_{n}
	)}{|\pi ( v_{n}^{\mathsf r}+\beta_n)|^M}.
\label{eq:active}
\end{aligned}
\end{eqnarray}
$\phi(\hat{\boldsymbol r}_{n})$ is the estimate of the active probability of device $n$. $v_{n}^{\mathsf{x}}$ can be obtained by differentiating $\hat{\boldsymbol r}_{n}$ in $\hat{\boldsymbol x}_{n}$.

\section{ Proof of Theorem 1}
\label{appendix:t1}

Eq. (\ref{xtau}) can be written as
\begin{eqnarray}
\begin{aligned}
\bm{\Gamma}(t)& 
=\mathbb{E}[(\bm y_l-{\hat {\boldsymbol p}_{l}(t)})(\bm y_l-{\hat {\boldsymbol p}_{l}(t)})^H]\\& =\mathbb{E}[(\bm z_l-{\hat {\boldsymbol p}_{l}(t)}+\bm w_l)(\bm z_l-{\hat {\boldsymbol p}_{l}(t)}+\bm w_l)^H]\\&=\mathbb E_{\mathbf p}\mathbb E_{\mathbf z|\mathbf p=\hat{\bm p}}[(\bm z_l-{\hat {\boldsymbol p}_{l}(t)})(\bm z_l-{\hat {\boldsymbol p}_{l}(t)})^H] +\sigma^2\mathbf I \\&=\mathbb E_{\mathbf p}[\bm v^{\mathsf p}_l(t)]+\sigma^2\mathbf I\overset{(b)}{=}( v^{\mathsf p}(t)+\sigma^2)\mathbf I,
\end{aligned}\label{eq:se}
\end{eqnarray}
where the expectation $\mathbb E_{\mathbf z|\mathbf p=\hat{\bm p}}[\cdot]$ is taken over $\mathbf z_l|\mathbf p_l\sim \mathcal{CN}(\hat{\bm p}_l,\bm v^{\mathsf p}_l)$ according to (\ref{eq:pp}). Due to the Onsager correction, it is approximated that $\bm w_l$ and $\bm z_l - \hat{\bm p}_l(t)$ are independent \cite{RN251,RN45}. (b) follows {\it Assumption 2}. Defining $\tau(t)\triangleq v^{\mathsf p}(t)+\sigma^2$, $\tau(t)$ is consistent with the state evolution in \cite{7457269}.

To guarantee that the algorithm converges, there must be ${\tau}(t+1)<{\tau}(t)$, i.e., $v^{\mathsf p}(t+1)<v^{\mathsf p}(t)$. In the asymptotic regime, $v^{\mathsf p}(t+1)$ can be approximated as (\ref{eq:vp}),
\begin{figure*}[!t]
	\vspace*{-8pt}
	\normalsize
\begin{eqnarray}
\begin{aligned}
v^{\mathsf p}(t+1)&\approx \frac{K}{L}\frac{\bar{\beta} v^{\mathsf r}(t)}{\bar{\beta} + v^{\mathsf r}(t)}+K\frac{v^{\mathsf q}(t)}{1+Lv^{\mathsf q}(t)}+K\frac{\bar{\beta} v^{\mathsf r}(t)}{\bar{\beta} + v^{\mathsf r}(t)}\frac{v^{\mathsf q}(t)}{1+Lv^{\mathsf q}(t)},
\label{eq:vp}
\end{aligned}
\end{eqnarray}
	\hrulefill
\vspace*{-8pt}
\end{figure*}
where $\bar \beta=\frac{1}{N}\sum_{ n=1}^{N}\beta_n$. From (\ref{eq:vp}), $v^{\mathsf p}(t+1)$  increases as $v^{\mathsf r}(t)$ and $v^{\mathsf q}(t)$ increase, which requires $v^{\mathsf r}(t+1)<v^{\mathsf r}(t)$ and $v^{\mathsf q}(t+1)<v^{\mathsf q}(t)$. In the asymptotic regime, according to (\ref{eq:xva}), (\ref{eq:qvar}), and {\it Assumption 2}, $v^{\mathsf r}(t+1)$ and $v^{\mathsf q}(t+1)$ are the function of $v^{\mathsf r}(t)$ and $v^{\mathsf q}(t)$ as follows
\begin{eqnarray}
\begin{aligned}
v^{\mathsf r}(t+1)&\approx\sigma^2 +v^{\mathsf p}(t+1)\triangleq\Phi_{\mathsf r}(v^{\mathsf r}(t),v^{\mathsf q}(t)),\\
v^{\mathsf q}(t+1)&\approx\frac{1}{M}\Phi_{\mathsf r}(v^{\mathsf r}(t),v^{\mathsf q}(t))\triangleq\Phi_{\mathsf q}( v^{\mathsf r }(t),v^{\mathsf q}(t)).
\end{aligned}
\end{eqnarray}
To ensure that $v^{\mathsf r}(t)$ and $v^{\mathsf q}(t)$ decrease as $t$ increases, there is
\begin{subequations}
\begin{align}
&\frac{\partial\Phi_{\mathsf r}(v^{\mathsf r }(t),v^{\mathsf q}(t))}{\partial v^{\mathsf r }(t)}\nonumber=\frac{K}{L}c_1<1, \\
&\frac{\partial\Phi_{\mathsf q}(v^{\mathsf r }(t),v^{\mathsf q}(t))}{\partial  v^{\mathsf q }(t)}\nonumber=\frac{K}{M}c_2<1,
\label{eq:vq1}
\end{align}
\end{subequations}
where constant $c_1$ and $c_2$ are
\begin{subequations}\label{eq:vr1}
	\begin{align}
	 c_1=\frac{\bar{\beta}^2}{(\bar{\beta} + v^{\mathsf r}(t))^2}+\frac{\bar{\beta}^2}{(\bar{\beta} + v^{\mathsf r}(t))^2}\frac{Lv^{\mathsf q}(t)}{1+Lv^{\mathsf q}(t)},\\ c_2=\frac{1}{(1+Lv^{\mathsf q}(t))^2}+\frac{\bar{\beta} v^{\mathsf r}(t)}{\bar{\beta} + v^{\mathsf r}(t)}\frac{1}{(1+Lv^{\mathsf q}(t))^2}.
	 \end{align}
 \end{subequations}
Without loss of generality, we assume $ v^{\mathsf r}(t), \bar \beta\le1$ and $v^{\mathsf q}(t) \ll 1$, then there is $\frac{1}{4}=\frac{\bar \beta^2}{(2\bar\beta)^2}<c_1<\frac{2\bar\beta^2}{(\bar\beta+v^{\mathsf r}(t))^2}<2$ and $\frac{1}{4}<\frac{1}{(1+Lv^{\mathsf q}(t))^2}<c_2<\frac{2}{(1+Lv^{\mathsf q}(t))^2}<2$. Therefore, the algorithm is convergent as
\begin{eqnarray}
\begin{aligned}
	L>c_1K, \quad \quad
	M>c_2K.
\end{aligned}
\end{eqnarray}

\section{ Proof of Theorem 2}
\label{appendix:t2}
Note that we omit iteration $t$ for simplification. According to \textbf{Definition 1}, the probability of $\hat\alpha_n = 0$ can be expressed as
\begin{eqnarray}
\begin{aligned}
P(\hat\alpha_n = 0)&=P(\phi(\bm r_n)\leq \varepsilon) 
= P({\boldsymbol r}_{n}^H{\boldsymbol r}_{n} \leq \theta),
\end{aligned}
\end{eqnarray}
where $\theta = M\log \frac{\beta_n+v_{n}^{\mathsf r}} { v_{n}^{\mathsf r}}/\frac{\beta_n}{v_n^{\mathsf r}(\beta_n + v_n^{\mathsf r})}$.
According to the definition of $\mathbf r_n$ in Appendix \ref{appendix:p2}, we have $\mathbf r_n \sim \mathcal{CN}(\bm 0, (v_n^\mathsf r + v_n^\mathsf x)\mathbf I)$ given $\alpha_n =1$ and $\mathbf r_n \sim \mathcal{CN}(\bm 0, v_n^\mathsf r \mathbf I))$ given $\alpha_n = 0$. Since $\mathbf r_n$'s real and imaginary modules are i.i.d., the random variables ${\mathbf r}_{n}^H{\mathbf r}_{n}/((v_n^\mathsf r + v_n^\mathsf x) /2)$ and ${\mathbf r}_{n}^H{\mathbf r}_{n}/(v_n^\mathsf r /2)$ follow $\chi^2$ distribution with $2M$ degree-of-freedom (DoF). Defining random variables $G_1=2{\mathbf r}_{n}^H{\mathbf r}_{n}/(v_n^\mathsf r + \beta_n)\sim \chi^2(2M)$ and $G_0=2{\mathbf r}_{n}^H{\mathbf r}_{n}/v_n^\mathsf r\sim \chi^2(2M)$, we have
\begin{subequations}
\begin{align}
P(\hat \alpha_n = 0|\alpha_n = 1)&=P({\boldsymbol r}_{n}^H{\boldsymbol r}_{n} \leq \theta|\alpha_n =1)\nonumber\\&=P(G_1\leq \frac{2\theta}{v_n^\mathsf r + \beta_n}) =\frac{\underline{\Gamma}(M,Mc_{n,t})}{\Gamma(M)},
\\
P(\hat \alpha_n = 1|\alpha_n = 0)&= P({\boldsymbol r}_{n}^H{\boldsymbol r}_{n} > \theta|\alpha_n =0)\nonumber\\&=P(G_0> \frac{2\theta}{v_n^\mathsf r + \beta_n})=\frac{\bar{\Gamma}(M,Mb_{n,t})}{\Gamma(M)}.
\end{align}
\end{subequations}
Therefore, the error probability of activity detection is
\begin{eqnarray}
\begin{aligned}
P_{n,t}^e(M)& =(1-\varepsilon)\frac{\bar{\Gamma}(M,Mb_{n,t})}{\Gamma(M)} + \varepsilon\frac{\underline{\Gamma}(M,Mc_{n,t})}{\Gamma(M)}.
\end{aligned}
\end{eqnarray}

\section{ Proof of Theorem 3}
\label{appendix:t3}

Substituting (\ref{eq:xhat}) into  (\ref{eq:esHA}), $\hat{\bm h}_{k,t}$ can be expressed as
\begin{eqnarray}
\begin{aligned}
\hat{\bm h}_{k,t} &= \phi(\hat{\boldsymbol r}_{k}(t))\frac{\beta_k}{\beta_k+ v_{k}^{\mathsf r}(t)}\hat{\boldsymbol r}_{k}(t)\\& \overset{(c)}{=}\phi_{k,t}
\frac{\beta_k}{\beta_k+ v_{k}^{\mathsf r}(t)}(\bm h_k + \bm w_k^{\mathsf r}(t)),
\label{eq:hath}
\end{aligned}
\end{eqnarray}
where (c) follows $\hat{\boldsymbol r}_{k}(t)=\bm h_k + \bm w_k^{\mathsf r}(t)$ and $\bm w_k^{\mathsf r}(t)$ is generated by $\mathcal{CN}(\bm 0, v_k^{\mathsf r}(t)\mathbf I)$, which is
illustrated in Appendix \ref{appendix:p2}.
For convenience, denote $\phi_{k,t}= \phi(\bm h_k + \bm w_k^{\mathsf r}(t))$. Then the error is
\begin{eqnarray}
\Delta{\bm h}_{k,t} = \phi_{k,t}\frac{\beta_k}{\beta_k+ v_{k}^{\mathsf r}(t)}(\bm h_k + \bm w_k^{\mathsf r}(t))-\bm h_k.
\label{eq:errorh}
\end{eqnarray}
Eq. (\ref{eq:hath}) and (\ref{eq:errorh}) indicate $\hat{\bm h}_k(t)$ and $\Delta{\bm h}_k(t)$ are random vectors. In the asymptotic regime, $\lim_{M\to \infty}\phi(\hat{\boldsymbol r}_{n}(t))$ is either $0$ or $1$ for any device $n$ according to (\ref{eq:thet}). Since device $k$ is active, i.e., $\hat{\alpha}_{k,t}=1$, there is $\lim_{M\to \infty}\phi_{k,t}=1$. Then Theorem 3 can be derived.

\section{Proof of Theorem 4}
\label{appendix:t4}
\begin{figure*}[!t]
	\vspace*{-8pt}
	\normalsize
	\begin{eqnarray}
	\begin{aligned}
	&Cov(\Delta \bm d_k^{n_{s}},\Delta \bm d_k^{n_{s}})= v^{\Delta\mathsf d}\mathbf I=\frac{1}{J}\mathbb{E}[(\frac{\bm d_{k}^{n_{s}}+\bm w^{\mathsf q}} {1+Lv^{\mathsf q}}-\bm d_k^{n_{s}})^H(\frac{\bm d_{k}^{n_{s}}+\bm w^{\mathsf q}} {1+Lv^{\mathsf q}}-\bm d_k^{n_{s}})]=\frac{v^{\mathsf q}}{1+Lv^{\mathsf q}}\mathbf I=v^{\mathsf a}\mathbf I.
	\end{aligned}\label{eq:cov}
	\end{eqnarray}
	\hrulefill
	\vspace*{-8pt}
\end{figure*}
Similar to the analysis of channel estimation, for active device $k$, 
according to (\ref{eq:avard}) and (\ref{eq:esHA}), the estimated $\hat{d}_{lk}$ can be expressed as
\begin{eqnarray}
\begin{aligned}
\hat{d}_{lk} = \frac{\hat q_{lk} } {1+Lv_{lk}^{\mathsf q}}=\frac{1 } {1+Lv_{lk}^{\mathsf q}}(d_{lk}+w_{lk}^{\mathsf q}),
\end{aligned}
\end{eqnarray}
where $w_{lk}^{\mathsf q}$ is generated by $\mathcal{CN}( 0, v_{lk}^{\mathsf q})$. Then there is
\begin{eqnarray}
\Delta{d}_{k} = \frac{d_{lk}+w_{lk}^{\mathsf q} } {1+Lv_{lk}^{\mathsf q}}-d_{lk}.
\label{eq:errora}
\end{eqnarray}
Considering { \it Assumption 2}, we have
\begin{eqnarray}
\begin{aligned}
\hat{\bm d}_{k}^{n_{s}}&=\frac{1 } {1+Lv^{\mathsf q}}(\bm d_{k}^{n_{s}}+\bm w^{\mathsf q}),\\\Delta {\bm d}_{k}^{n_{s}}&= \frac{\bm d_{k}^{n_{s}}+\bm w^{\mathsf q}} {1+Lv^{\mathsf q}}-\bm d_{k}^{n_{s}},
\end{aligned}
\end{eqnarray}
where $\bm w^{\mathsf q}$ is generated by $\mathcal{CN}(\bm 0, v^{\mathsf q}\mathbf I)$. Then the covariance matrix of the estimation error is (\ref{eq:cov}).
Thus, $\Delta{\bm d}_k^{n_{s}}$ can be interpreted as a random vector generated by $ \mathcal{CN}(\bm 0, v^{\mathsf a}\mathbf I)$. 
To prove Theorem 4, we consider the Gallager-type bound and let $\bm d_k^{'n_{s}}\in\mathcal{D} \backslash \{\bm d_k^{n_{s}}\}$.  Considering the signal detection in Section \ref{SER}, define error events $F(\bm d_k^{n_{s}},\bm d_k^{'n_{s}})\triangleq \{\|\bm d_k^{n_{s}}-\bm d_k^{'n_{s}}+\Delta\bm d_k^{n_{s}}\|_2<\|\Delta\bm d_k^{n_{s}}\|_2\}$ \cite{RN239} and $F(\bm d_k^{n_{s}})\triangleq\cup_{\bm d_k^{'n_{s}}\in\mathcal{D} \backslash \{\bm d_k^{n_{s}}\}}F(\bm d_k^{n_{s}},\bm d_k^{'n_{s}})$.
Next, given $\lambda>0$ and $\mathbf z\sim\mathcal{CN}(\bm 0, \mathbf I)$, the following identity holds.
\begin{eqnarray}
\begin{aligned}
\mathbb{E}[\exp (-\lambda\|\sqrt a\bm z + \bm u\|_2^2)]=\frac{\exp (\frac{-\lambda\| \bm u\|_2^2}{1+a\lambda})}{(1+a\lambda)^J},
\label{eq:ineq}
\end{aligned}
\end{eqnarray}
Using Chernoff bound and (\ref{eq:ineq}), there is
\begin{eqnarray}
\begin{aligned}
&\mathbb{P}(F(\bm d_k^{n_{s}},\bm d_k^{'n_{s}})|\bm d_k^{n_{s}},\Delta\bm d_k^{n_{s}})\\&=\mathbb{P}(\|\bm d_k^{n_{s}}-\bm d_k^{'n_{s}}+\Delta\bm d_k^{n_{s}}\|_2^2<\|\Delta\bm d_k^{n_{s}}\|_2^2|\bm d_k^{n_{s}},\Delta\bm d_k^{n_{s}})\\&=\mathbb{P}(\exp(-\lambda\|\bm d_k^{n_{s}}-\bm d_k^{'n_{s}}+\Delta\bm d_k^{n_{s}}\|_2^2)\\&\qquad\qquad>\exp(-\lambda\|\Delta\bm d_k^{n_{s}}\|_2^2)|\bm d_k^{n_{s}},\Delta\bm d_k^{n_{s}})\\&\leq \frac{\exp(\lambda\|\Delta\bm d_k^{n_{s}}\|_2^2)}{(1+\frac{\lambda}{L})^J}\exp(-\frac{\lambda\|\bm d_k^{n_{s}}+\Delta\bm d_k^{n_{s}}\|_2^2}{1+\frac{\lambda}{L}}).
\end{aligned}
\end{eqnarray}
Then we invoke Gallager's $\rho$-trick, i.e. $\mathbb{P}\left[\cup_d A_d\right]\leq\left(\sum_d \mathbb{P}[A_d]\right)^\rho$ for any $\rho\in [0,1]$, to get
\begin{eqnarray}
\begin{aligned}
\mathbb{P}(F(\bm d_k^{n_{s}})&|\bm d_k^{n_{s}},\Delta\bm d_k^{n_{s}})\\&\leq (D-1)^{\rho}\frac{\exp(\lambda\rho(\|\Delta\bm d_k^{n_{s}}\|_2^2-\frac{\|\bm d_k^{n_{s}}+\Delta\bm d_k^{n_{s}}\|_2^2}{1+\frac{\lambda}{L}}))}{(1+\frac{\lambda}{L})^{\rho J}}.\nonumber
\end{aligned}
\end{eqnarray}
Employing (\ref{eq:ineq}) twice to take expectation over $\bm d_k^{n_{s}}$ and $\Delta\bm d_k^{n_{s}}$, we get
\begin{eqnarray}
\begin{aligned}
\mathbb{P}(F(\bm d_k^{n_{s}}))\leq (D-1)^{\rho}\frac{1}{(1+\frac{\lambda}{L})^{\rho J}}\frac{1}{(1+\frac{\mu}{L})^J}\frac{1}{(1-\mu_1 v^{\mathsf a})^J},
\end{aligned}
\end{eqnarray}
where $\mu = \rho\lambda/(1+\frac{\lambda}{L})$ and $\mu_1=\rho\lambda-\mu/(1+\frac{\mu}{L})$. Therefore, we have
\begin{eqnarray}
\begin{aligned}
P_{d}^e&\leq
\mathbb{P}(F(\bm d_k^{n_{s}}))
\leq\exp(-JE_{\rho,\lambda}),
\label{eq:pe}
\end{aligned}
\end{eqnarray}
where
\begin{eqnarray}
\begin{aligned}
E_{\rho,\lambda}=&\frac{\rho}{J}\ln(D-1)+\rho\ln (1+\frac{\lambda}{L})\\&\qquad+\ln(1+\frac{\mu}{L})+\ln(1-\mu_1 v^{\mathsf a}),
\label{eq:Erho}
\end{aligned}
\end{eqnarray}
with $1-\mu_1 v^{\mathsf a}>0$. The optimum value of $\lambda$ which maximizes $E_{\rho,\lambda}$ is given by $\lambda = \frac{1}{v^{\mathsf a}(1+\rho)}$. Plugging $\lambda = \frac{1}{v^{\mathsf a}(1+\rho)}$  into (\ref{eq:Erho}) and (\ref{eq:pe}), we have
\begin{eqnarray}
\begin{aligned}
P_{d}^e\leq\exp\left({-\rho\ln\left(D-1\right)-J\rho\ln \left(1+\frac{1}{Lv^{\mathsf a}(1+\rho)}\right)}\right).
\end{aligned}
\end{eqnarray}


\ifCLASSOPTIONcaptionsoff
  \newpage
\fi

\bibliographystyle{IEEEtran}
\bibliography{IEEEabrv}

\begin{thebibliography}{10}
\providecommand{\url}[1]{#1}
\csname url@samestyle\endcsname
\providecommand{\newblock}{\relax}
\providecommand{\bibinfo}[2]{#2}
\providecommand{\BIBentrySTDinterwordspacing}{\spaceskip=0pt\relax}
\providecommand{\BIBentryALTinterwordstretchfactor}{4}
\providecommand{\BIBentryALTinterwordspacing}{\spaceskip=\fontdimen2\font plus
\BIBentryALTinterwordstretchfactor\fontdimen3\font minus
  \fontdimen4\font\relax}
\providecommand{\BIBforeignlanguage}[2]{{%
\expandafter\ifx\csname l@#1\endcsname\relax
\typeout{** WARNING: IEEEtran.bst: No hyphenation pattern has been}%
\typeout{** loaded for the language `#1'. Using the pattern for}%
\typeout{** the default language instead.}%
\else
\language=\csname l@#1\endcsname
\fi
#2}}
\providecommand{\BIBdecl}{\relax}
\BIBdecl

\bibitem{RN175}
X.~Chen, D.~W.~K. Ng, W.~Yu, E.~G. Larsson, N.~Al-Dhahir, and R.~Schober,
  ``Massive access for {5G} and beyond,'' \emph{IEEE J. Sel. Areas Commun.},
  vol.~39, no.~3, pp. 615--637, 2021.

\bibitem{RN56}
L.~Liu, E.~G. Larsson, W.~Yu, P.~Popovski, C.~Stefanovic, and E.~d. Carvalho,
  ``Sparse signal processing for grant-free massive connectivity: A future
  paradigm for random access protocols in the internet of things,'' \emph{IEEE
  Signal Process. Mag.}, vol.~35, no.~5, pp. 88--99, 2018.

\bibitem{chap4_mtm}
M.~Hasan, E.~Hossain, and D.~Niyato, ``Random access for machine-to-machine
  communication in {LTE-advanced} networks: issues and approaches,'' \emph{IEEE
  Commun. Mag.}, vol.~51, no.~6, pp. 86--93, 2013.

\bibitem{chap4_mimo}
E.~Björnson, E.~de~Carvalho, J.~H. Sørensen, E.~G. Larsson, and P.~Popovski,
  ``A random access protocol for pilot allocation in crowded massive {MIMO}
  systems,'' \emph{IEEE Trans. Wireless Commun.}, vol.~16, no.~4, pp.
  2220--2234, 2017.

\bibitem{3gpp.36.331}
3GPP, ``Uplink multiple access schemes for {NR}: R1-165174,'' Tech. Rep., May
  2016.

\bibitem{RN244}
L.~Dai, B.~Wang, Y.~Yuan, S.~Han, I.~Chih-lin, and Z.~Wang, ``Non-orthogonal
  multiple access for {5G}: solutions, challenges, opportunities, and future
  research trends,'' \emph{IEEE Commun. Mag.}, vol.~53, no.~9, pp. 74--81,
  2015.

\bibitem{RN186}
F.~Wei, W.~Chen, Y.~Wu, J.~Ma, and T.~A. Tsiftsis, ``Message-passing receiver
  design for joint channel estimation and data decoding in uplink grant-free
  {SCMA} systems,'' \emph{IEEE Trans. Wireless Commun.}, vol.~18, no.~1, pp.
  167--181, 2019.

\bibitem{RN307}
F.~Wei, W.~Chen, Y.~Wu, J.~Li, and Y.~Luo, ``Toward {5G} wireless interface
  technology: Enabling nonorthogonal multiple access in the sparse code
  domain,'' \emph{IEEE Veh. Technol. Mag.}, vol.~13, no.~4, pp. 18--27, 2018.

\bibitem{RN245}
R.~Tibshirani, ``Regression shrinkage and selection via the lasso,'' \emph{J.
  Roy. Stat. Soc. Ser. B}, vol.~58, no.~1, pp. 267--288, 1996.

\bibitem{RN281}
Z.~Qin, K.~Scheinberg, and D.~Goldfarb, ``Efficient block-coordinate descent
  algorithms for the group lasso,'' \emph{Mathematical Programming
  Computation}, vol.~5, no.~2, pp. 143--169, 2013.

\bibitem{RN133}
S.~Boyd, N.~Parikh, E.~Chu, B.~Peleato, and J.~Eckstein, ``Distributed
  optimization and statistical learning via the alternating direction method of
  multipliers,'' \emph{Found. Trends Mach. Learn.}, vol.~3, pp. 1--122, 2011.

\bibitem{9374476}
A.~Fengler, S.~Haghighatshoar, P.~Jung, and G.~Caire, ``Non-bayesian activity
  detection, large-scale fading coefficient estimation, and unsourced random
  access with a massive {MIMO} receiver,'' \emph{IEEE Trans. Inf. Theory},
  vol.~67, no.~5, pp. 2925--2951, 2021.

\bibitem{RN164}
L.~Liu and W.~Yu, ``Massive connectivity with massive {MIMO}—{Part I}: Device
  activity detection and channel estimation,'' \emph{IEEE Trans. Signal
  Process.}, vol.~66, no.~11, pp. 2933--2946, 2018.

\bibitem{RN106}
M.~Ke, Z.~Gao, Y.~Wu, X.~Gao, and R.~Schober, ``Compressive sensing-based
  adaptive active user detection and channel estimation: Massive access meets
  massive {MIMO},'' \emph{IEEE Trans. Signal Process.}, vol.~68, pp. 764--779,
  2020.

\bibitem{RN114}
Y.~Cheng, L.~Liu, and L.~Ping, ``Orthogonal {AMP} for massive access in
  channels with spatial and temporal correlations,'' \emph{IEEE J. Sel. Areas
  Commun.}, vol.~39, no.~3, pp. 726--740, 2021.

\bibitem{RN251}
S.~Rangan, P.~Schniter, and A.~K. Fletcher, ``Vector approximate message
  passing,'' in \emph{Proc. IEEE ISIT}, Jun. 2017, pp. 1588--1592.

\bibitem{2010arXiv1010.5141R}
S.~{Rangan}, ``{Generalized Approximate Message Passing for Estimation with
  Random Linear Mixing},'' p. arXiv:1010.5141, Oct. 2010.

\bibitem{8550778}
Y.~Yang, J.~Sun, H.~Li, and Z.~Xu, ``{ADMM-CSNet}: A deep learning approach for
  image compressive sensing,'' \emph{IEEE Trans. Pattern Anal. Machine
  Intell.}, vol.~42, no.~3, pp. 521--538, 2020.

\bibitem{RN148}
Y.~Cui, S.~Li, and W.~Zhang, ``Jointly sparse signal recovery and support
  recovery via deep learning with applications in {MIMO}-based grant-free
  random access,'' \emph{IEEE J. Sel. Areas Commun.}, pp. 1--1, 2020.

\bibitem{RN225}
W.~Zhu, M.~Tao, X.~Yuan, and Y.~Guan, ``Deep-learned approximate message
  passing for asynchronous massive connectivity,'' \emph{IEEE Trans. Wireless
  Commun.}, vol.~20, no.~8, pp. 5434--5448, 2021.

\bibitem{RN166}
L.~Liu and W.~Yu, ``Massive connectivity with massive {MIMO}—{Part II}:
  Achievable rate characterization,'' \emph{IEEE Trans. Signal Process.},
  vol.~66, no.~11, pp. 2947--2959, 2018.

\bibitem{RN108}
S.~Jiang, X.~Yuan, X.~Wang, C.~Xu, and W.~Yu, ``Joint user identification,
  channel estimation, and signal detection for grant-free {NOMA},'' \emph{IEEE
  Trans. Wireless Commun.}, vol.~19, no.~10, pp. 6960--6976, 2020.

\bibitem{RN270}
K.~Senel and E.~G. Larsson, ``Grant-free massive {MTC-Enabled} massive {MIMO}:
  A compressive sensing approach,'' \emph{IEEE Trans. Commun.}, vol.~66,
  no.~12, pp. 6164--6175, 2018.

\bibitem{RN69}
Z.~Chen, F.~Sohrabi, Y.~Liu, and W.~Yu, ``Covariance based joint activity and
  data detection for massive random access with massive {MIMO},'' in
  \emph{Proc. IEEE ICC}, Shanghai, China, May 2019, pp. 1--6.

\bibitem{RN157}
J.~T. Parker, P.~Schniter, and V.~Cevher, ``Bilinear generalized approximate
  message passing—{Part I}: Derivation,'' \emph{IEEE Trans. Signal Process.},
  vol.~62, no.~22, pp. 5839--5853, 2014.

\bibitem{RN83}
T.~Ding, X.~Yuan, and S.~C. Liew, ``Sparsity learning-based multiuser detection
  in grant-free massive-device multiple access,'' \emph{IEEE Trans. Wireless
  Commun.}, vol.~18, no.~7, pp. 3569--3582, 2019.

\bibitem{RN45}
M.~Bayati and A.~Montanari, ``The dynamics of message passing on dense graphs,
  with applications to compressed sensing,'' \emph{IEEE Trans. Inf. Theory},
  vol.~57, no.~2, pp. 764--785, 2011.

\bibitem{RN42}
D.~L. Donoho, A.~Maleki, and A.~Montanari, ``Message passing algorithms for
  compressed sensing: {II}. analysis and validation,'' \emph{Proc. IEEE ITW},
  pp. 1--5, 2010.

\bibitem{7457269}
Y.~Kabashima, F.~Krzakala, M.~Mézard, A.~Sakata, and L.~Zdeborová, ``Phase
  transitions and sample complexity in bayes-optimal matrix factorization,''
  \emph{IEEE Trans. Inf. Theory}, vol.~62, no.~7, pp. 4228--4265, 2016.

\bibitem{RN286}
Y.~Polyanskiy, ``A perspective on massive random-access,'' in \emph{Proc. IEEE
  ISIT}, 2017, pp. 2523--2527.

\bibitem{RN287}
I.~Zadik, Y.~Polyanskiy, and C.~Thrampoulidis, ``Improved bounds on gaussian
  {MAC} and sparse regression via gaussian inequalities,'' in \emph{Proc. IEEE
  ISIT}, 2019, pp. 430--434.

\bibitem{RN239}
S.~S. Kowshik and Y.~Polyanskiy, ``Fundamental limits of many-user {MAC} with
  finite payloads and fading,'' \emph{IEEE Trans. Inf. Theory}, vol.~67, no.~9,
  pp. 5853--5884, 2021.

\bibitem{RN313}
S.~M. Kay, ``Fundamentals of statistical signal processing: estimation
  theory.''\hskip 1em plus 0.5em minus 0.4em\relax Prentice-Hall, Inc., 1993.

\bibitem{RN221}
B.~J. Frey and D.~MacKay, ``A revolution: Belief propagation in graphs with
  cycles,'' in \emph{Proc. Neural. Inf. Process. Syst. Conf.}, 1997, pp.
  479--485.

\bibitem{RN199}
F.~R. Kschischang, B.~J. Frey, and H.~Loeliger, ``Factor graphs and the
  sum-product algorithm,'' \emph{IEEE Trans. Inf. Theory}, vol.~47, no.~2, pp.
  498--519, 2001.

\bibitem{10.5555/2073796.2073849}
K.~P. Murphy, Y.~Weiss, and M.~I. Jordan, ``Loopy belief propagation for
  approximate inference: An empirical study,'' in \emph{Proc. Uncertainty in
  AI}, San Francisco, CA, USA, 1999, p. 467–475.

\bibitem{RN43}
D.~L. Donoho, A.~Maleki, and A.~Montanari, ``Message passing algorithms for
  compressed sensing: I. motivation and construction,'' in \emph{Proc. IEEE
  ITW}, Cairo, Jan. 2010.

\bibitem{RN140}
J.~T. Parker, P.~Schniter, and V.~Cevher, ``Bilinear generalized approximate
  message passing—{Part II}: Applications,'' \emph{IEEE Trans. Signal
  Process.}, vol.~62, no.~22, pp. 5854--5867, 2014.

\bibitem{RN209}
M.~I. Jordan and M.~J. Wainwright, ``Graphical models, exponential families,
  and variational inference,'' \emph{Foundations and Trends in Machine
  Learning}, vol.~1, no. 1–2, pp. 1--305, 2007.

\bibitem{RN218}
D.~J. MacKay, ``Information theory, inference, and learning algorithms.''\hskip
  1em plus 0.5em minus 0.4em\relax Cambridge University Press, 2003.

\bibitem{Gautschi98theincomplete}
W.~Gautschi, ``The incomplete gamma functions since tricomi,'' \emph{Atti dei
  Convegni Linci}, no. 1998, pp. 203--237, 2011.

\bibitem{9940392}
S.~Zhang, Y.~Cui, and W.~Chen, ``Joint detection for massive grant-free access
  via bigamp,'' in \emph{Proc. IEEE ISWCS}, 2022, pp. 1--6.

\bibitem{RN442}
G.~Kaddoum, Y.~Nijsure, and H.~Tran, ``Generalized code index modulation
  technique for high-data-rate communication systems,'' \emph{IEEE Trans. Veh.
  Technol.}, vol.~65, no.~9, pp. 7000--7009, 2016.

\bibitem{RN105}
L.~You, X.~Gao, X.-G. Xia, N.~Ma, and Y.~Peng, ``Pilot reuse for massive {MIMO}
  transmission over spatially correlated rayleigh fading channels,'' \emph{IEEE
  Trans. Wireless Commun.}, vol.~14, no.~6, pp. 3352--3366, 2015.

\end{thebibliography}

\end{document}